\documentclass[twocolumn]{emulateapj}
\usepackage{color}
\usepackage{amsmath,graphicx}
\usepackage{graphics}
\usepackage{epsf}
\usepackage{epstopdf}
\input{epsf}
\usepackage{epsfig}
\usepackage{color}
\DeclareGraphicsExtensions{.jpg,.pdf,.png,.eps,.ps}
\usepackage{hyperref}
\usepackage{cleveref}
\usepackage{ulem}
\usepackage{wrapfig}
\usepackage{enumitem}
\hypersetup{
    bookmarks=true,         
    unicode=false,          
    pdftoolbar=true,        
    pdfmenubar=true,        
    pdffitwindow=false,     
    pdfstartview={FitH},    
    pdftitle={My title},    
    pdfauthor={Author},     
    pdfsubject={Subject},   
    pdfcreator={Creator},   
    pdfproducer={Producer}, 
    pdfkeywords={keyword1} {key2} {key3}, 
    pdfnewwindow=true,      
    colorlinks=true,       
    linkcolor=red,          
    citecolor=blue,        
    filecolor=magenta,      
    urlcolor=cyan           
}

\def\red#1 {\textcolor{red}{#1}\ }   
\def\green#1 {\textcolor{green}{#1}\ }   
\def\cyan#1 {\textcolor{cyan}{#1}\ }

\def\gs{\mathrel{\raise0.35ex\hbox{$\scriptstyle >$}\kern-0.6em \lower0.40ex\hbox{{$\scriptstyle \sim$}}}}
\def\ls{\mathrel{\raise0.35ex\hbox{$\scriptstyle <$}\kern-0.6em \lower0.40ex\hbox{{$\scriptstyle \sim$}}}}
\newcommand{\Msolar}{\mbox{$M_{\odot}\,$}}
\newcommand{\Lsolar}{\mbox{$L_{\odot}\,$}}
\newcommand{\arcmins}{\mbox{$^{\prime}$}}
\newcommand{\arcsecs}{\mbox{$^{\prime\prime}$}}

\shorttitle{\textit{HIFI} observations of Water in nuclei of actively star forming galaxies}
\shortauthors{Liu et al.}

\begin{document}


\title{\textit{HIFI} Spectroscopy of ${\rm H_2O}$ submm Lines in Nuclei of Actively Star Forming Galaxies}


\author{
 L.~Liu\altaffilmark{1,2,3}, 
 A.~Wei$\ss$\altaffilmark{2},
 J. P.~Perez-Beaupuits\altaffilmark{2,4}, 
 R.~G\"{u}sten\altaffilmark{2}, 
 D.~Liu\altaffilmark{1,3}, 
 Y.~Gao\altaffilmark{1}, 
 K. M.~Menten\altaffilmark{2},
 P.~van der Werf\altaffilmark{5},
 F. P.~Israel\altaffilmark{5},
 A.~Harris\altaffilmark{6},
 J.~Martin-Pintado\altaffilmark{7},
 M. A.~Requena-Torres\altaffilmark{2,6},
 J.~Stutzki\altaffilmark{8}
}
\altaffiltext{1}{Purple Mountain Observatory, Key Lab of Radio Astronomy, 2 West Beijing Road, 210008 Nanjing, PR China}
\altaffiltext{2}{Max-Planck-Institut f\"ur Radioastronomie, Auf dem H\"ugel 69, D-53121 Bonn, Germany}
\altaffiltext{3}{University of Chinese Academy of Sciences, 19A Yuquan Road, PO Box 3908, 100039 Beijing, PR China}
\altaffiltext{4}{European Southern Observatory, Santiago, Chile}
\altaffiltext{5}{Sterrewacht Leiden, Leiden University, PO Box 9513, 2300 RA,
Leiden, The Netherlands}
\altaffiltext{6}{Department of Astronomy, University of Maryland, College Park,
MD 20742, USA}
\altaffiltext{7}{Consejo Superior de Investigaciones Cientificas, Spain}
\altaffiltext{8}{Physikalisches Institut der Universit{\"a}t zu K{\"o}ln, Z{\"u}lpicher Stra{\ss}e 77, D-50937 K{\"o}ln, Germany}
\email{ljliu@pmo.ac.cn, aweiss@mpifr-bonn.mpg.de}

\begin{abstract}
We present a systematic survey of multiple velocity-resolved H$_2$O spectra using \textit{Herschel}/{\it HIFI}
towards nine nearby actively star forming galaxies. 
The ground-state and low-excitation lines (E$_{\rm up}\,\le$130\,K) show 
profiles with emission and absorption blended together, while absorption-free 
medium-excitation lines (130\,K\,$\le\,$E$_{\rm up}\,\le$350\,K) typically display line 
shapes similar to CO. 
We analyze the \textit{HIFI} observation together with archival \textit{SPIRE}/\textit{PACS} 
H$_2$O data using a state-of-the-art 3D radiative transfer code which includes 
the interaction between continuum and line emission.
The water excitation models are combined with information on the dust- and CO 
spectral line energy distribution to determine the physical structure of the interstellar 
medium (ISM). We identify two ISM components
that are common to all galaxies:  A warm ($T_{\rm dust}\,\sim\,40-70$\,K), dense
($n({\rm H})\,\sim\,10^5-10^6\,{\rm cm^{-3}}$) phase which dominates the emission 
of medium-excitation H$_2$O lines. This gas phase also dominates the FIR emission 
and the CO intensities for $J_{\rm up} > 8$. In addition a cold 
($T_{\rm dust}\,\sim\,20-30$\,K), dense ($n({\rm H})\sim\,10^4- 10^5\,{\rm cm^{-3}}$)
more extended phase is present. It outputs the emission in the low-excitation H$_2$O 
lines and typically also produces the prominent line absorption features. 
For the two ULIRGs in our sample (Arp 220 and Mrk 231) 
an even hotter and more compact (R$_s\,\le\,100$ pc) region is present which is 
possibly linked to AGN activity.  We find that collisions dominate the water excitation 
in the cold gas and for lines with $E_{\rm up}\le300$\,K and $E_{\rm up}\le800$\,K in 
the warm and hot component, respectively. Higher energy levels are mainly 
excited by IR pumping. 
\end{abstract}



\section{Introduction}
Galactic nuclei play a key role in our understanding of galactic evolution.
An important method to determine their physical and chemical conditions
is the analysis of molecular emission lines 
from the interstellar medium (ISM).
Of particular interest is the water molecule, which has been
demonstrated to have uniquely powerful potential of deriving information on
the ISM of external galaxies \citep[e.g.,][]{GA2010}.
The abundance of water in the gas phase ($[{\rm H_2O}]/[{\rm H_2}]$, $X({\rm H_2O}$)) 
in quiescent molecular clouds is quite low
as suggested by studies in the Milky Way \citep[e.g., $X(\rm {H_2O}) < 1 
\times 10^{-9}$,][]{caselli2010}. 
But water becomes one of the most (third) abundant species in 
the shock-heated regions \citep[e.g.,][]{bergin2003, GA2013}
and in the dense warm regions in which radiation from newly-formed
stars raises the dust temperature above the ice evaporation temperature
\citep[e.g.,][]{cernicharo2006b}.
Therefore, unlike other molecular gas tracers traditionally used 
to study the dense, star-forming (SF) ISM in extragalactic systems 
(such as CO and HCN), water probes the gas exclusively 
associated with SF regions or heated in the extreme 
environment of active galactic nuclei (AGN).
Because of its complex energy level structure and large level spacing, 
${\rm H_2O}$ possesses a large number of rotation lines that lie mostly
in the submillimeter (submm) and far-infrared (FIR) wavelength regime.
These lines can be very prominent 
in actively star-forming galaxies with intensities comparable to
those of CO lines - much more prominent than other dense gas tracers 
such as HCN \citep[e.g.,][]{vanderwerf2011}.
The water lines do not only probe the physical conditions of the 
gas phase ISM (such as gas density and 
kinetic temperature), but also provide important clues on
the dust IR radiation density as both collision with hydrogen molecule
and IR pumping are important for their excitation 
\citep[e.g.,][]{weiss2010, GA2012, GA2014}.
The high-excitation water lines can even be used to reveal the presence of
extended infrared-opaque regions in galactic nuclei and
probe their physical conditions \citep{vanderwerf2011}.
This offers a potential diagnostic to distinguish AGN from starburst activity.
Observations of water also shed light on the dominant chemistry in
nuclear regions \citep[e.g.,][]{bergin1998, bergin2000, melnick2000}
as water could be a major reservoir of gas-phase interstellar oxygen
\citep[e.g.,][]{cernicharo2006a}.
Overall, water provides a unique tool to probe the 
physical and chemical processes occurring 
in the galaxy nuclei and their surroundings 
\citep[e.g.,][]{vanderwerf2011, GA2014}.

However, previous observations of water in nearby extragalactic systems
suffered great limitations.
Ground-based observations of water in nearby galaxies have
been limited to radio maser transitions (such as the famous 22 GHz 
water line) or to a few systems with significant redshift 
\citep[e.g.,][]{combes1997, cernicharo2006b, menten2008}
due to the absorption by terrestrial atmospheric water vapour.
Earlier satellite missions, such as ODIN and SWAS, did not have enough 
collecting area to detect the relatively faint ground transitions of water in 
external galaxies. ISO and, more recently, Spitzer have provided the 
first systematic studies of water in the far-infrared regime 
\citep[e.g.,][]{fischer1999, GA2004}.
These missions, however, did not cover the frequencies of the molecule's
ground-state transitions and other low-excitation\footnote{Throughout this paper, 
we use the term low-excitation for H$_2$O lines with upper level energies
$E_{\rm up} \le 130$~K,  medium-excitation for lines with $130 < E_{\rm up} \le 350$\,K 
and high-excitation for lines with $E_{\rm up}> 350$~K.} lines.
These low-excitation water transitions provide crucial information
on the widespread diffuse medium in galaxies 
\citep{weiss2010, vandertak2016}.
Only with the launch of \textit{Herschel}\footnote{\textit{Herschel} is an ESA 
space observatory with science instruments provided by European-led 
Principal Investigator consortia and with important participation from NASA.},
with its large collecting area,  have these transitions become accessible 
in the nearby universe \citep[e.g.,][]{GA2010, weiss2010}.
Yet, \textit{SPIRE} (and also \textit{PACS}) on-board \textit{Herschel} does not provide the 
spectral resolution to obtain velocity resolved spectra and only the
integrated line intensities (or barely resolved spectra) can be obtained
from these observations.

High velocity resolution spectroscopy with \textit{Herschel's} Heterodyne
Instrument for the Far Infrared (\textit{HIFI}), however, 
allows us to derive detailed information on the shapes of H$_2$O lines, 
which is critical because emission and absorption are often mixed in 
water line profiles \citep{weiss2010}. 
This implies that the modest spectral resolution of the 
\textit{Herschel/SPIRE} spectroscopy results in severe limitations for the detections of
low-excitation lines 
and limits the construction
of excitation models, since emission and absorption from different ISM 
components along the line of sight are averaged.
Recently, water has been detected in high-$z$ sources
with both high spectral and spatial resolution
afforded by \textit{ALMA} and \textit{NOEMA} 
\citep[e.g.,][]{omont2011,omont2013,combes2012,yang2016}.
The results confirm that ${\rm H_2O}$ lines are among the strongest 
molecular lines in high-$z$ ultra-luminous starburst galaxies, with intensities almost 
comparable to those of the high-$J$ CO lines \citep[e.g.,][]{omont2013, yang2016}.
In order to obtain a better understanding of observed water spectra
in the early universe, a comprehensive analysis of water line shapes
in the local universe is required.
Only with \textit{HIFI}, we are able to investigate multiple 
water transitions resulting from levels with a wide range of energies 
in nearby galaxies in more detail than ever before.

The observed water line profiles provide crucial 
information on the geometry, dynamics and physical structure of the ISM.
However, retrieving these information is not straightforward, 
because most water lines have high optical depth 
\citep[e.g.,][]{emprechtinger2012,poelman2007a,poelman2007b}
so that column densities cannot 
be accurately derived from the observed line intensities alone.
The excitation of water is also more complicated than other traditional
gas tracers (e.g., CO, CS) as IR pumping has to be taken into account.
The gas-phase ${\rm H_2O}$ could be a major coolant of the dense,
star-forming ISM in case it is mainly collisionally excited.
Yet, the relative importance of collision and IR pumping on the
excitation of water in extragalactic sources
has not achieved a full understanding.
Interstellar chemistry will benefit from an accurate knowledge of water
abundances, the derivation of which requires detailed modelling of 
H$_2$O's excitation of the rotational levels.
Hence, to extract the underlying physical properties of the ISM (both gas 
and dust), to investigate the relative contribution of the two 
excitation channels and derive chemical abundances,
a detailed modelling of the water excitation is required.

In this paper we present velocity-resolved \textit{HIFI} spectroscopy
of multiple FIR ${\rm H_2O}$ lines (with upper energy $E_{\rm up}
\sim 50 - 450$ K) in a sample of nine local galaxies with different
nuclear environments.  We analyse the data using a 3D, non-LTE
radiative transfer code. Our main goal is to deepen our understand of
the water excitation and to explore ${\rm H_2O}$ as a diagnostic tool
to probe the physical and chemical conditions in the nuclei of active
star-forming galaxies.  We present our sample, observations and data
reduction in Section\ \ref{section:Observation}. A discussion of the
line shapes is presented in Section\ \ref{section:spectral analysis}.
A description of our modelling method and a summary of our general
model results is given in Section\ \ref{section:modelling}. In Section
\ref{section: Discussion} we discuss the contributions from collisions
and IR pumping on the excitation of water
as well as the resulting shape of the ${\rm H_2O}$
SLEDs,  and establish a $L_{\rm H_2O} - L_{\rm FIR}$ luminosity
relation. Our conclusions are summarised in Section\ \ref{section: summary and conclusion}.

\section{Observation} \label{section:Observation}

Our sample is selected from the HEXGAL (\textit{Herschel} ExtraGALactic) key project
(PI: G\"{u}sten). 
HEXGAL is a project that aims to study the physical 
and chemical composition of the ISM in galactic nuclei, 
utilising the very high spectral resolution of the \textit{HIFI} instrument.
Our sample consists of a total of nine galaxies and has been selected 
to cover a diversity of nuclear environments ranging from pure 
nuclear starburst galaxies (such as M82, NGC 253) to starburst nuclei that 
also host an AGN (such as NGC 4945) to
AGN dominated environments (such as Mrk 231) and to
major mergers with even higher IR luminosity (such as
Arp 220).
The source names, systemic velocities, distances, 
FIR \citep[$40 - 120 ~\mu$m, ][]{helou1985}  
luminosities and galaxy types are given
in Table\ \ref{table:sample_galaxies}.
The FIR luminosities are computed by 
integrating our fitted SEDs over the wavelength range 
$40 - 120~\mu$m (see Section\ \ref{section: IR and submm data} 
for more details on the dust SED fitting).

\begin{deluxetable*}{lcccccl} 
\tabletypesize{\small}
\renewcommand{\arraystretch}{0.4}
\setlength{\tabcolsep}{1pt}
\tablecolumns{6}
\tablewidth{0pc}
\tablecaption{Sample galaxies} 
\tablehead{ \colhead{Galaxy} & \colhead{v$_{\rm LSR}$} & 
\colhead{distance} & \colhead{$L_{\rm FIR}$(FWHM=40\arcsec)} 
& \colhead{RA} & \colhead{Dec} & type \\
& \colhead{$[{\rm km~s^{-1}}]$} & \colhead{[Mpc]}  & 
\colhead{[$Log~\Lsolar$]} & \colhead{h m s.s} 
& \colhead{\ensuremath{\mathrm{deg}}~\arcmins~\arcsecs}&  }
\startdata
M82 & 203 & 3.9 & 9.74 & 09 55 52.2 & +69 40 46 & SB \\
NGC 253 & 243 & 3.2 & 9.47 & 00 47 33.1 & $-$25 17 17 & SB \\
NGC 4945 & 563 & 3.9 & 10.70 & 13 05 27.4 & $-$49 28 05& SB/AGN \\
NGC 1068 & 1137 & 12.6 & 10.32 & 02 42 40.7  & $-$00 00 47 & AGN/SB \\
Cen A & 547 & 3.7 & 9.23 & 13 25 27.6 & $-$43 01 08  & AGN/SB \\
Mrk 231 & 12642 & 186 & 12.19 & 12 56 14.2 & +56 52 25 & AGN/SB \\
Antennae & 1705 & 21.3 & 9.69 & 12 01 54.8 & $-$18 52 55 & SB, Major Merger \\
NGC 6240 & 7339 & 106 & 11.81 & 16 52 58.8 & +02 24 03 & AGN/SB, Major Merger \\
Arp 220 & 5434 & 78.7 & 11.98 & 15 34 57.2  & +23 30 11 & SB/AGN, Major Merger 
\enddata
\tablecomments{The FIR luminosities are computed by 
integrating our fitted SEDs over the wavelength range 
$40 - 120~\mu$m. 
The last column indicates whether the IR luminosity of a galaxy is dominated by 
starburst (SB), AGN or both, and whether the galaxy is a major merger.}
\label{table:sample_galaxies}
\end{deluxetable*}

\begin{figure}[t]
\includegraphics[height=5.8cm]{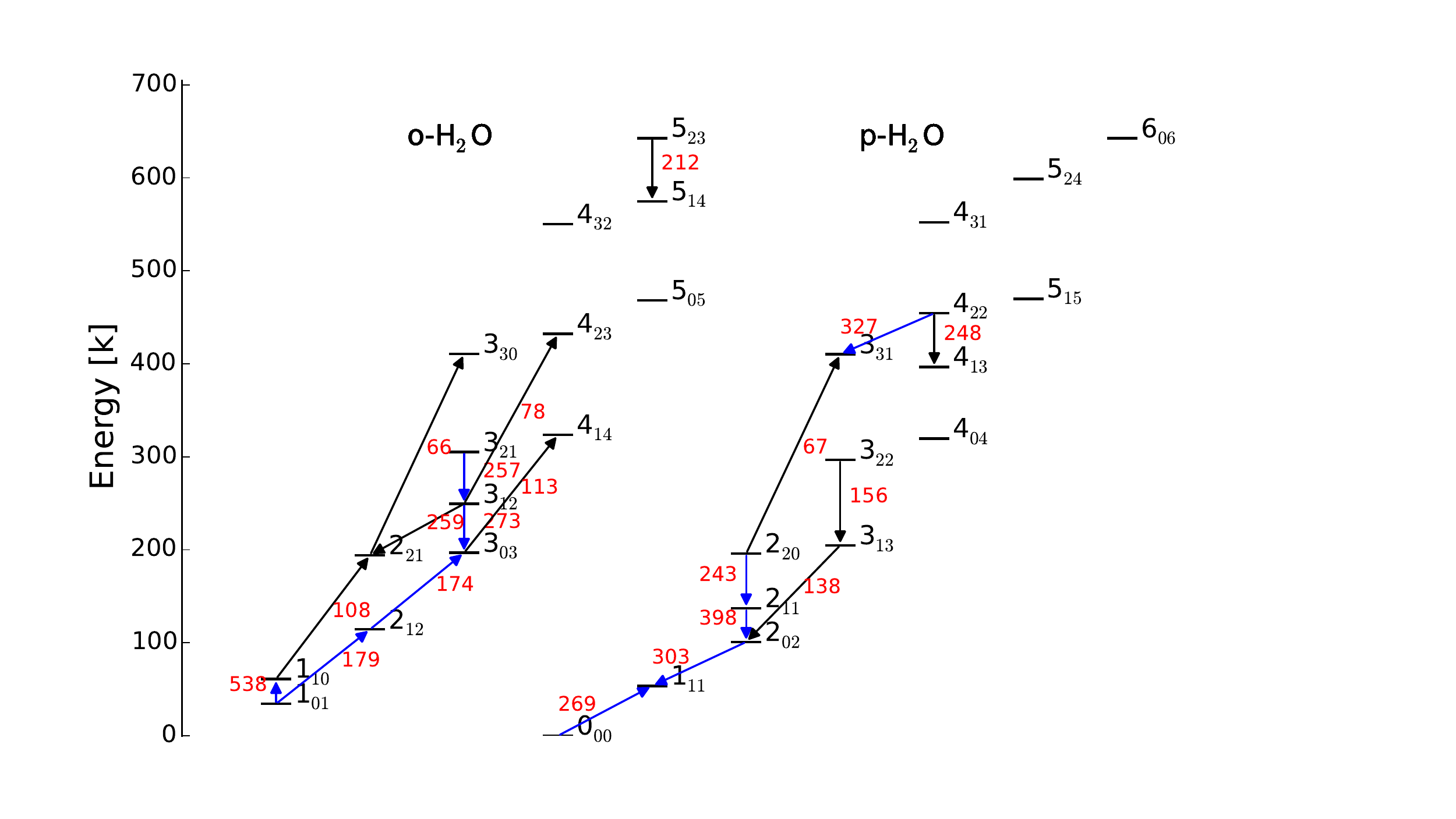}
\caption{Energy level diagrams of ${\rm H_2O}$ (ortho and para). 
Blue arrows indicate lines observed with \textit{HIFI} and
black arrows denote lines observed with \textit{SPIRE}/\textit{PACS}
(data taken from literature).
The downward arrows indicate the lines that are always detected in emission,
while the upward arrows indicate the lines that are often observed in absorption.
The red number denotes the wavelength (in $\mu$m) of each transition.}
\label{figure:energy_diagram}
\end{figure}

We have utilised \textit{HIFI} to observe five to ten carefully selected 
(both ortho- and para-) water transitions.
Fig.\ \ref{figure:energy_diagram} shows the water energy diagram. Transitions observed 
with \textit{HIFI} are indicated by blue arrows whereas black arrows denote 
additional ${\rm H_{2}O}$ lines covered by \textit{Herschel} \textit{SPIRE} and \textit{PACS} 
that are also included in our modelling (more details in Section\ \ref{section:spire_paces h2o data}).
Our observed lines cover a wide energy range, from low-excitation transitions
(with $E_{\rm up} \le 130$ K) to medium-excitation transitions (with $130 < E_{\rm up} \le 350$ K)
to high-excitation transitions ($E_{\rm up} \sim 350 - 450$ K).
Table.\ \ref{table:selected_lines} reports our selected water transitions, 
the line frequencies, the energies of upper levels, the corresponding \textit{HIFI} 
beam sizes, the galaxies toward which each line has been observed,
whether emission or absorption is found and the detection rate. 
The frequencies of our selected lines almost span the full \textit{HIFI}
frequency coverage of Bands 1-5 ($480-1250$ GHz) and Band 6 ($1410-1910$ GHz).
The angular resolutions changes from $\sim 40 \arcsec$ for the o-${\rm H_{2}O}$ 
557 GHz line to $\sim 13 \arcsec $ for the o-${\rm H_2O}$ 1717 GHz line.
We observed each galaxy towards a single position given in Table\ \ref{table:sample_galaxies}.
Thus, except for the most distant sources (Arp 220, Mrk 231 and NGC 6240),
only the nuclear region is covered by our pointed observations. 

\begin{deluxetable*}{crrcclc} 
\tabletypesize{\small}
\renewcommand{\arraystretch}{0.4}
\setlength{\tabcolsep}{1pt}
\tablecolumns{6}
\tablewidth{0pc}
\tablecaption{Selected Water Transitions} 
\tablehead{ \colhead{Line} & \colhead{Freq.} & 
\colhead{$E_{\rm up}$} & \colhead{$FWHM$} 
 & \colhead{observed galaxies} & Emission or Absorption$^{a}$ & detection rate$^{b}$ \\
 & \colhead{[GHz]} & \colhead{[K]}  & 
\colhead{[$^{\prime\prime}$]} & & & }
\startdata
p-${\rm H_2O}$ (1$_{11}$-0$_{00}$) & 1113 & 53.4 & 19 & all & absorption, emission & 7/9 \\
o-${\rm H_2O}$ (1$_{10}$-1$_{01}$) & 557 & 61.0 & 40 & all & absorption, emission & 8/9 \\
p-H$_2$O (2$_{02}$-1$_{11}$) & 988 & 100.8 & 22 & all & emission & 8/9 \\
o-H$_2$O (2$_{12}$-1$_{01}$) & 1670 & 114.4 & 13 & NGC 253, NGC 4945 & absorption & 2/2 \\
p-H$_2$O (2$_{11}$-2$_{02}$) & 752 & 136.9 & 28 & all & emission & 6/9 \\
p-H$_2$O (2$_{20}$-2$_{11}$) & 1229 & 195.9 & 17 & NGC253, CenA & emission & 1/2 \\
o-H$_2$O (3$_{03}$-2$_{12}$) & 1717 & 196.8 &  12 & NGC 253, NGC 4945 & absorption, emission & 2/2 \\
o-H$_2$O (3$_{12}$-3$_{03}$) & 1097 & 249.4 & 19 & all but NGC 1068 & emission & 6/8 \\
o-H$_2$O (3$_{21}$-3$_{12}$) & 1163 & 305.3 & 18 & NGC 4945/253/6240, CenA & emission & 2/4 \\
p-H$_2$O (4$_{22}$-3$_{31}$) & 916 & 454.3 & 23 & all &  emission & 1/9
\enddata
\tablecomments{$a -$  whether a line has been detected in emission, 
absorption, or both in our sample galaxies; $b-$ the number of detected galaxies divided by 
the number of observed galaxies. }
\label{table:selected_lines}
\end{deluxetable*}

The data was obtained between March 2010 and September 2012, in a
total of 124 hours of integration time.  The dual beamswitch mode was
used with a wobbler throw of 3$\arcmin$ for all observations.  The
data was recorded using the wide-band acousto-optic spectrometer,
consisting of four units with a bandwidth of 1 GHz each, covering the
4 GHz intermediate frequency band (IF) for each polarization with a
spectral resolution of 1 MHz.  Our spectra were calibrated using
HIPE\footnote{Version 10.0.0. \textit{HIPE} is a joint development by the
\textit{Herschel} Science Ground Segment Consortium, consisting of
ESA, the NASA \textit{Herschel}.}  and then exported to
CLASS\footnote{http://www.iram.fr/IRAMFR/GILDAS} format with the
shortest possible pre-integration. For each scan we computed the
underlying continuum using the line-free channels of a combined 4 GHz
spectrum from the four sub-bands.  Then a first-order baselines was
subtracted from each individual sub-band (in the cases where the
signal spans over more than one sub-band, the nearby sub-bands were
merged before subtracting the baselines).  Next the
baseline-subtracted sub-bands in each scan were combined and the
continuum level was added again.  The noise-weighted spectra from two
polarizations (H and V) were thereafter averaged.  Note that the
continuum radiation enters the receiver through both sidebands while
the line is only in one sideband.  Therefore the continuum used in our
analysis (and for our figures) represents half of the value actually
measured by \textit{HIFI}.  The o-${\rm H_{2}O}$
($3_{21}-3_{12}$) 1163 GHz line of NGC 4945 was found to partly blend
with CO ($J=10-9$) line, we therefore have estimated the CO ($J=10-9$)
line profile from the APEX CO ($J=3-2$) line and subtracted it from
the spectra.

\section{Spectral Results and Analysis} \label{section:spectral analysis}

We detected strong water emission and absorption in all galaxies
except for the Antennae, which has no detection in any ${\rm H_{2}O}$
line.  Our \textit{HIFI} ${\rm H_{2}O}$ spectra are presented in
Figs.\ \ref{figure:m82_spectra}, \ref{figure:ngc253_spectra},
\ref{figure:ngc4945_spectra}, \ref{figure:ngc1068_spectra},
\ref{figure:cena_spectra}, \ref{figure:mrk231_spectra}, 
\ref{figure:ngc6240_spectra},  \ref{figure:arp220_spectra}.
The velocity scale on each panel is relative to the systemic
velocity listed in Table\ \ref{table:sample_galaxies}.  
Except for a few sources (Mrk 231,
NGC 1068 and NGC 6240), a wide variety of line shapes is observed for
most galaxies in our sample (e.g., NGC 4945, NGC 253, M82). 
In the latter cases emission and absorption features are often blended.
Unlike line profiles from multiple transitions of other molecules
(such as CO), the line profiles of water cannot be assumed to be similar.

\subsection{Line Shapes} \label{}
\subsubsection{Emission Lines} \label{} 

\begin{figure*}[t]
\includegraphics[scale=0.40]{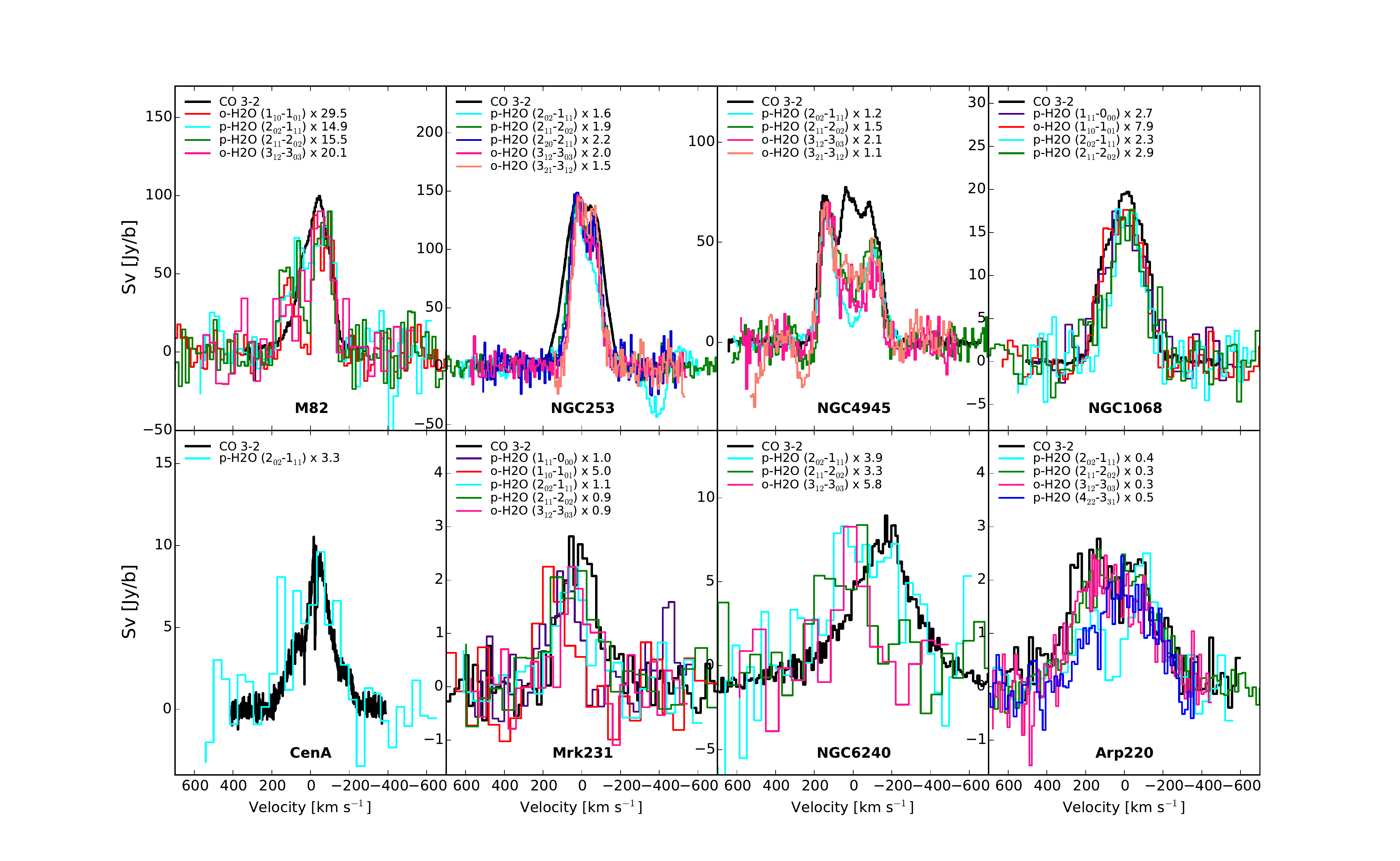}
\caption{Water emission line profiles are superposed on the CO($J = 3-2$)
	line.
The velocity scale is relative to the systemic velocity of each galaxy. 
The water line profiles are scaled to provide closest matches to the peak value of
CO profile. }
\label{figure:Comparison_CO}
\end{figure*}

A few of the lines (indicated by blue downward solid arrows in Fig.\
\ref{figure:energy_diagram}) are always detected in emission.  
They include a low-excitation line (p-${\rm H_{2}O}$
($2_{02}-1_{11}$)), four medium-excitation lines
(o-${\rm H_{2}O}$ ($3_{12}-3_{03}$), o-${\rm H_{2}O}$
($3_{21}-3_{12}$), p-${\rm H_{2}O}$ ($2_{11}-2_{02}$), 
and p-${\rm H_{2}O}$ ($2_{20}-2_{11}$)) and a
high-excitation line (p-${\rm H_{2}O}$ ($4_{22}-3_{31}$)). 
These emission lines display similar line shapes among each other and also show
a good correspondence to the line profile of CO.
Fig.\ \ref{figure:Comparison_CO} presents the CO ($J=3-2$) line obtained 
by APEX\footnote{This publication is based in part on data acquired with the Atacama 
Pathfinder Experiment (APEX). APEX is a collaboration between the 
Max-Planck-Institut f\"ur Radioastronomie, the European Southern Observatory, 
and the Onsala Space Observatory.}  ($FWHM \sim 20 \arcsec$) or
JCMT ($FWHM \sim 14 \arcsec$) overlaid on the HIFI detected ${\rm
H_{2}O}$ emission lines. 
All H$_{2}$O line profiles in  Fig.\ \ref{figure:Comparison_CO} have been scaled to 
the peak of the CO line for better visualisation of the line shapes. 
One can see that, except for NGC
253 whose water line profile is slightly narrower than the CO ($3-2$)
profile (see Appendix\ \ref{appendix: detailed models for individual
galaxies} for more discussions on this), water is often detected over
the full velocity range of CO.  This suggests that water is as widespread as CO 
and likely traces the bulk of the molecular gas in the central region of galaxies.

The closest resemblance is found between the CO and
four medium-excitation ${\rm H_{2}O}$ lines which have 
$E_{\rm up} \simeq 130 - 305$\,K above the ground-state.  
The high-excitation emission line -
p-${\rm H_{2}O}$ ($4_{22}-3_{31}$) (with $E_{\rm up} \simeq 450$ K) -
which has been detected only in Arp 220, displays a narrower velocity
dispersion ($235 \pm 18~{\rm km~s^{-1}}$) compared with that of the CO 
and medium-excitation ${\rm H_{2}O}$ lines ($412 \pm
32~{\rm km~s^{-1}}$) (see Fig.\ \ref{figure:Comparison_CO}).  The
low-excitation emission line p-${\rm H_{2}O}$ ($2_{02}-1_{11}$) (with
$E_{\rm up} \simeq 100$ K) often exhibits diminished emission compared
to CO at the velocities where ground-state absorptions are detected,
implying that the line is partly absorbed at
the same velocities.

\subsubsection{Absorption Lines} \label{}

We have found four lines with absorption features in at least one
galaxy of our sample. These are the p-${\rm H_{2}O}$ ground-state
($1_{11}-0_{00}$) line, the o-${\rm H_{2}O}$
ground-state ($1_{10}-1_{01}$) line, the o-${\rm H_{2}O}$
($2_{12}-1_{01}$) line and the o-${\rm H_{2}O}$
($3_{03}-2_{12 }$) line (see blue upward arrows in Fig.\
\ref{figure:energy_diagram}).
Except for the o-${\rm H_{2}O}$ ($3_{03}-2_{12}$) line that has
$E_{\rm up} \simeq 195$ K, all other absorption lines occur between
low energy levels ($50 \le E_{\rm up} \le 115$ K).
The two ground-state ${\rm H_{2}O}$ lines show absorptions towards all galaxies
except for Mrk 231, NGC 1068 and NGC 6240.  We further find that the
absorption depth of the o-${\rm H_{2}O}$ ground-state line is usually
much weaker ($10\%-25\%$) than that of the p-${\rm H_{2}O}$
ground-state line.  The other two absorption lines (o-${\rm H_{2}O}$
($2_{12}-1_{01}$) and o-${\rm H_{2}O}$ ($3_{03}-2_{12}$)) have only
been observed towards NGC 253 and NGC 4945. Their line shapes are
similar to the absorption feature of the p-${\rm H_{2}O}$ ground-state line.

The observed absorption features can appear to be either broad and deep
(e.g., Arp220 and NGC 4945),  or narrow and shallow (e.g., Cen A).
In some galaxies (e.g., NGC 253 and NGC 4945), 
the low-excitation absorption feature covers a velocity range matching
that of medium-excitation ${\rm H_{2}O}$ emission lines,
while in some other galaxies (e.g., M82) the absorption feature
occurs at a velocity that does not show emission in other lines.

Absorption and emission features are often found to be blended.
Especially for the o-${\rm H_{2}O}$ ground-state line, strong emission
is detected in all of our sample galaxies, in particular towards the
high and low velocity wings of the line profile.  
 Conspicuous emission features also show up in the p-${\rm H_{2}O}$ ground-state 
 line in a few galaxies (e.g., NGC 253 and NGC 1068), although they appear to
be much weaker.  Finally, we find the observed global line profiles
with absorption and emission blended together are best explained by an
emission profile similar to the medium-excitation ${\rm
H_{2}O}$ lines modified by absorption components from foreground gas.
It is therefore tempting to speculate that the lack of absorption at
certain velocities has a geometrical origin, i.e., gas at these
velocities is located outside of sightline of the continuum
\citep{weiss2010}.

\subsection{Gaussian Decomposition of Line Profiles} \label{Gauss-decomp}

The complex water line shapes found in our sample galaxies suggest an ISM
structure with several different physical components. In order to
separate the individual contributions of multiple physical regions and
to disentangle absorption from emission, we have performed a Gaussian
decomposition of the observed ${\rm H_{2}O}$ line profiles.  We first
decompose the absorption free medium-excitation ${\rm H_{2}O}$ emission lines
and the CO(J$=3-2$) line, which typically requires two or
three Gaussian components.

We next fit the remaining ${\rm H_{2}O}$ lines but constrain their
line centroids and widths to narrow ranges centered on 
the thus derived Gaussian fit parameters.
The intensity of each component is then free to vary (from negative to positive). 
This procedure works well
for the galaxies that show only emissions (Mrk 231, NGC 1068 and NGC
6240), and NGC 4945 where the width and velocity centroid of the absorption
feature matches one of the medium-excitation emission components.

For the remaining galaxies, however, one or two additional Gaussian
components are required to fit the profile of the low-excitation
and/or high-excitation lines.  Specifically, we added a component for
M82 and Cen A to match the narrow absorption feature seen at the
galaxy systemic velocity, and a component for NGC 253 to fit
the red-shifted broader emission seen only in the two ground-state
lines. We added two additional components for Arp 220 to match
the absorption feature in the low-excitation lines and the narrower
emission feature evident in the higher excitation p-${\rm H_{2}O}$
($4_{22}-3_{31}$) line, respectively.

The IDL package MPFIT \citep{markwardt2009} was used in the fitting
analysis.  In most cases,  we allow the position of each Gaussian 
component to change by  $\pm~2-3$ ${\rm km~s^{-1}}$ and the line width 
by $\pm~5\%$.  The line 
centroids and widths of Arp 220 are allowed to change by 
$\pm~5$ ${\rm km~s^{-1}}$ and $\pm~8\%$, respectively.
The resulting parameters from Gaussian decomposition are given in 
Table\ \ref{table:gaussian_decomposition}.

\section{Line modelling} \label{section:modelling}

\subsection{Additional Observation Data} \label{}

In order to better constrain the physical parameters of our model, and
also to check the reliability of our final model results, we have
gathered IR and (sub)millimeter wavelength spectroscopy and continuum
data from the literature.  This supplementary data includes
\textit{SPIRE}/\textit{PACS} ${\rm H_{2}O}$ data,  IR and
(sub)millimeter wavelength continuum data ($\nu \sim 1 \times 10^2 -
10^5$ GHz), as well as ground-based and \textit{SPIRE}/\textit{HIFI} 
CO $1 \le J_{\rm up} \le 13$ fluxes.  The dust continuum data is required to constrain the
intrinsic IR radiation field and its effect on water
excitation. The inclusion of the CO data allows us to investigate
to which level the gas traced by water emission is related to the
shape of CO SLEDs.
For extended sources, all data has been scaled to a uniform beam 
size of $40\arcsec$ by applying the correction factors derived 
from IR images (see Sect.\ \ref{section: IR and submm data} for more 
details on the IR images).

\subsubsection{\textit{SPIRE}/\textit{PACS} ${\rm H_{2}O}$ Data} 
\label{section:spire_paces h2o data}
In addition to our \textit{HIFI} ${\rm H_{2}O}$ data, published ${\rm H_{2}O}$ data 
observed by \textit{Herschel}/\textit{SPIRE} and \textit{PACS} 
from literature has also been incorporated into our 
line modelling, with the aim of studying the overall water excitation
across a large number of energy levels.
\textit{SPIRE} combines a 3 color photometer and a low to
medium resolution Fourier Transform Spectrometer (FTS), with
continuous spectral coverage from 190 to 670 $\mu$m 
($\nu \approx 450 - 1550$ GHz) and a spectral resolving power of 
$\approx 1400$.
\textit{SPIRE} ${\rm H_{2}O}$ data was used for transitions
that were not observed/detected by \textit{HIFI}.
\textit{PACS} detects ${\rm H_{2}O}$ lines with frequencies 
higher than those covered by \textit{SPIRE} and \textit{HIFI}
($\nu \sim 1500 - 5000$ GHz), most of which have high excitation
($E_{\rm up} \sim 300 - 650$ K).

H$_2$O transitions, for which we only have \textit{SPIRE} and \textit{PACS}
data are labeled by black arrows in Fig.\ \ref{figure:energy_diagram}. 
Again, downward arrows indicate emission lines, and upward
arrows indicate the transitions that are often detected in absorption 
(some of them appear in emission occasionally).
From Fig.\ \ref{figure:energy_diagram}, we can see that 
the high-excitation lines with frequencies close to the peak of the
dust continuum SED in star-forming/starburst galaxies 
($\lambda \sim 60 - 150~{\rm \mu m}$ or $\nu \sim 2000 - 4500$ GHz) usually 
appear in absorption.
While the lower-frequency  ($\lambda \ge 150~{\rm \mu m}$ or 
$\nu \leq 2000$ GHz) high-excitation lines are often detected in emission.
Note that for the \textit{SPIRE} and \textit{PACS} ${\rm H_{2}O}$ data only
integrated intensities are available. When information on their line shape
is required we use our \textit{HIFI} ${\rm H_{2}O}$ line profiles as a proxy.

\subsubsection{IR and Submm Data} \label{section: IR and submm data}

The submm to IR imaging of our target galaxies is used in two ways.
First the observed distribution of the dust continuum has been used
for our extended sources to compute aperture corrections to compensate
for the different beam sizes of our {\it HIFI} observation as well as
for the other line data used in our analysis. To derive the
aperture corrections we have smoothed the highest spatial resolution
map (typically PACS observations near the peak of the dust SED but in
some cases also 350$\mu$m maps from the Submillimetre Apex Bolometer Camera
(SABOCA, $FWHM=7.5''$)) to different spatial resolutions up to 40$''$
which corresponds to the largest HIFI beam size (in our data set, that
of the H$_2$O (1$_{10}-1_{01}$) 557 GHz line). For each smoothed map we derive the
aperture correction from the ratio of the peak flux relative to the
flux at 40$''$ resolution which allows us to scale all observations to a
common aperture of 40$''$.

Secondly, the observed dust SEDs are used to constrain the dust
continuum models for our target galaxies, which is a crucial
ingredient for the modelling of water excitation. Apart from the IR
fluxes measured by our \textit{HIFI} observations at the ${\rm H_2O}$
line frequencies, we collect submm to IR fluxes in the
frequency range of $\sim 3 \times 10^2 - 10^5$ GHz ($\lambda \sim 3 -
1000~\mu m$) from the observations by
$Spitzer$, $WISE$, $IRAS$,  \textit{Herschel} \textit{PACS}/\textit{SPIRE} and 
$ISO$, as well as APEX 870$\mu$m LABOCA and 350$\mu$m 
SABOCA observations. 
The submm data on the long wavelength (Rayleigh-Jeans) tail enables us to 
better constrain the far-IR SED and the
properties of cold dust in the galaxy \citep[e.g.][]{weiss2008}.
The submm and IR images are gathered for the extended sources.

We compute for each model the full dust SED and compared
it to the observed dust flux densities (see Appendix \ref{appendix:
dust SED model} for detailed description on our approach of dust SED
modelling). Since we cannot be sure that all dust continuum emission
is physically associated with water line emission, we consider a model
more reliable if the predicted dust SED does not exceed the observed dust
continuum intensities.

\subsubsection{CO Data} \label{}

In order to verify that our H$_2$O models are also consistent with
other ISM tracers and to investigate to which level the H$_2$O
emitting volume contributes to the line intensity of other molecules,
we also incorporate CO into our models.  The CO molecule is a good
tracer of overall gas content and excitation because it is mainly
collisionally excited. In addition it is the best studied ISM tracer
in extragalactic sources. The ground-state and low$-J$ ($J_{\rm up} \le 3$) CO data 
has been collected from various sources in the literature \citep[e.g.,][and references
therein]{papadopoulos2012,greve2014}. 
The $J_{\rm up}=4$ to 13 CO line intensities (CO SLED) have been extracted from
archival \textit{SPIRE}/FTS observations
\citep[][]{vanderwerf2010,panuzzo2010,rangwala2011,spinoglio2012,meijerink2013,
papadopoulos2014,rosenberg2014}. For some sources (e.g., M82, NGC 253 and Cen A)
high$-J$ ($J_{\rm up}=5$ to 13) CO lines observed with \textit{HIFI} have
also been collected \citep[see e.g.][]{loenen2010, israel2014}.
The velocity resolved \textit{HIFI}  CO observations allow us to model in detail the CO SLED
for each Gaussian velocity component present in the \textit{HIFI} ${\rm H_2O}$ profiles.

We have calculated the fluxes of CO transitions with $J_{\rm up} = 1$ to 13 for each of our
models and compared them to the observed values. As for the modelling 
of the dust continuum, we require that the model predicted CO intensities from the
H$_2$O emitting volume shall not exceed the observed CO SLED.

\subsection{Basic Model Description} \label{}
\subsubsection{The $\beta$3D code} \label{}

An updated version of the non-LTE 3D radiative transfer code
`$\beta$3D' is used to calculate the excitation and radiative
transfer of the molecular gas species (${\rm H_{2}O}$ and CO in our
work).  $\beta$3D was first developed by \citet{poelman2005,
poelman2006}.  The main advantages of the code are its dimensionality
and speed.  It is not limited to spherical or axis symmetric
problems but allows to model arbitrary 3D structures where a unique
gas and dust temperature, density, and abundance value can be
attributed to every position (i.e., 3D grid cell).  
The code does not suffer from convergence
problems at high optical depth which reduces the computing time,
as it adopts the escape probability method.
The use of a multi-zone formalism, in contrast to a one-zone approach,
allows to calculate excitation gradients within opaque sources.
We here use a modified version of $\beta$3D, where the molecular and 
atomic line intensities and profiles are calculated within a line tracing 
approach for an arbitrary viewing angle \citep{perez-beaupuits2011}.
Numerical results from $\beta$3D have extensively been tested against benchmark 
problems \citep[see][]{vanzadelhoff2002,vandertak2005}.

In the work presented here, we further extended 
the code by implementing the dust emission and absorption in the line 
radiative transfer by adopting the extended escape 
probability method developed by \citet{takahashi1983}. This allows us to 
take the interaction between dust and molecular gas into account.
The dust grains are assumed to be mixed evenly with hydrogen gas 
(assuming a gas to dust mass ratio of $100 : 1$),
and the radiation field from the thermal dust emission is computed from each grid
cell. More details on the extended escape probability method and our default parameter 
setting in $\beta$3D (e.g., dust grain property) are given in
Appendix\  \ref{appendix: the extended escape probability method}.
The resulting global line profile is computed using our newly developed 
ray tracing approach, where the photons at various velocity channels
are integrated through the dust and gas column
along a line of sight within multiple ISM components (for more details on our ray 
tracing approach see Appendix\ \ref{appendix: the ray tracing approach}).

\subsubsection{Applying $\beta$3D to a galaxy using multiple ISM Components} \label{}
Modelling a galaxy as a whole still turns out to be impractical at present,
because building up a galaxy with a detailed 3D geometry structure (e.g., arms, 
rings, disks) and kinematics (e.g., rotation, outflow, inflow) requires a huge 
cube which will result in heavy memory usage and extremely slow computation speed.
With the angular resolution of \textit{HIFI}, our main goal is to investigate the 
observed different ${\rm H_{2}O}$ line shapes and the underlying properties of 
different physical regions, i.e., ISM components.
Therefore, we model a galaxy by utilizing several different ISM components,
assuming each ISM component is an ensemble of molecular clumps with 
identical physical properties.
The equilibrium temperature and level populations of the gas,
however, are calculated within only a single clump based on our 
assumption that the excitation of the molecular gas at a given location
should be connected mainly with gas and dust of the same clump
and barely related to external clumps.
This assumption is reasonable given that the contribution of an external clump to
the local radiation intensity $\langle J_{\nu}\rangle$ at a test point
depends on its spanned solid angle seen by the point (as suggested 
by Equations \ref{equation:Jv} - \ref{equation:eta} 
in Appendix\ \ref{appendix: the extended escape probability method}), 
which is usually negligible considering the small volume filling 
factor of molecular clumps in galaxies.
This assumption is similar to the approach in other radiative transfer
calculations such as large velocity gradient (LVG) models where the
velocity gradient of a clump provides an intrinsic escape mechanism for
photons by Doppler-shifting the frequencies out of the line. Thereby the 
radiative trapping is generally confined to the local region, i.e., the molecular 
clump \citep{takahashi1983}.

A clump has been assumed to be a homogeneous, isothermal cube (grid
size is $20 \times 20 \times 20$) whose main constituents are the
hydrogen molecular gas (${\rm H_{2}}$), dust grains and the molecular
species of interest (${\rm H_{2}O}$ and CO in our work).  Hydrogen is
assumed to be totally molecular in our model because the main part of
the dissociating UV radiation is already absorbed in regions 
where ${\rm H_{2}O}$ is present \citep{poelman2005}.
The thermodynamic equilibrium statistical value of 3 is 
adopted to the water ortho-to-para ratio (OPR).
In order to examine the water excitation under different 
physical conditions, we have generated a grid of clump models by varying 
five free parameters: hydrogen column density of clump 
$N_{\rm clump}({\rm H})$, hydrogen density n(H), gas kinetic
temperature $T_{\rm K}$, dust temperature $T_{\rm dust}$ and ${\rm H_{2}O}$
abundance $X({\rm H_{2}O})$.
However, for simplicity, we fixed the CO abundance to the value of 
$1 \times 10^{-4}$, because it has been found to vary very little
in different molecular clouds in nearby galaxies
\citep[e.g.,][]{elmegreen1980, tacconi1985, france2014, bialy2015}.
In fact, most of the CO lines are found to be optically thick in our sample
galaxies and thereby the modelled CO fluxes are not very sensitive to the 
adopted CO abundance.

With the level populations of ${\rm H_{2}O}$ and CO calculated for
a clump, we next built an ISM component from an 
ensemble of clumps.
The emergent line profile and
dust continuum flux from an ISM component is calculated by our newly
developed ray tracing program (see Appendix\ \ref{appendix: the ray
tracing approach}), which integrates both the line and dust continuum
photons over all the overlapping clumps along a line of sight.  This
procedure is crucial, as the resulting global line profile is not just
a simple superposition of intrinsic line profiles of individual clumps,
especially when the line is optically thick or dust continuum at line
frequency becomes non-negligible.  For example, the gas of foreground
clumps will absorb the emission from clumps in the background
and thereby their contributions to the final line
profile will significantly deviate from the sum of their intrinsic line profiles.
The problem of a simple superposition of line profiles from overlapping
clumps in optically-thick region has been pointed out by several
authors \citep[e.g.,][]{downes1993, aalto2015}.  As a result, the total
column density of ISM component has
significant influence on not only integrated line intensity but
also the line shape (including whether a modelled line appears in 
emission or absorption),
and has therefore been introduced as the sixth free parameter (N(H)).
Since we do not want to involve the detailed galaxy
dynamics in our model, a random normal distribution of velocities was
attributed to the clumps of each ISM component as a
statistical approximation.  The velocity distributions follow the
properties derived from our Gaussian decomposition of \textit{HIFI}
${\rm H_{2}O}$ spectra (see Sect.\,\ref{Gauss-decomp}).

We model each Gaussian-decomposed velocity component of ${\rm H_{2}O}$ 
spectra separately. 
 However, even for a single velocity component, we fail to
fit all observed ${\rm H_{2}O}$ line intensities by using only 
one ISM component.  This implies that different ${\rm H_{2}O}$
lines (at a certain velocity) arise from different physical regions. 
Therefore, we model each velocity component of ${\rm H_{2}O}$ spectra 
with multiple ISM components (i.e. the combination
of different clump properties).  The final model of a galaxy is then
built by adding up the sets of ISM components at various velocities.
If different ISM components do not spatially overlap, the emergent
global line profile is derived by simply adding individual line
profiles of each ISM component together.  Otherwise, the emergent line
and continuum intensities are integrated along all the overlapped ISM
components using the ray tracing approach mentioned above (see
Appendix\ \ref{appendix: the ray tracing approach}), where the foreground
ISM component will absorb both line and dust photons generated by the
background ISM component.

\subsection{General Modelling Results} \label{general_modelling_results}
For each velocity component, we derived the first ISM component by
fitting only the medium-excitation ${\rm H_{2}O}$ emission
lines, because they are most likely to arise from the same physical
region given by their similarities in both energies and line shapes (which
also show a good correspondence with CO line).  Then a second ISM
component is added to fit the ground-state and low-excitation emission line
features which are not accounted for in the first ISM component.  
In the presence of ground-state and low-excitation
absorption features, we utilize the ISM component dominating FIR luminosity
(normally the first component derived by fitting medium-excitation lines) as
the background continuum source and add an front absorbing ISM component
whose physical size is allowed to vary from zero to maximum coverage.  
For the galaxies which have been detected
in the high-excitation lines (Mrk 231 and Arp 220), we require an additional 
ISM component to match the high-excitation line features.  
Our best-fit models are obtained based on fitting to
the \textit{HIFI}-detected ${\rm H_{2}O}$ line profiles and the
\textit{SPIRE}/\textit{PACS} ${\rm H_{2}O}$ data, with additional constraints from 
fitting the observed dust SED and CO SLED.  The detailed fitting procedures are
given in Appendix\ \ref{appendix: select best models}.

We have derived multi-component ISM model for each of the eight
galaxies with ${\rm H_{2}O}$ line detections. 
The best-fit values for each galaxy are presented in Table\
\ref{table: model parameters}, which lists the individual results for
each ISM component at different velocities ($\delta v$).  
The column density of clumps $N_{\rm clump}$(H) (which is allowed to vary from
$1 \times 10^{22} - 1 \times 10^{24}$ ${\rm cm^{-2}}$ during the fitting)
is usually found to be of the order  
$1 \times 10^{23}$ ${\rm cm^{-2}}$.
Details on the model of each galaxy are given in Appendix\
\ref{appendix: detailed models for individual galaxies}.

\subsubsection{Common ISM phases among galaxies}
Although the derived models differ in some details between galaxies, 
we found a few general results that apply to most systems.
First of all, our results reveal two typical ISM components that are shared 
by all galaxies - a warm component and a cold extended region (ER).
The warm component has typical parameters with density
of the order of $\sim 1 \times 10^5 - 10^6~{\rm cm^{-3}}$, gas and dust 
temperature between $40 - 70$\,K and column density around a few times 
$10^{24}~{\rm cm^{-2}}$, while the cold component has density 
of the order of $\sim 1\times 10^4~{\rm cm^{-3}}$, gas and dust temperature 
of $20 - 30$\,K and column density of a few times $10^{23}~{\rm cm^{-2}}$. 
The cold component is found to be much more widespread than the warm component
(see Table\ \ref{table: model parameters}).

\begin{figure*}[h]
\includegraphics[width=\textwidth]{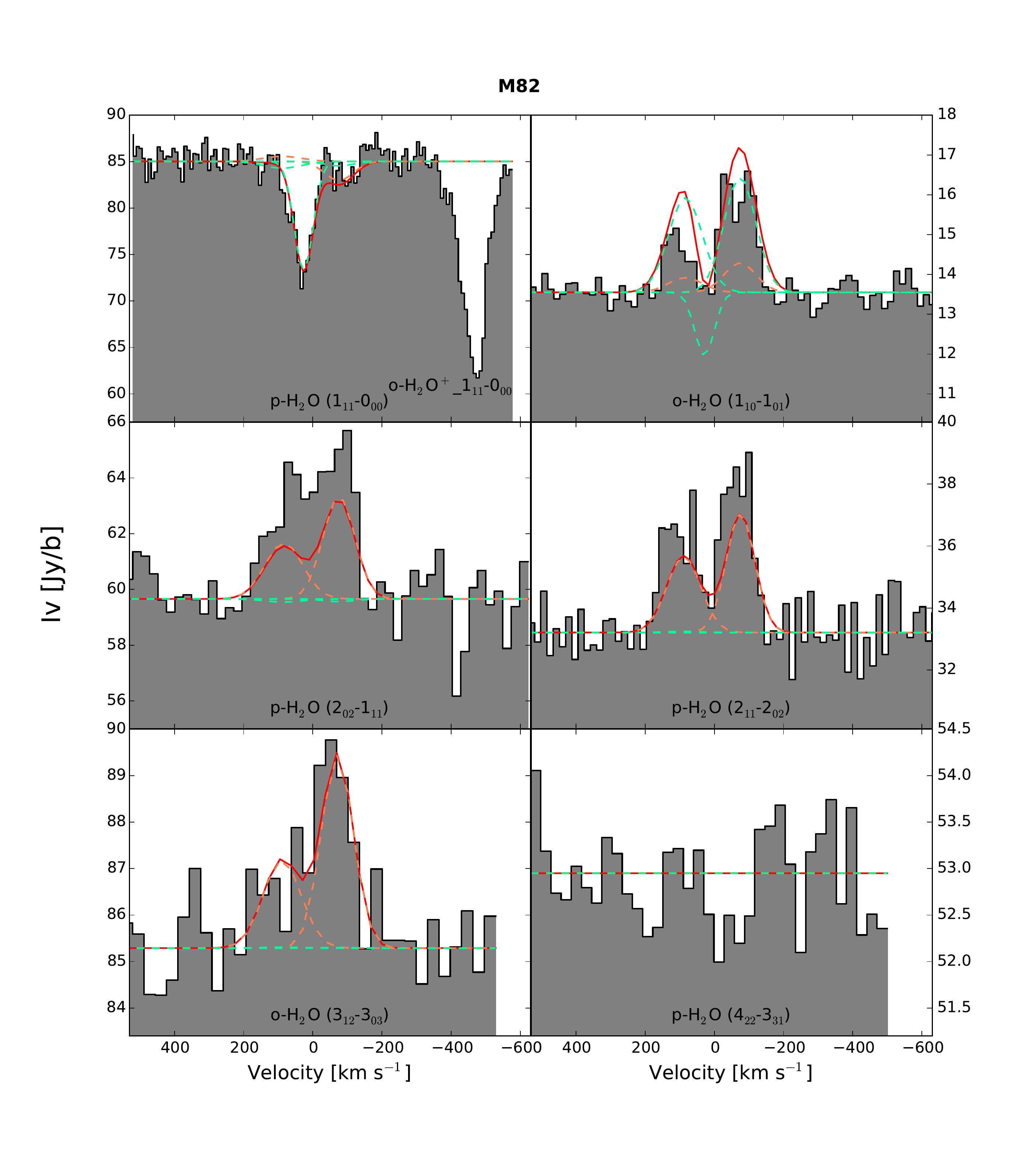}
\caption{The observed ${\rm H_2O}$ spectra (in grey colour) and the 
modelling results (in red solid curve) of M82.
The individual contributions from the warm component and cold ER
are displayed by orange and green colours, respectively.
The horizontal lines indicate the observed continuum flux level.
The panels are ordered according to the upper level energies
of the lines.
On each panel, we shift the systemic velocity of galaxy 
to $v_{\rm LSR}=0~{\rm km~s^{-1}}$.
\label{figure:m82_spectra}}
\end{figure*}

\begin{figure*}[h]
\includegraphics[width=\textwidth]{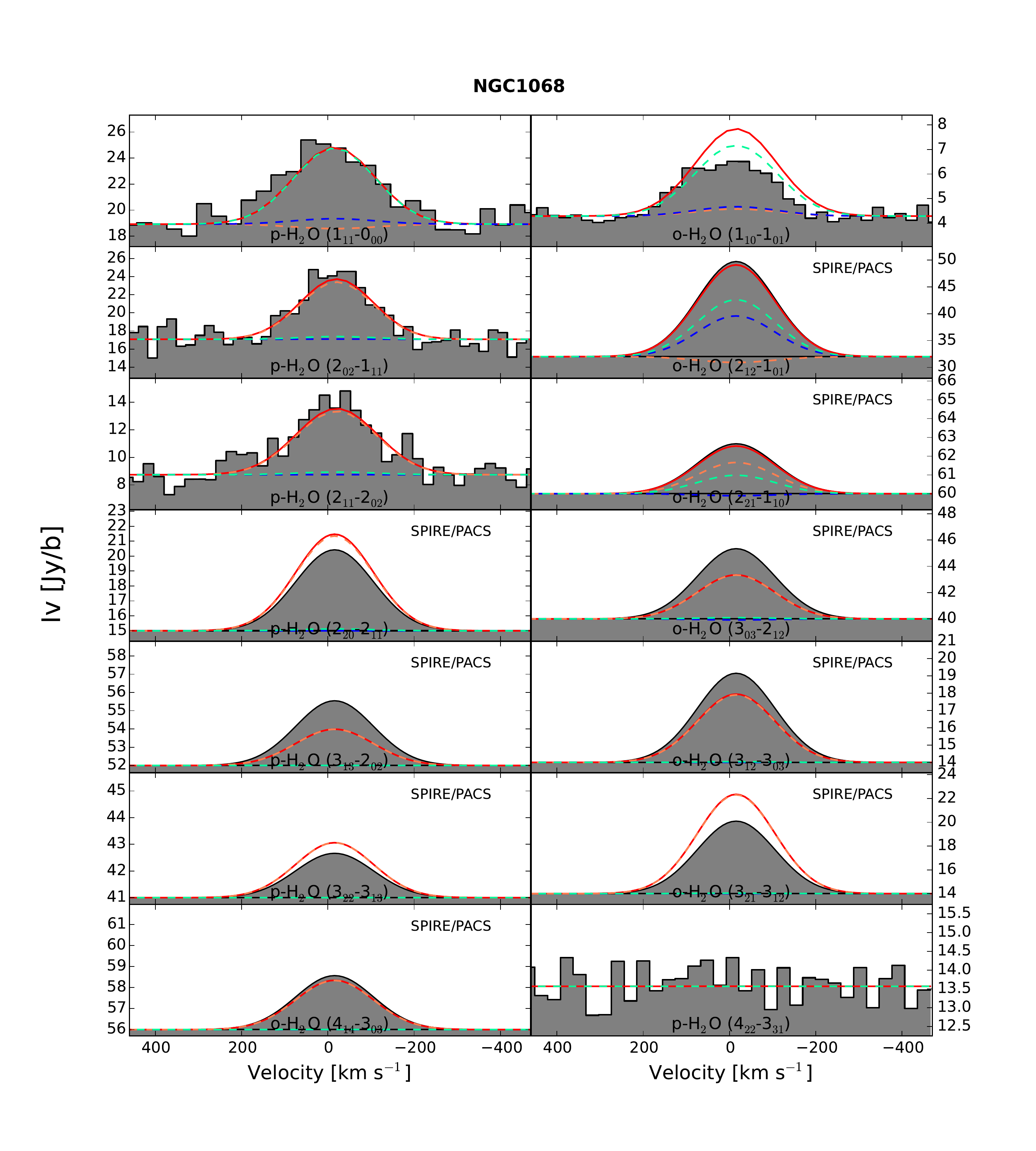}
\caption{The observed ${\rm H_2O}$ spectra (in grey colour) and the 
modelling results (in red solid curve) of NGC 1068.
The black-solid lines in the subpanels with text `SPIRE/PACS'
present the estimated line profiles 
for \textit{SPIRE}/\textit{PACS}-detected
lines, whose line shapes are assumed to be same as
the \textit{HIFI}-detected medium-excitation
lines.
The blue dashed lines denote the modelling results of the 
shocked outflow gas in NGC 1068.
For explanation of the other symbols used in the plots, 
see Fig.\ \ref{figure:m82_spectra}.
\label{figure:ngc1068_spectra}}
\end{figure*}

\begin{figure*}[h]
\includegraphics[width=\textwidth]{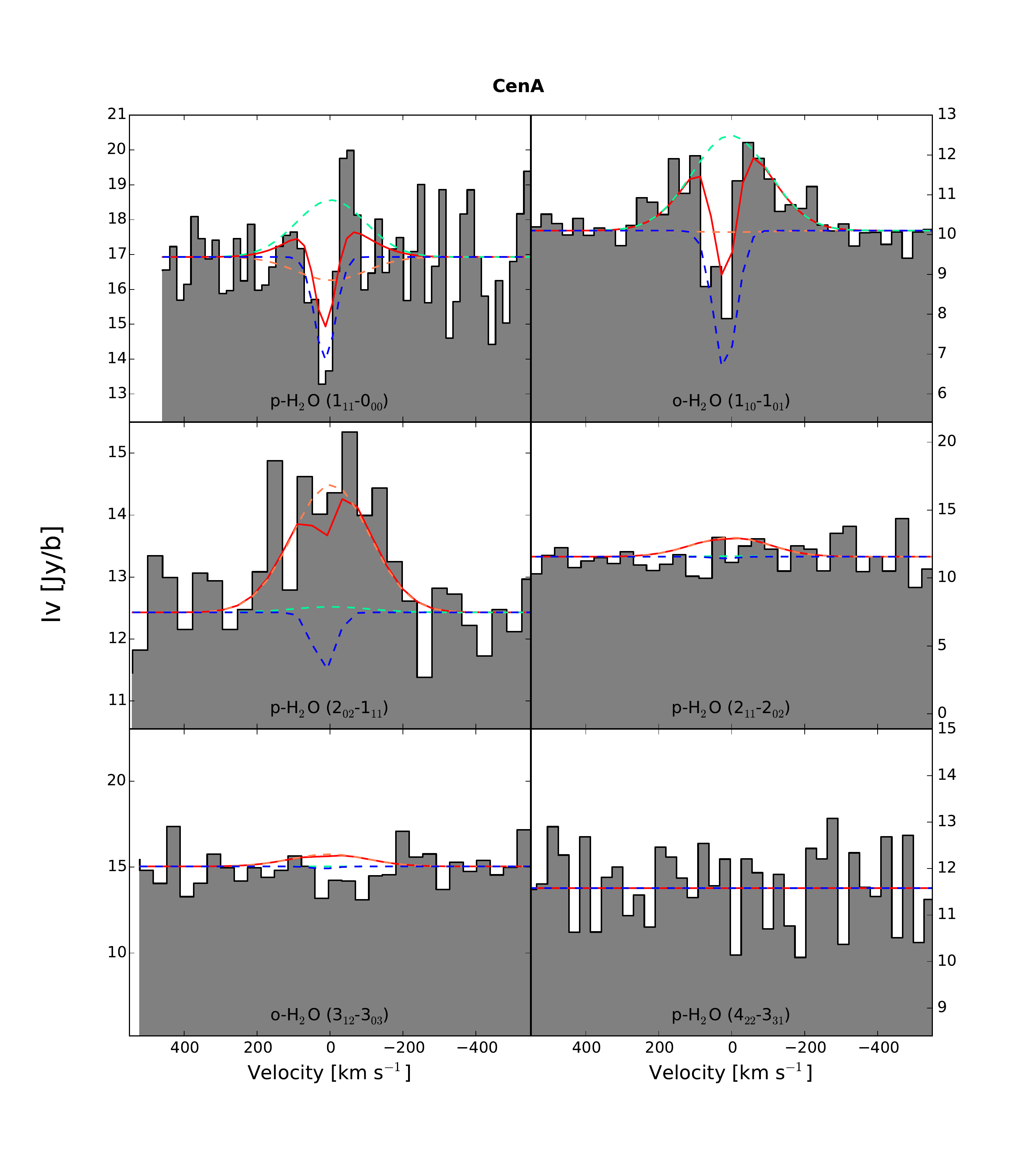}
\caption{The observed ${\rm H_2O}$ spectra (in grey colour) and the 
modelling results (in red solid curve) of Cen A.
The blue dashed curves denote the absorption of shocked gas
towards the power-law continuum of radio core.
For explanation of other symbols used in the plots, 
see Fig.\ \ref{figure:m82_spectra}.
\label{figure:cena_spectra}}
\end{figure*}

\begin{figure*}[h]
\includegraphics[width=\textwidth]{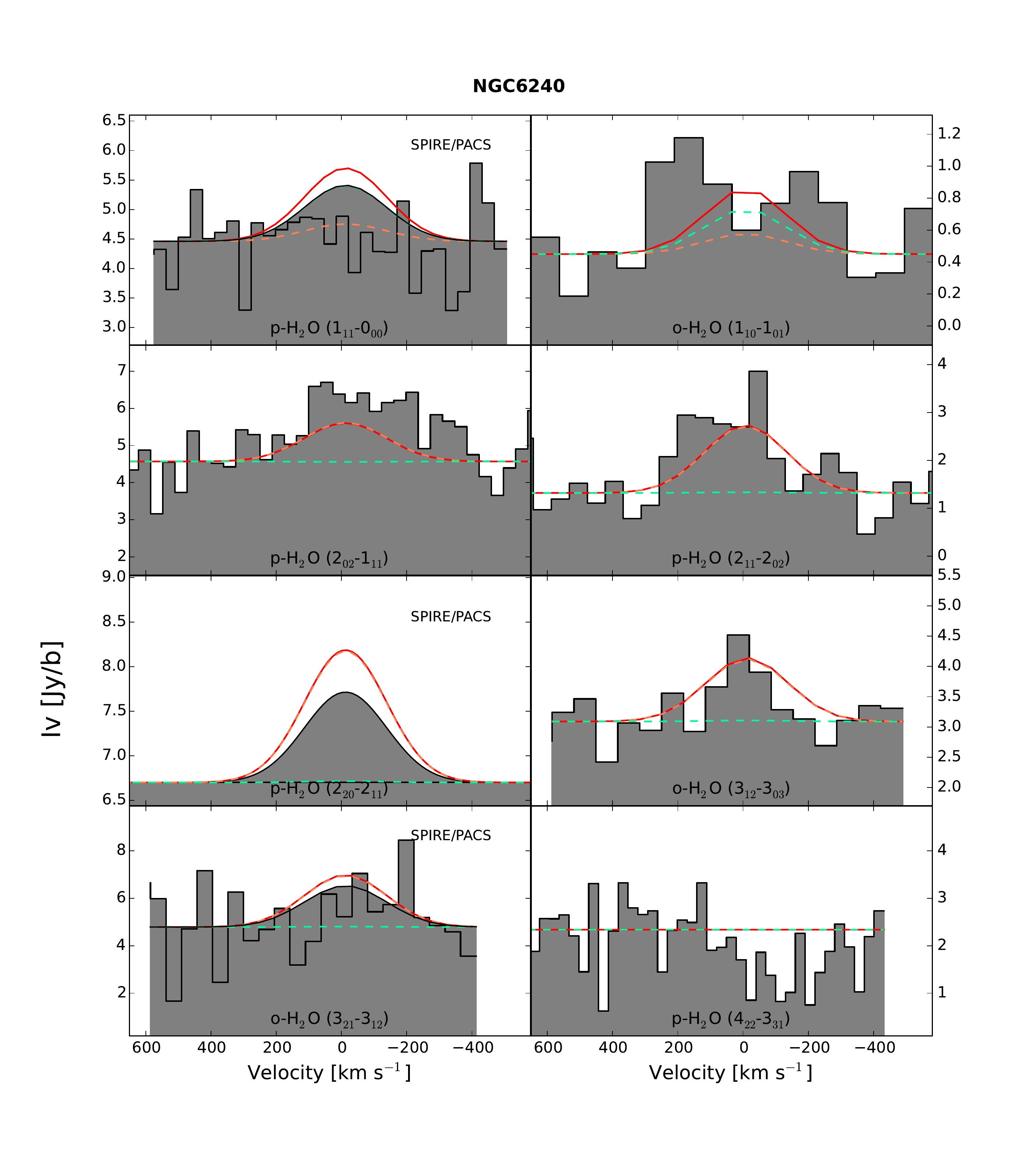}
\caption{The observed ${\rm H_2O}$ spectra (in grey colour) and the 
modelling results (in red solid curve) of NGC 6240.
For explanation of the symbols used in the plots, 
see Fig.\ \ref{figure:m82_spectra}.
The p-${\rm H_{2}O}$ ($1_{11}-0_{00}$) and o-${\rm H_{2}O}$ 
($3_{21}-3_{12}$) lines have been not detected by \textit{HIFI}, 
so their integrated fluxes  detected by \textit{Herschel}/\textit{SPIRE} 
were  utilised.
\label{figure:ngc6240_spectra}}
\end{figure*}

With these two components, we are able to explain almost all of the
water emission detected by \textit{HIFI}.  Fig.\
\ref{figure:m82_spectra} - \ref{figure:arp220_spectra} present in
grey colour the \textit{HIFI}-detected ${\rm H_{2}O}$ spectra and the
\textit{SPIRE}/\textit{PACS} ${\rm H_{2}O}$ data with line widths
assumed to be identical to the medium-excitation \textit{HIFI}
line shapes. In the figures we also show our line modelling results
(red solid curve) together with individual contributions from the warm
component (orange dashed curve) and cold ER (green dashed curve).
From the figures it is obvious that the warm component and cold ER
contribute differently to the ${\rm H_{2}O}$ line intensities.
The cold ER produces observable line emission only in the ground-state
transitions and in some cases in the o-H$_2$O (2$_{12}$-$1_{01}$) line. 
The warm gas on the other hand emits almost all power in the 
medium-excitation lines, but contributes little to the intensity of the 
ground transitions.

This finding is further illustrated in Fig.\
\ref{figure:multi_components} where we show the most prominent ${\rm
H_{2}O}$ lines predicted by our models for each ISM component.  The
black downward and red upward arrows denote emission and absorption,
respectively. The relative line strength is coded in the line style
with solid arrows indicating the strongest ($>$70\% of the
maximum intensity, I$_{\rm max}$), dash arrows medium ($70-10$\% of I$_{\rm max}$) 
and dotted arrows weak lines with less that 10\% of the maximum line intensity.
Fig.\ \ref{figure:multi_components} implies that the excitation of water
is very sensitive to the underlying physical conditions.
In Fig.\ \ref{figure:level_populations} we present the partition functions
(the fractional population of each level as a function of its energy)
for each ISM component. The figure shows that only the first two or three 
levels (with energies below $\sim 150$ K) are 
populated in the cold ER, while water can efficiently be populated 
to levels with energies up to $\sim 500$\,K in the warm gas component.
Another useful feature to distinguish between the two ISM components
is that the two ground-state lines seen in cold ER
become invisible in the warm component.
The different pattens shown in the Fig.\ \ref{figure:multi_components} suggest that 
${\rm H_{2}O}$ is powerful diagnostic tool to distinguish multiple ISM
components with different physical conditions in galaxies.

\subsubsection{Relation to the dust continuum and CO emission}
The two gas components also account for the major part of observed
dust SED and CO SLED, although their generated dust SED and CO SLED
are found to be very distinct.  Fig.\
\ref{figure:galaxy_co_IR_summary} presents the observed dust SED and
CO SLED (in black points) as well as the results predicted from our
best-fit models (red lines).  Overall, most
of IR luminosity is generated by the warm component (its dust SED
peaks at FIR wavelength), while the dust continuum emission on the long
wavelength Rayleigh-Jeans side arises mainly from the cold ER.  The CO
SLED of the warm component typically peaks at $8 \leq J_{\rm up} \leq
10$ transitions and dominates the emission of middle/high$-$J CO lines
($7 \leq J_{\rm up} \leq 12$) lines. The CO gas of the warm component
almost approaches LTE in levels  $J_{\rm up} \le 6$.
On the other hand, the cold ER generates most of CO line emissions in the 
low$-J$ CO transitions and its SLED peaks at $4 \leq J_{\rm up} \leq 6$.
Our models on average recover around 70\% of the observed dust continuum 
and CO line intensities.
It is worth mentioning that we find a good match between the physical sizes 
of warm component (derived from model) and the starburst regions 
measured from high-resolution molecular gas or IR observations.
This fact, in combination with its relatively high excitation of
${\rm H_{2}O}$ and CO, suggests that the warm component is associated with 
the nuclear starburst region in our sample galaxies. The cold ER, however, is 
likely associated with the more widespread quiescent ISM and may partly trace gas 
in the outer disks and spiral arms in some of our targets (e.g. in NGC253 and NGC4945).

\subsubsection{H$_2$O absorption}
Another typical ISM component in our models is the bulk of the
absorbing gas which normally locates in front of the warm component
that dominates the FIR luminosity.  Unlike the other two ISM 
components that exist in all
galaxies, we have only detected absorption in ground-state and other
low-excitation lines within five sources.  The existence of the
absorbing gas seems to depend on both the galaxy orientation and
geometry structure.  We find that the absorbing gas is more likely to
be detected in the edge-on galaxies, as in the case of NGC 4945, NGC
253 and M82 in our sample.  This is not surprising given that most of
the cold material in the disk of a galaxy will not be located in front of
the warm dust continuum if the galaxy is seen face-on (e.g, such as in NGC 1068).
The absorbing gas is very likely partly associated with the cold ER given the
similar physical parameters we find for the cold ER and the absorbing 
material in e.g. NGC4945 and NGC253. Furthermore the partition function 
of the cold gas (with a significant population of the first few energy
levels of water only, see Fig.\ \ref{figure:level_populations} {\it left}) 
naturally explains why absorption is usually only detected in the ground-state 
and the low-excitation ${\rm H_{2}O}$ lines. 
The medium-excitation ${\rm H_{2}O}$ lines from the background warm
component can thereby pass through the cold ER almost without being absorbed.
The profile of a resulting absorption line 
depends on how the cold ER is distributed relative to the warm component.  
If a large part of the warm component is covered by the cold ER, 
as suggested by our models for NGC 4945 and NGC 253, the absorption will appear 
to be very broad and deep. If only a small part of the warm component is covered, the
absorption will be narrow and shallow as the case for M82.  

Note that the absorbing gas does not always has to be associated with the
cold ER.  For example, the ground-state/low-excitation ${\rm H_{2}O}$
absorption detected in Arp 220 is found to arise from warm gas (possibly 
associated with molecular outflows driven by the nuclear activity) against
the even warmer background radiation from the hot component.

\subsubsection{Hot water in Mrk231 and Arp220}
\begin{figure*}[t]
\includegraphics[width=1.00\textwidth]{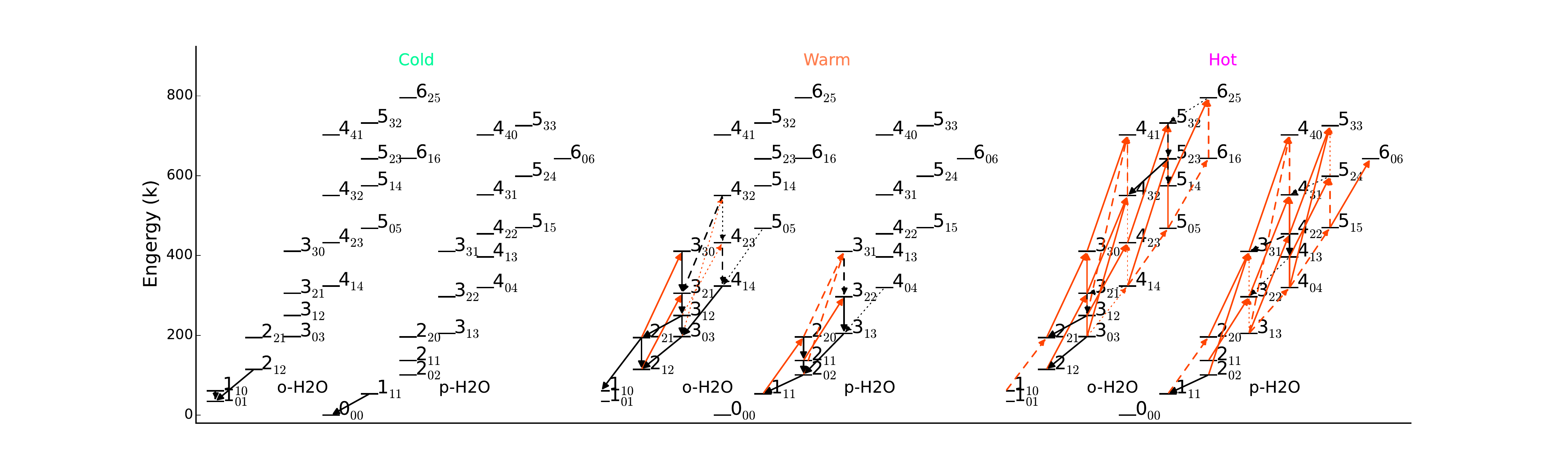}
\caption{The prominent ${\rm H_{2}O}$ line features from 
multiple ISM components predicted by our model.
The black downward and red upward arrows denote the emissions and absorptions, 
respectively. The solid arrows in the figure indicate those strongest lines (with 
intensities larger than 70\% of the highest value), while the dash arrows 
show the weaker lines and the dotted arrows show the 
weakest lines of all (with intensities less than $70\%$ and $10\%$ of the 
highest value, respectively).}
\label{figure:multi_components}
\end{figure*}

\begin{figure}[t]
\includegraphics[width=0.5\textwidth]{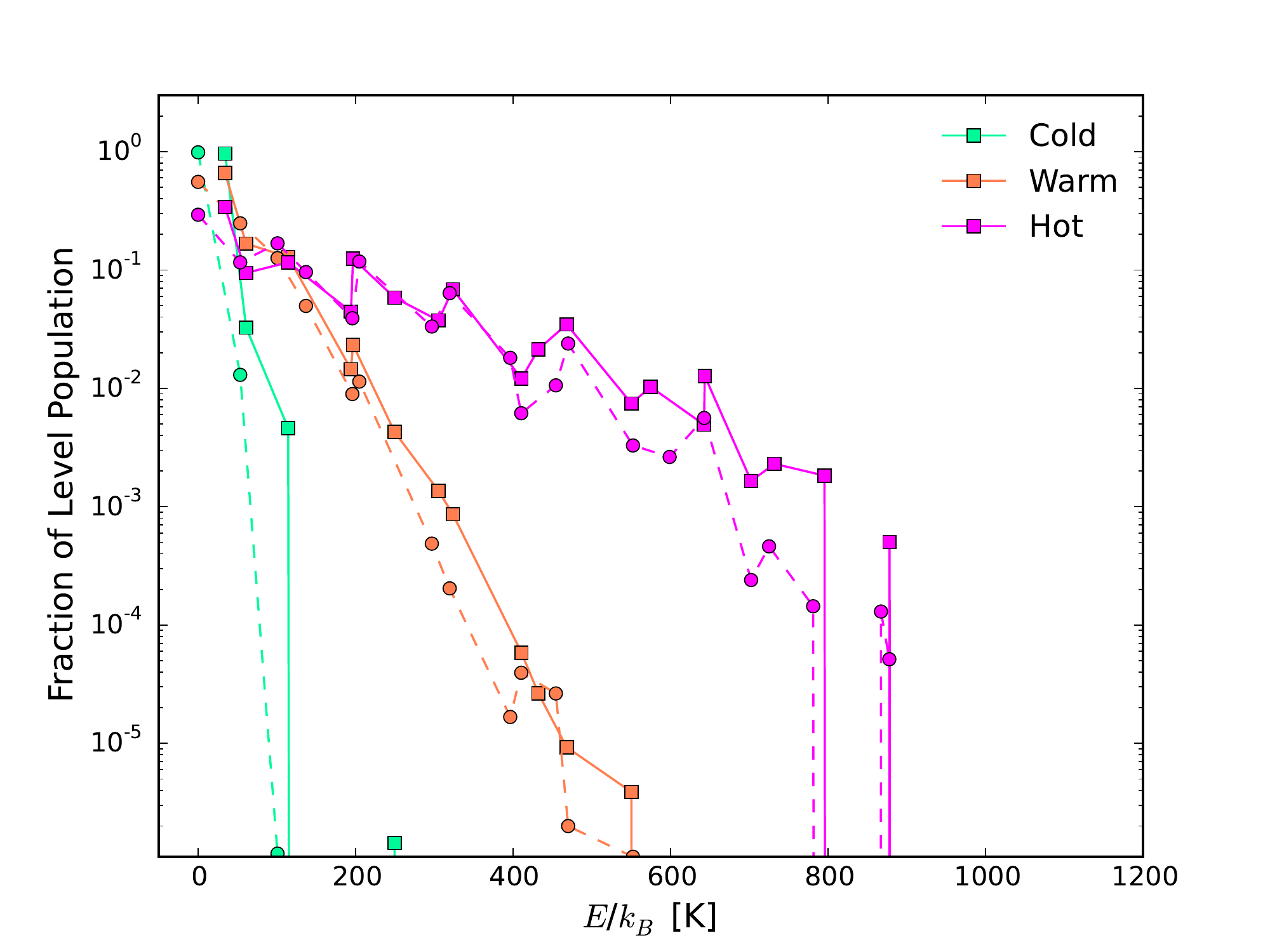}
\caption{The distributions of water level populations of multiple ISM components.
The $X-$ axis indicates the energy of the level and the $Y-$ axis
indicates the fraction of water gas residing at each level.
The squares denote level populations for o-${\rm H_2O}$ and 
the circles denote those for p-${\rm H_2O}$.}
\label{figure:level_populations}
\end{figure}

For the two most IR luminous sources in our sample, Mrk231 and
Arp220, our models require another physical component contributing
significantly to ${\rm H_{2}O}$ line intensities.  This component is
required to explain the high-excitation transitions exclusively 
detected towards both sources (e.g the p-H$_2$O ($4_{22}-4_{13}$) line). 
The component contains a substantial amount of
hot ($T_{\rm k}$ and $T_{\rm dust} \sim 100 - 200$ K) and dense ($n({\rm H})
\geq 1 \times 10^6$) gas with high column density ($N_{\rm H} \geq 5
\times 10^{24}~ {\rm cm^{-2}}$). The physical size of this hot
component is found to be very compact ($60 - 100$ pc).  
As we can see from the Fig.\ \ref{figure:multi_components} ({\it
right}), the high-excitation transitions from the hot gas are seen
both in emission and absorption, most of which are not excited by the
warm component.  In the hot component, water can be populated to
extremely high-excited levels with energies up to $800 \sim 1000$ K
(see Fig.\ \ref{figure:level_populations}). This is due to efficient 
collisional excitation in this gas phase and, more importantly, due to 
a combination of the strong IR emission associated with the hot gas 
and the large number of mid-IR water transitions which allow for 
efficient pumping of ${\rm H_2O}$ rotational levels to high energy 
states. 
A significant fraction of the continuum emission from the hot component,
however, may be attenuated by foreground material at short wavelength ($\lambda<200\mu$m) as the hot component is 
usually deeply buried in the galaxy nuclei \citep{downes2007}.
Due to its small size, the hot gas component has only a small 
contribution to the low$-J$ CO transitions but it becomes increasingly 
important for higher $J$ transitions and may 
dominate the CO SLED for CO lines with  $J_{\rm up} \ge 10$ (see 
Fig.\ \ref{figure:galaxy_co_IR_summary}).

Finally, we find that the gas phase abundance of water varies from 
$10^{-9} - 10^{-8}$ in the cold extended region, to $10^{-8} - 10^{-7}$ in the warm component 
and jumps to $10^{-6} - 10^{-5}$ in the hot gas due to the efficient release 
of H$_2$O from dust grain into the gas phase at these high temperatures.


\bigskip
\section{Discussion} 
\label{section: Discussion}
\subsection{Water Excitation in a Multi-Phase ISM} 
\label{section: water excitation in a multi-phase ISM}

\begin{figure*}[t]
\includegraphics[width=1.00\textwidth]{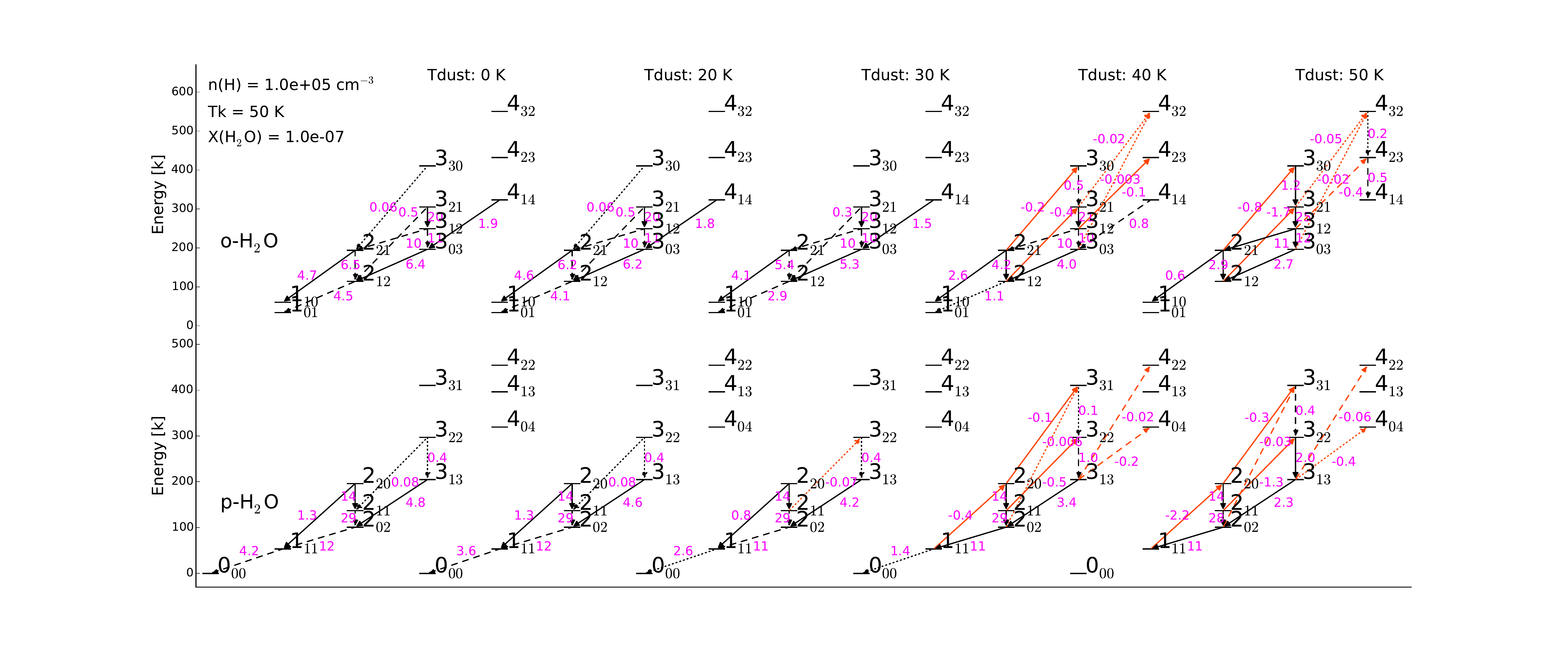}
\caption{The water excitation of a set of clumps 
that have characteristic parameters of a typical warm component 
with fixed $n({\rm H})$, $T_{\rm K}$ and $X({\rm H_{2}O})$,
but with $T_{\rm dust}$ varying from 0 to 50 K.
The clump with $T_{\rm dust} = 0$\,K shows the result without considering 
the IR pumping, and the clumps that have $T_{\rm dust} \sim 40 - 50$\,K 
display results under typical physical conditions of warm component.
The black downward arrows indicate emissions,
while the red upward arrows denote absorptions which can be regarded as
where IR pumping occurs. 
The different line styles represent different strengths of line intensities 
(same as Fig.\ \ref{figure:multi_components}).
The magenta number indicates the peak brightness temperature of each transition.}
\label{figure:warm_Tdust_water_excitation}
\end{figure*}

\begin{figure*}[t]
\begin{center}
\includegraphics[width=0.8\textwidth]{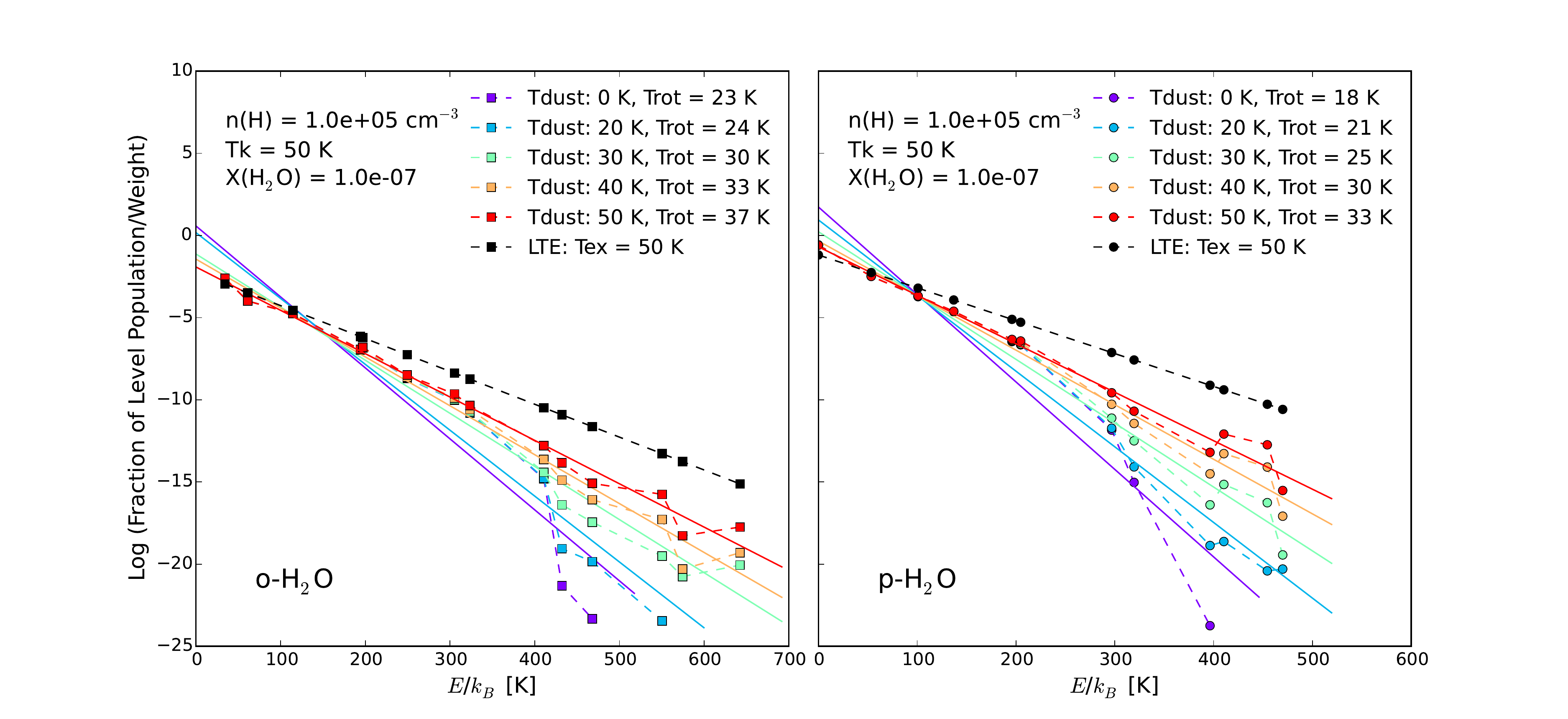}
\caption{The level populations of water for the clumps shown in
Fig.\ \ref{figure:warm_Tdust_water_excitation}, whose dust temperatures
are labeled by different colors.
The $X-$ axis indicates the energy of each level and the $Y-$ axis
indicates the ratio of the fraction of water gas to
statistical weight of level.
The solid straight line presents the best fit to the data points,
whose slope equals to -1/$T_{\rm rot}$. 
The black points show the level populations of water in LTE whose
rotational temperature is assumed to be the same as 
$T_{\rm k}$. }
\label{figure:warm_Tdust_level_populations}
\end{center}
\end{figure*}

In order to explore the relative importance of collisions and IR
pumping on the water excitation under typical conditions derived for
our galaxy sample, we compute the water excitation in the warm gas
component ($n({\rm H}) = 10^5~{\rm cm^{-3}}$, $T_{\rm K} = 50$ K,
$X({\rm H_{2}O}) = 10^{-7}$) with varying the dust temperature from 0
to 50 K.  The resulting emission and absorption lines from this
calculation are shown in Fig.\ \ref{figure:warm_Tdust_water_excitation} 
while the underlying level populations are shown in form of Boltzmann diagrams 
in Fig.\ \ref{figure:warm_Tdust_level_populations}.
From the first panel of Fig.\ \ref{figure:warm_Tdust_water_excitation}
, which ignores the effect of IR pumping ($T_{\rm dust} = 0$ K), one
can see that water can be excited by collision to levels with energies
up to 250 - 350\,K (250\,K for p-H$_2$O and 350\,K for o-H$_2$O). 
This picture does not change significantly as long as the
dust temperature stays below $\sim\,30$\,K. Only when the dust temperatures
reaches 40--50\,K, IR pumping starts to play a dominant role by populating ${\rm H_{2}O}$ 
levels with $250 - 350~{\rm K} \le E/k_{\rm B} \le 500 - 700$ K. 
The magenta numbers in Fig.\ \ref{figure:warm_Tdust_water_excitation} indicate
the peak brightness temperature of each transition in the warm gas component 
(averaged over the surface of a single clump). From these numbers one can see that 
the line strength of transitions between levels $E_{\rm up} > 250 - 350$ K
increases rapidly with dust temperature (e.g, o-${\rm H_{2}O}$ 
($3_{30}-3_{21}$) and p-${\rm H_{2}O}$ ($3_{31}-3_{22}$)).
In contrast, line intensities for transitions with E$_{\rm up} < 250 - 350$ K
depend little on the dust temperature (e.g., o-${\rm H_{2}O}$ 
($3_{21}-3_{12}$)) and some of them (e.g., the lines with 
E$_{\rm up} < 200$ K) even decrease 
for an increasing dust temperature due to the increased continuum levels
(e.g, o-${\rm H_{2}O}$ ($2_{21}-2_{12}$)).

This behavior is more quantitatively shown in the Boltzmann diagrams
in Fig.\ \ref{figure:warm_Tdust_level_populations}. The collisional
excitation drives the o-H$_2$O (p-H$_2$O)  populations 
with $E/k_{\rm B} \le 200$ (100)\,K  towards a
Boltzmann distribution at the kinetic temperature (i.e. these levels
are thermalised),
and dominates the equilibrium  population of the levels 
with $E/k_{\rm B} \le 350$ (250)\,K for o-H$_2$O (p-H$_2$O) 
independent of the dust temperature. 
Note that this happens already at densities much below the critical 
density of the water lines ($\sim10^8-10^9~{\rm cm^{-3}}$) 
due to the high optical depth of these lines (see Equation \ref{equation:ncr} 
in Appendix). The population of levels with energies
above 250 - 350\,K, however, is almost exclusively determined by radiative
pumping.  Fig.\ \ref{figure:warm_Tdust_level_populations} shows that
the dust radiation field tends to drive the population of these
levels (which are poorly populated by collisions) towards a
Boltzmann distribution close to the dust temperature. 
When the dust temperature is about equal to the kinetic temperature (red points), 
the overall level population (from ground-state to most highly excited levels) can be
fit well with a single rotational temperature (red solid line).
However, we also notice that the rotational temperature can only 
rise to 50\% - 70\% of the dust temperature even at the highest dust temperature
 ($T_{\rm dust} = T_{\rm K}$). We attribute this effect to the dust optical depth, 
i.e., it is due to the modified blackbody shape,
rather than a blackbody shape, of our dust radiation field.

The dust radiation energy is absorbed by the short-wavelength, far-IR
transitions (see red upward arrows in Fig.\ \ref{figure:warm_Tdust_water_excitation}) 
and then emitted by the long-wavelength, far-IR/submm high-excitation lines.
Under the typical conditions in the warm gas derived for our galaxy sample, 
the most efficient pumping
occurs at the strong $2_{12} \rightarrow 3_{21}$/$2_{21} \rightarrow 3_{30}$/$3_{21} 
\rightarrow 4_{32}$/$3_{03} \rightarrow 4_{32}$/$3_{12} \rightarrow 4_{23}$
lines ($2_{11} \rightarrow 3_{22}$/$2_{20} \rightarrow 3_{31}$/$2_{02} \rightarrow
3_{31}$ lines for p-${\rm H_{2}O}$) through absorptions of 
far-IR photons at 75/66/58/40/79 $\mu$m (90/67/46 $\mu$m 
for p-${\rm H_{2}O}$), which greatly enhance the intensities of high-excitation 
$3_{21} \rightarrow 3_{12}$/$3_{30} \rightarrow 3_{21}$/$3_{30} 
\rightarrow 3_{21}$/$4_{32} \rightarrow 4_{23}$/$4_{23} \rightarrow
4_{14}$ ($3_{22} \rightarrow 3_{13}$/$3_{31} \rightarrow 3_{22}$ for p-${\rm
H_{2}O}$) emission lines.

We have limited the discussion here to the typical warm environment derived for our galaxy sample.
Whether a transition is excited mainly by IR pumping or collision in the general case
is, however, very sensitive to the ambient ISM conditions. 
We show the Boltzmann diagrams for other cases in Fig.\ \ref{figure:level_populations_summary} in Appendix\ \ref{appendix: water
excitation under different physical conditions}.
In summary our models suggest, that our observed low-excitation 
lines ($E_{\rm up} < 130$ K) are collisionally excited,
and most of the observed medium-excitation lines ($130 < E_{\rm up} < 250 - 350$ K) 
are also predominantly excited by collision. 
IR pumping starts to populate the medium-excitation transitions with 
$250 - 350~K < E_{\rm up}$ for the typical dust fields derived in our sample and becomes increasingly important for higher transitions. 
For the high-excitation transitions with 
$E_{\rm up} \ge 600 - 800$\,K, IR pumping is the dominant 
excitation channel even for the extreme ISM conditions derived for our hot gas in Arp 220 and Mrk 231.


\subsection{${\rm H_{2}O}$ SLED} \label{section: h2o sled}
\begin{figure}[h]
\includegraphics[scale=0.4]{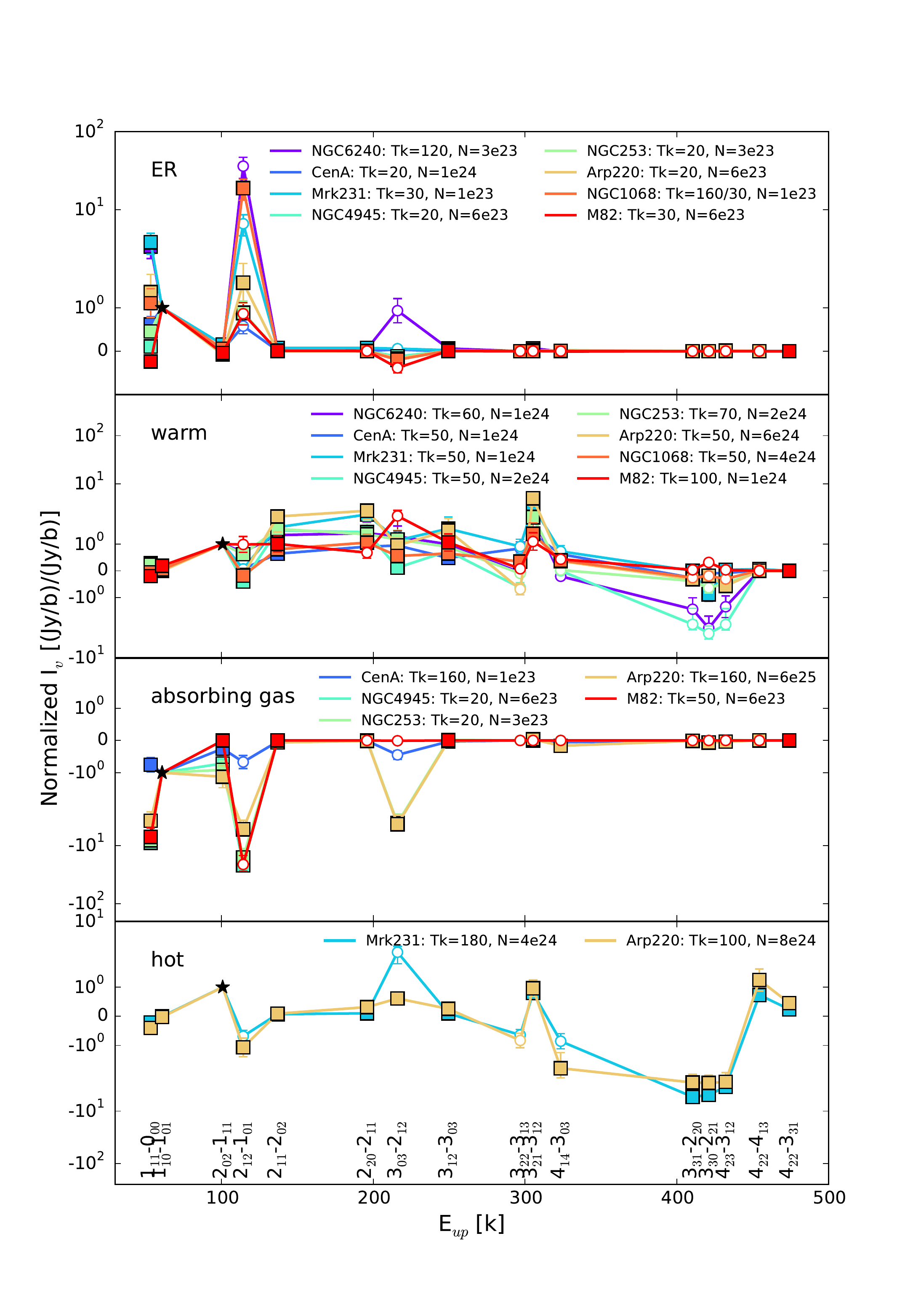}
\caption{The normalized ${\rm H_{2}O}$ SLED for multiple ISM
components. The black star indicates the ${\rm H_2O}$ transition 
which is used to normalize all the other ${\rm H_2O}$ lines.
The solid squares indicate the observed values  
for each ISM component which are derived from Gaussian decomposition.
The open circles denote the transitions without observations
whose intensities are predicted by our models.
The best-fit values for gas kinetic temperature $T_{\rm k}$ (in unit of K)
and hydrogen column density $N$ (in unit of ${\rm cm^{-2}}$) for each ISM
component are given in the legends. 
Note that the upper levels of $3_{03} - 2_{12}$, $3_{30} - 2_{21}$ and 
$4_{22} - 3_{31}$ are shifted to the right by 20, 10 and 20 K, respectively,
for clarity. }
\label{figure:h2o_sled}
\end{figure}
To investigate the spectral line energy distribution of ${\rm H_{2}O}$
within various ISM components and discuss the implications for
the observed various line ratios, we present normalized ${\rm H_{2}O}$ SLEDs 
for our sample galaxies in Fig.\ \ref{figure:h2o_sled}.
The solid squares in Fig.\ \ref{figure:h2o_sled} indicate the observed 
values (for each velocity components as derived from the Gaussian decomposition), the open circles indicate the model predicted values for the 
lines without observations and the black star indicates the ${\rm H_2O}$ transition which has been used
for the normalization.
The kinetic temperature and the column density for each component
are given in the plot legend, while the values for other parameters are presented in 
Table\ \ref{table: model parameters}.

\subsubsection{Elevated  $2_{12}-1_{01}$/$1_{10}-1_{01}$ ratios in the ER}
The most remarkable feature in the first panel of Fig.\ \ref{figure:h2o_sled}, 
which shows $1_{10}-1_{01}$ normalized ${\rm H_{2}O}$ SLED 
for the extended region (note that ER for NGC 1068 also includes the 
component of outflow
listed in Table\ \ref{table: model parameters}), 
is that the $2_{12}-1_{01}$/$1_{10}-1_{01}$ line ratios are strongly boosted
in galaxies with high kinetic temperatures.
The $2_{12}-1_{01}$/$1_{10}-1_{01}$ line ratio varies from $0.5-2.0$ in 
galaxies with kinetic temperature around $20 - 30$\,K (e.g., CenA, NGC
4945 and NGC 253), to the extreme large values ($\sim 10 -40$) in galaxies with 
$T_{\rm k} \ge 80 - 100$ K, which are well known to harbor ``shocked gas" 
\citep[e.g., NGC 6240 and NGC 1068,][]{meijerink2013,papadopoulos2014, mullersanchez2009, wang2012, wang2014}. 
As such, we suggest that the $2_{12}-1_{01}$/$1_{10}-1_{01}$ line ratio can be 
utilized as a good indicator of ``shock condition" where gas is heated to high 
kinetic temperatures.
The strong collisional excitation in shocks allows water to be 
populated above the ground states efficiently, and thereby
strongly enhances the line intensity of $2_{12}-1_{01}$.

\subsubsection{Line ratios in the warm gas}
The second panel in Fig.\ \ref{figure:h2o_sled} presents $2_{02}-1_{11}$ normalized 
${\rm H_{2}O}$ SLED for warm components of our galaxies.
The $2_{02}-1_{11}$ line is one of the visible lines
from the warm component that can be most easily
excited by collisions (therefore least affected by 
radiative pumping),
thereby the line ratios of other transitions to it can directly 
reflect the effect of radiative pumping.
The first notable feature inferred from the plot is that the ${\rm H_{2}O}$ SLED 
seen in medium-excitation lines (from $2_{11}-2_{02}$ to $3_{22}-3_{13}$) appears to be nearly flat, with their emitted line intensities comparable 
within a factor of $\sim 2$ in our sample.
This implies that the upper levels of these medium-excitation lines tend to 
approach statistical equilibrium in agreement with our findings in Sect.\ 
\ref{section: water excitation in a multi-phase ISM} and with the analysis by \citet{GA2014}.

However, all ${\rm H_{2}O}$ SLEDs show a peak at
$3_{21}-3_{12}$ line, demonstrating that $3_{21}-3_{12}$ line
has relatively stronger intensities since these medium-excitation lines
have similar line frequencies ($\nu \sim 1000 - 1200$ GHz). 
The underlying reason is that the upper level $3_{21}$ is not only approximately
thermalized by collisions (as the other lower levels), but it is also one of the levels
that can be most easily excited by IR pumping (see Fig.\ \ref{figure:warm_Tdust_water_excitation} 
and Fig.\ \ref{figure:warm_Tdust_level_populations}).
So the $3_{21}-3_{12}$ line usually exhibits
higher excitation temperature.
The $3_{21}-3_{12}$ line is found to be optically thinner 
($\tau_{\rm line} \simeq 1$) than the lower-excited medium-excitation lines
($\tau_{\rm line} > 1$) due to its required higher excitation energy.
Therefore, the $3_{21}-3_{12}$/$2_{02}-1_{11}$ line ratios  
 are mostly sensitive to the total ${\rm H_{2}O}$ column density 
($N({\rm H_2O})$) and are enhanced in galaxies with higher $N({\rm H_2O})$.
The $3_{21}-3_{12}$/$2_{02}-1_{11}$ ratio exhibits the largest value ($\sim 5$) 
in Arp 220 which has highest $N({\rm H_2O})$ ($\sim 6 \times 18~{\rm
cm^{-2}}$), moderate values ($\sim 2$) in galaxies with $N({\rm H_2O}) 
\sim 1 \times 17 ~{\rm cm^{-2}}$ (e.g., NGC 4945, NGC 253, NGC 6240) and the 
smallest value ($\sim 1.2$) in M82 which has the lowest $N({\rm H_2O})$ 
($\sim 1 \times 16~{\rm cm^{-2}}$).
Yet, not all galaxies (e.g., CenA and NGC 1068) follow this trend,
indicating that other parameters, such as the dust temperature as suggested by
\citet{GA2014}, may also influence this ratio.
A comparable trend is also seen in the $2_{20}-2_{11}$ line,  which 
locates in a position of para-H$_2$O energy diagram similar to that of
the $3_{21}-3_{12}$ line in ortho-H$_2$O energy diagram.
The $2_{20}-2_{11}$ line also exhibits
smaller line optical depth ($\tau_{\rm line} \simeq 1$) and slightly 
higher excitation temperature than other para-${\rm H_2O}$ medium-excitation lines.
We have not found a strong variance of  ${\rm H_{2}O}$ SLEDs over dust
temperatures, possibly due to the small range of dust temperatures in our 
warm components or physically suggesting that radiative pumping does not play
a dominant role in exciting these medium-excitation lines.
Finally, we find that the lines that are most sensitive to collision 
($3_{03}-2_{12}$ and $2_{12}-1_{01}$) have 
larger intensities in galaxies with higher kinetic temperatures (e.g., M82).

\subsubsection{The line ratios in the absorbing gas}
The third panel in Fig.\ \ref{figure:h2o_sled} shows 
the ${\rm H_{2}O}$ SLEDs
for the absorbing gas of five galaxies in our sample,
which are normalized to the absolute line intensity
of the $1_{10}-1_{01}$ line.
The first conclusion inferred from this plot is that
the absorption is often observed only up to the $2_{12}-1_{01}$ line. 
This is because higher levels are usually not populated for this component.
We further find that the ${\rm H_{2}O}$
SLED of absorbing gas is closely related to the dust SEDs ($T_{\rm dust}$
and $\tau_{\rm dust}$) of both the background continuum 
source and the absorbing gas. 
If the dust continuum from the absorbing gas is negligible,
then the ratios of absorption lines depend mainly on the 
flux ratios of background continuum sources at the line frequencies,
as the low-excitation absorptions ($1_{11}-0_{00}$
$1_{10}-1_{01}$, $2_{12}-1_{01}$) are often found to be optically thick.
We find that the galaxies whose absorbing gas contributes little 
to the dust continuum 
(e.g., $T^{\rm abs}_{\rm dust} \simeq  20 - 30$ K, 
$\tau^{\rm abs}_{100 \mu m} \simeq 0.2$ in NGC 4945, M82 and NGC 253) 
have the $1_{11}-0_{00}$/$1_{10}-1_{01}$ ratios 
around $8-10$ and the $2_{12}-1_{01}$/$1_{10}-1_{01}$ ratios around $20 - 30$, 
which are very close to the dust continuum
1113/557 GHz ($S_{\rm 1113GHz}/S_{\rm 557GHz} 
\simeq 8.5$) and 1670/557 GHz ($S_{\rm 1670GHz}/S_{\rm 557GHz} \simeq 23$) 
flux ratios of the background source (i.e., the warm component).
However, once the dust continuum from the absorbing gas becomes significant
(e.g., $T^{\rm abs}_{\rm dust} \simeq  60 - 80$ K,
$\tau^{\rm abs}_{100 \mu m} \geq 1.5$ in Arp 220), 
the $1_{11}-0_{00}$/$1_{10}-1_{01}$ and $2_{12}-1_{01}$/$1_{10}-1_{01}$
line ratios will decrease as increasing column density of the absorbing gas 
(as the dust continuum $S_{\rm 1113GHz}/S_{\rm 557GHz}$ and 
$S_{\rm 1670GHz}/S_{\rm 557GHz}$ 
flux ratios are smaller at lower dust temperature
and higher column density).
In the cases where the absorption lines are optically thin
(in this case, $N({\rm H_2O}) \le 1 \times15 ~{\rm cm^{-2}}$), however,
our modelling suggests that the $1_{11}-0_{00}$/$1_{10}-1_{01}$ and 
$2_{12}-1_{01}$/$1_{10}-1_{01}$ ratios can be significantly
larger than the values estimated from the background dust
continuum.
We do not find optically-thin absorption lines in our work.
The $1_{11}-0_{00}$/$1_{10}-1_{01}$ and $2_{12}-1_{01}$/$1_{10}-1_{01}$ 
ratios should be always larger than one, if the background continuum is a 
modified blackbody spectrum.
Otherwise, it suggests that the background continuum may be a 
radio source with a power-law spectrum just as in the case of Cen A
(see Appendix \ref{appendix: detailed models for individual galaxies} 
for more details on Cen A).
Thus, the ratios of ${\rm H_2O}$ absorption lines are able to provide 
valuable hints on the properties of background continuum 
sources even without spatially resolving the nuclear regions.

\subsubsection{The H$_2$O SLED of the hot gas}
The last panel in Fig.\ \ref{figure:h2o_sled} presents $2_{02}-1_{11}$ normalized 
${\rm H_{2}O}$ SLED of hot component which
has only been found in the ULIRGs in our sample (Arp 220 and Mrk 231).
The ${\rm H_{2}O}$ SLED of hot component displays similar features with the 
warm component in medium-excitation lines but displays quite distinct behavior
in high-excitation lines which show strong detections in 
both emission and absorption. 
Considering its high dust temperature and other extreme physical conditions, 
the hot component is possibly associated with a dusty toroid heated by an AGN 
\citep[e.g.,][]{downes2007, weiss2007}.

\subsection{Dependence of the water line intensities on the FIR continuum properties} 
\label{section: the Lh2o - LFIR relation}
\subsubsection{The $L'_{\rm H_2O }$ - $L_{\rm FIR}$ correlation}
\begin{figure}[h]
\includegraphics[scale=0.3]{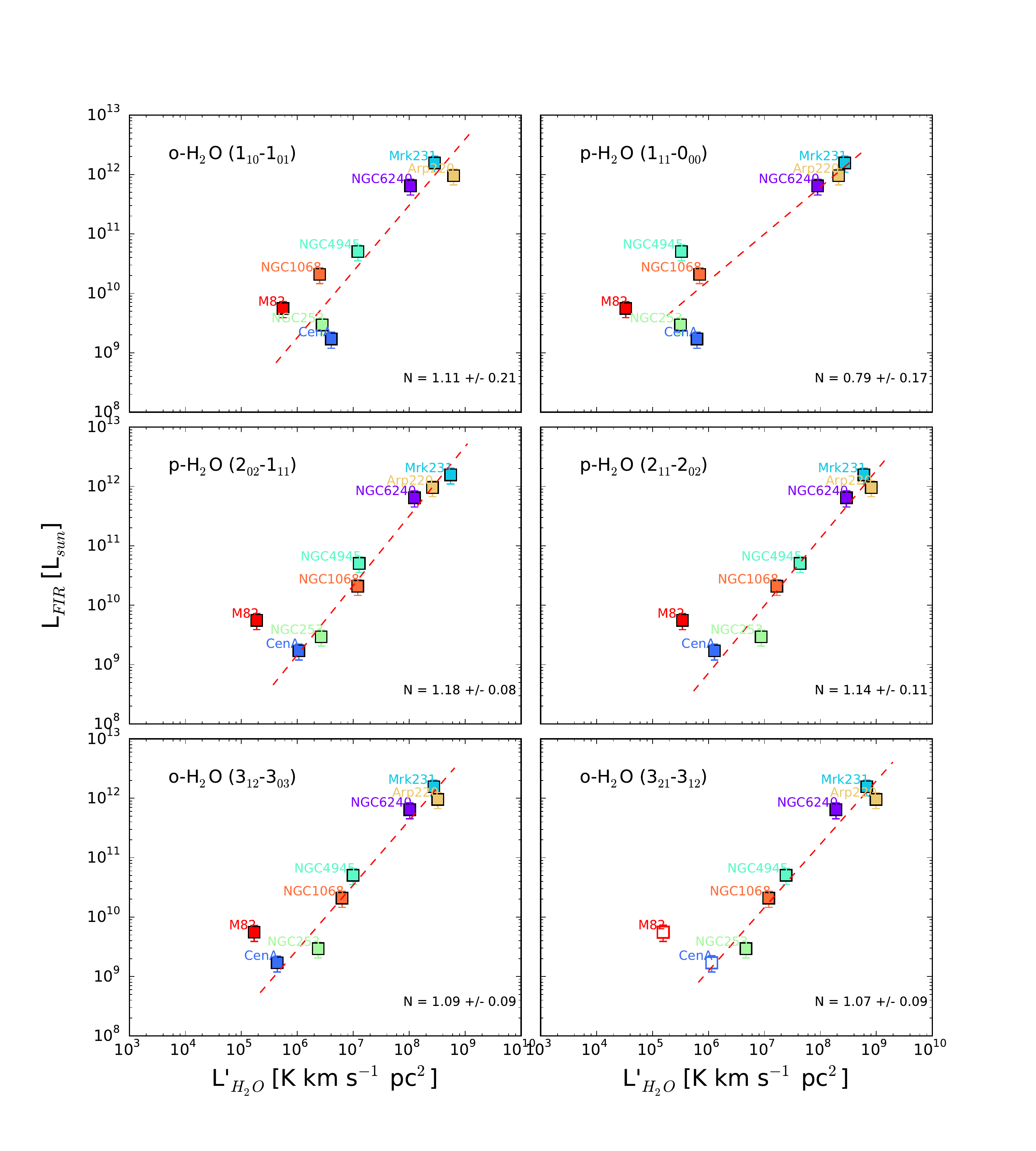}
\caption{The correlations between $L'_{\rm H_2O }$ and $L_{\rm FIR}$
for our sample galaxies. The line luminosities of o-${\rm H_2O }$
($1_{10}-1_{01}$) and p-${\rm H_2O }$ ($1_{11}-0_{00}$) are calculated without
considering the absorptions.
The open squares in last panel indicate the galaxies whose $L'_{\rm H_2O 
(3_{21}-3_{12}) }$ are estimated from our model as no observations 
have been obtained. 
The red dashed line denotes the 
best-fitting correlation whose slope 
is given by $N$.}
\label{figure:lh2o_lir}
\end{figure}
\begin{figure}[h]
\includegraphics[scale=0.30]{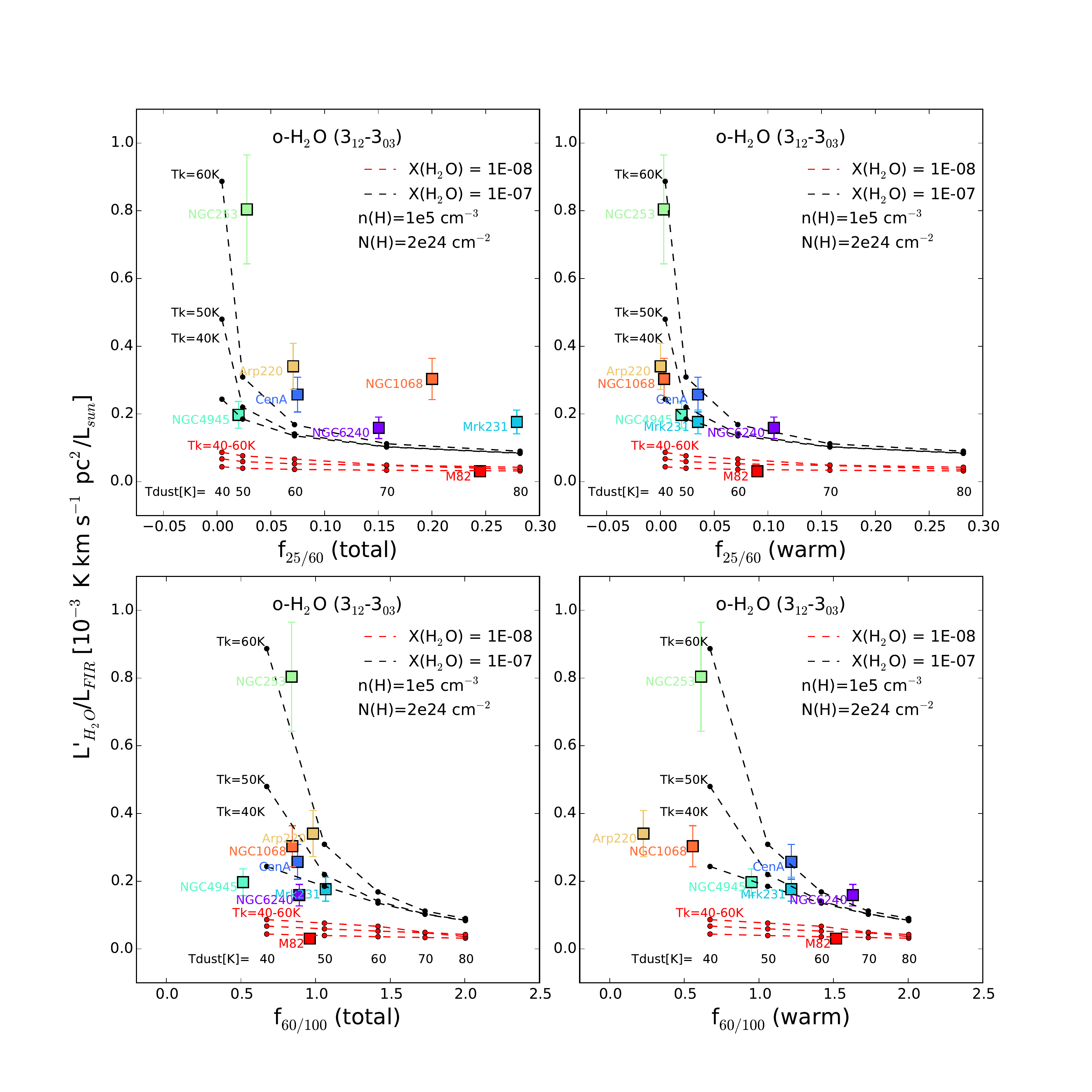}
\caption{$L'_{\rm H_2O (3_{12}-3_{03}) }$/$L_{\rm FIR}$  vs. $f_{25}/f_{60}$ (top)
and $f_{60}/f_{100}$ (bottom), respectively.
The left column panels show the observed $L'_{\rm H_2O}/L_{\rm FIR}$ ratios
versus the observed total IR colors, while the right column panels present
the observed $L'_{\rm H_2O}/L_{\rm FIR}$ versus the IR colors 
derived from the warm component only.
The dashed curve presents our modelled $L'_{\rm H_2O}/L_{\rm FIR}$ 
- IR color correlation for a set of warm components with fixed $T_{\rm K}$ 
and $X({\rm H_2O})$, where different solid circles indicate various dust temperatures whose values are labeled at the bottom of the plot. 
The values of $T_{\rm K}$ are labelled in the text, while various 
$X({\rm H_2O})$ are given by different colors.
The gas density and column density of all models are fixed to the values
labelled in the plots.
 }
\label{figure:ir_color}
\end{figure}
\begin{figure*}[t]
\begin{center}
\includegraphics[scale=0.40]{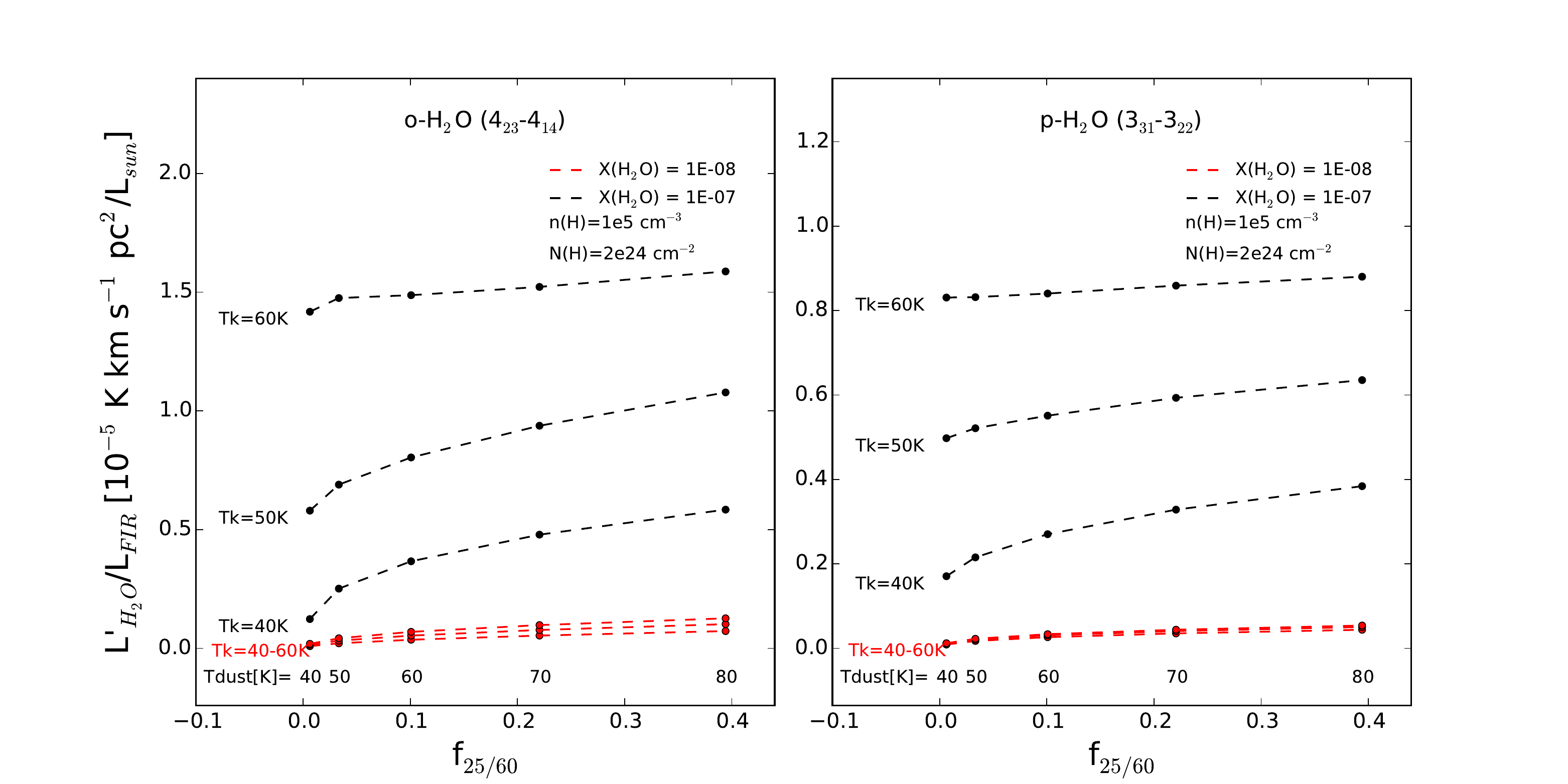}
\caption{The modelled $L'_{\rm H_2O }$/$L_{\rm FIR}$ - $f_{25}/f_{60}$ for two 
${\rm H_2O}$ lines that are supposed to be excited dominantly by IR pumping.
We have no observations for these two lines in all our sample galaxies.
For explanation of the symbols used in the plots, 
see Fig.\ \ref{figure:ir_color}.
}
\label{figure:ir_color_model}
\end{center}
\end{figure*}

We present the correlations between $L_{\rm FIR}$ 
($40 - 120 ~\mu$m)
and $L'_{\rm H_2O}$ for the two ground transitions and four
medium-excitation lines in Fig.\ \ref{figure:lh2o_lir}. Note
that we adopted only the emission luminosity, rather than total line
luminosity, for the two ground-state lines to avoid biases due to
foreground absorption. Fig.\ \ref{figure:lh2o_lir} shows, that M82
deviates significantly from the overall correlations in particular for
the medium-excitation lines. We attribute this finding to its
lower water abundance (more details on M82 see Appendix \ref{appendix:
detailed models for individual galaxies}) and excluded M82 from our
correlation analysis between $L'_{\rm H_2O }$ and $L_{\rm FIR}$.

We find that all medium-excitation transitions show a tight
nearly linear $L'_{\rm H_2O }$ - $L_{\rm FIR}$ correlation, which
agrees well with the linear correlation found in larger galaxy sample
observed with the \textit{SPIRE-FTS} \citep[]{yang2013}.  As our model
suggests that collision plays a significant role in exciting most of
the medium-excitation transitions (see Fig.\
\ref{figure:warm_Tdust_water_excitation} and Fig.\
\ref{figure:warm_Tdust_level_populations}), this linear correlation is
not simply a consequence of IR pumping as suggested by
\citet{yang2013}. Our models show that the medium-excitation
transitions probe the same physical regions of galaxies where most of FIR 
emission is generated (i.e., the warm component, see Section
\ref{general_modelling_results}), suggesting that the observed 
correlations between FIR and the medium-excitation line is
mostly driven by the sizes of the FIR and water emitting regions in these
lines.

This view is also supported by the $L'_{\rm H_2O }$ - $L_{\rm FIR}$
correlations of the two ground-state lines which we present here for
the first time. While the $L'_{\rm H_2O }$ - $L_{\rm FIR}$ correlation
for o-${\rm H_2O}$ ($1_{10}-1_{01}$) line has an approximately linear
slope ($N=1.11 \pm 0.21$), the p-${\rm H_2O}$ ($1_{11}-0_{00}$)
line shows a slope slightly below unity ($N=0.79 \pm 0.17$). In
addition, the correlations for ground-state lines have smaller Spearman
rank correlation coefficients ($\sim 0.82$) compared with those for
medium-excitation lines ($\sim 0.97$), suggesting that the
former are less correlated with $L_{\rm FIR}$ compared with the
latter. This is not surprising given that the ground-state lines are found to 
arise mostly from the cold extended regions of galaxies which contribute 
less to FIR luminosity.

\subsubsection{Dependence on $L'_{\rm H_2O}/L_{\rm FIR}$ on the dust color}

In order to investigate the variations of luminosity ratios ($L'_{\rm
H_2O}/L_{\rm FIR}$) with $T_{\rm dust}$, we have analyzed the
correlation between $L'_{\rm H_2O}/L_{\rm FIR}$ and the IR colors
($f_{25}/f_{60}$ and $f_{60}/f_{100}$) for the
medium-excitation lines.  All medium-excitation lines
display a similar trend and we show as an example the relation for the
o-${\rm H_{2}O}$ ($3_{12}-3_{03}$) line in Fig.\
\ref{figure:ir_color}.  The left column of Fig.\ \ref{figure:ir_color}
shows the observed $L'_{\rm H_2O}/L_{\rm FIR}$ ratios versus the
observed total IR colors for our sample galaxies (with $f_{25}/f_{60}$
in the top row and $f_{60}/f_{100}$ in the bottom row), while the
right column presents the observed $L'_{\rm H_2O}/L_{\rm FIR}$ versus
the IR colors derived from the warm component only.

The first conclusion inferred from Fig.\ \ref{figure:ir_color} is that 
we do not find a clear trend where the observed $L'_{\rm H_2O}/L_{\rm FIR}$
ratios vary with the observed total IR colors (for both $f_{25}/f_{60}$ 
and $f_{60}/f_{100}$), however, we find that the observed 
$L'_{\rm H_2O}/L_{\rm FIR}$ ratios decrease with the increasing values
of IR colors estimated from the warm component only.
The first fact could be due to the different physical origins of the 
observed $L'_{\rm H_2O}/L_{\rm FIR}$ ratios and IR colors.
The former arises mainly from the warm component, while the 
observed $f_{25}/f_{60}$ and $f_{60}/f_{100}$ will be contaminated by
other ISM components. 
The $f_{25}/f_{60}$ will be greatly enhanced 
if a strong AGN is present (e.g., Mrk231, NGC 1068 and Arp 220),
while the $f_{60}/f_{100}$ will be decreased largely
if the cold ER contributes significantly to the total IR luminosity.
We can see from the right column of Fig.\ \ref{figure:ir_color} that once the 
contamination in IR colors from other components is removed,
$L'_{\rm H_2O}/L_{\rm FIR}$ ratios start to show a correlation with IR colors.
Within a larger sample of galaxies that are detected by the \textit{SPIRE-FTS}, however, a
slight trend that the $L'_{\rm H_2O}/L_{\rm IR}$ ratio decreases with
increasing $f_{25}/f_{60}$ has been observed \citep{yang2013}.

To further understand how the $L'_{\rm H_2O}/L_{\rm FIR}$ ratios vary
with IR colors, we have modeled the relations between $L'_{\rm
H_2O}/L_{\rm FIR}$ and IR colors for a set of warm components with
variable dust temperatures, kinetic temperatures and water abundances
but with a fixed gas density ($n({\rm H})=1 \times 10^5~{\rm
cm^{-3}}$) and column density ($N({\rm H})=2 \times 10^{24}~{\rm
cm^{-2}}$).  The dashed curves in Fig.\ \ref{figure:ir_color} present
the model results for the warm components.  One can see that our model
predicts a strong decrease of $L'_{\rm H_2O}/L_{\rm FIR}$ with
increasing $f_{25}/f_{60}$ and $f_{60}/f_{100}$ (i.e., $T_{\rm dust}$), which 
is in very good correspondence with the trend seen in the warm component only.

We further find that the constant $T_{\rm k}$ lines drop much faster
at the low-$T_{\rm dust}$ end where most of our sample galaxies
reside, while they decrease very little at the high-$T_{\rm dust}$ end.
This is because the o-${\rm H_{2}O}$ ($3_{12}-3_{03}$) line (along
with the other medium-excitation lines in our work) is excited
largely by collisions at low-$T_{\rm dust}$ ($T_{\rm dust} \le
T_{\rm k}$), and therefore its line intensity does not increase
significantly with $T_{\rm dust}$, but is greatly enhanced for
increasing kinetic temperature.  That is why NGC 253 stands out
significantly in the plot, because it has a relatively high $T_{\rm k}$
compared to its $T_{\rm dust}$.  At the high-$T_{\rm dust}$ end
($T_{\rm dust} > T_{\rm k}$) where IR pumping becomes more important,
the line intensity $L'_{\rm H_2O }$ increases rapidly with $T_{\rm
dust}$ and therefore its ratio to $L_{\rm FIR}$ remains almost
constant for increasing $T_{\rm dust}$ (i.e, IR colors), and it will
show no dependence on $T_{\rm k}$. The $L'_{\rm H_2O}/L_{\rm FIR}$
ratio of o-${\rm H_{2}O}$ ($3_{12}-3_{03}$) is found to be always
larger than $1 \times 10^{-4}$ if $X({\rm H_2O}) \ge 1\times 10^{-7}$,
and decreases rapidly with decreasing water abundance.

For the lines that are excited dominantly by IR pumping (such as the
p-${\rm H_{2}O}$ ($3_{31}-3_{22}$) and the o-${\rm H_{2}O}$
($4_{23}-4_{14}$) lines which have both not been observed in our work),
our model predicts the $L'_{\rm H_2O}/L_{\rm FIR}$ ratios have a
different dependence on the IR color than the lines mainly excited by
collisions. As shown in Fig.\ \ref{figure:ir_color_model} these lines
do not show a decreasing but an increasing $L'_{\rm H_2O}/L_{\rm FIR}$
ratio for an increasing dust temperature.  
A similar trend is found by \citet{GA2014}, which suggests that the 
$L'_{\rm H_2O}/L_{\rm FIR}$ ratio is expected to decrease with increasing 
$T_{\rm dust}$ for all lines with $E_{\rm up} < 300$\,K but increases with 
$T_{\rm dust}$ for the high-excitation ${\rm H_2O}$ lines with $E_{\rm up} > 400$ K.
The modelling results shown
in Fig.\ \ref{figure:ir_color} and Fig.\ \ref{figure:ir_color_model}
imply that the observed correlation between $L'_{\rm H_2O}/L_{\rm FIR}$
and IR colors can be utilized as a diagnostic tool to distinguish
between different excitation channels (i.e., collisional excitation
and IR pumping) of ${\rm H_2O}$ lines.  That is if the $L'_{\rm
H_2O}/L_{\rm FIR}$ ratio is found to decrease with increasing IR color,
then the line is possibly excited mainly by collision within the
sample, otherwise the line may be excited dominantly by IR pumping.

\subsection{The Implication of Line Profiles} 
\label{section:implication of line profiles}

Unlike the line profiles of other molecules (such as CO or HCN), the
${\rm H_2O}$ line shapes strongly depend on the involved energy
levels. The absorption free medium-excitation ${\rm H_2O}$
transitions show line profiles in agreement with or slightly narrower
than CO.  Our model suggests that the medium-excitation ${\rm
H_2O}$ lines stem from the same volume that gives rise to the FIR
emission. This leads to a tight linear correlation between the
luminosities of medium-excitation lines and FIR. 
We therefore conclude, that the medium-excitation
${\rm H_2O}$ lines are good probes to study the kinematics of the
starburst/star-forming gas, e.g., they can serve as kinematic probes
of the FIR emission region.

The line profiles of ground-state and low-excitation lines are found
to differ from galaxy to galaxy. Some of these lines show pure
absorption, while the others show pure emission or a mix of both.  The
emission of the ground-state lines often spans a larger velocity range
than the medium-excitation transitions (e.g, NGC 4945, NGC 253 and
NGC 1068), implying that the ground-state emission arises from a more
extended physical region.  This is most prominently seen in NGC 253
where the o-${\rm H_{2}O}$ ground-state line is $\sim 60~{\rm km~s^{-1}}$ 
wider than the medium-excitation lines.  
The absorption line shapes and centroids depend on the dynamics of the
foreground gas and its location related to the background continuum
source.  Although detailed kinematics of galaxies are not considered
in our work, some important information can still be
obtained by comparing the line profiles of ${\rm H_2O}$ with those of
other species (e.g, CO, HF) and the high spacial-resolution CO
observations.

We noticed that most ${\rm H_2O}$ absorption profiles have line shapes 
that deviate from pure Gaussian profiles and often show P-Cygni or inverse
P-Cygni profiles indicative for non-circular motions driven by gas outflow
or infall. For example, the absorption of NGC 4945 is better fitted by a broad 
Gaussian component centered close to the systemic velocity plus a 
redshifted (by $\sim 80~{\rm km~s^{-1}}$) narrower Gaussian component. 
The latter is possibly related to an infall of gas in the molecular ring or non-circular 
motion \citep[for instance in a barred potential,][]{vandertak2016}.
The absorption of NGC 253 shows a P-Cygni profile which has also been 
observed in other gas species \citep[e.g., H{\small I} and HF,][]{koribalski1995, monje2014}, 
and is likely connected to the molecular outflow observed in CO \citep[][]{bolatto2013}.
By comparing the absorptions of M82 with high-resolution CO maps,
\citet{weiss2010} found the absorption of M82 only displays a good 
agreement with the CO profiles from a small region towards the galaxies center,
implying that the absorption may arise from a small strip orthogonal 
to the molecular disk.
 
In the case of Arp 220 the ground-state (and low-excitation)
absorption feature display a velocity dispersion which is much
narrower than those of the medium-excitation emission lines
($FWHM \sim 226~{\rm km~s^{-1}}$ compared to FWHM $\sim 412~{\rm
km~s^{-1}}$), but in good agreement with the width of the
high-excitation lines ($FWHM \sim 235~{\rm km~s^{-1}}$).  
High-spatial resolution ALMA imaging by \citet[][]{koenig2016} find that
most of the high-excitation H$_2$O line emission stems from the
western nucleus which is also suspected to harbor a buried AGN
\citep{downes2007}. The velocity centroid of the H$_2$O absorption is
consistent with that arising from the western source. It is therefore
likely that the low-excitation absorption of Arp 220 arises within the
compact western nucleus in a region that could be about the same size
as the region where high-excitation lines are generated from.

The o-${\rm H_{2}O}$ ground transition profile of Cen A is similar to
those of low$-J$ CO lines observed by \citet{israel2014} showing a
broad emission profile arising from the circumnuclear disk with a
narrow absorption close to the systemic velocity against the compact
nuclear source.  In Cen A, the p-${\rm H_{2}O}$ ground-state line 
shows an inverse P-Cygni profile, which could be a sign of infall 
\citep[e.g.,][]{vandertak2016}.  

Another interesting finding comes from a comparison between the line
profiles of o-${\rm H_{2}O}$ and p-${\rm H_{2}O}$. 
We find that parts of the
o-${\rm H_{2}O}$ and p-${\rm H_{2}O}$ energy ladders are very similar, 
where the o-${\rm H_{2}O}$
$1_{01}$/$2_{12}$/$3_{03}$/$3_{12}$ /$3_{21}$ levels correspond to the p-${\rm
H_{2}O}$ $0_{00}$/$1_{11}$/$2_{02}$/$2_{11}$/ $2_{20}$ levels, 
respectively (see Fig.\ \ref{figure:energy_diagram}).  Therefore, the
p-${\rm H_{2}O}$ ground-state ($1_{11}$-$0_{00}$) transition shows 
a good correspondence in the line shape to the o-${\rm H_{2}O}$ ($2_{12}-1_{01}$) 
line rather than the o-${\rm H_{2}O}$ ($1_{10}$-$1_{01}$) 
ground-state transition.
This is also why the o-${\rm H_{2}O}$
($3_{21}-3_{12}$) and ($3_{12}-3_{03}$) lines have very similar shapes
compared to the p-${\rm H_{2}O}$ ($2_{20}-2_{11}$) and
($2_{11}-2_{02}$) lines, respectively, even though the upper levels of the
former are around $100$\,K higher than those of the latter.

\section{Summary and Conclusion} \label{section: summary and conclusion}

Using {\it HIFI} on-board of {\it Herschel} we have obtained for the first 
time velocity-resolved ${\rm H_2O}$ spectra for a sample of nearby actively 
starforming galaxies. Our observations include the ground-state transitions of 
ortho and para H$_2$O and cover transitions with E$_{\rm up}\le450$\,K.
The main observational results from our spectroscopic survey are:

\begin{itemize}
\item The ${\rm H_2O}$ spectra show a diversity of line shapes.
The medium-excitation ${\rm H_2O}$  lines ($130<\,$E$_{\rm up}\,\le350$\,K) 
are always detected in emission. Their line profiles typically resemble those
of CO indicating that water is widespread in the warm nuclear regions of active galaxies.

\item The line profiles of ground-state and low-excitation ${\rm H_2O}$ transitions
 (E$_{\rm up}\,\le130$\,K)  often show a mix of emission and absorption.
The emission features of these low-excitation lines usually display a
wider velocity range than the medium-excitation lines,
while the absorption features are often found to show more complex line 
profiles that differ from galaxy to galaxy.
\end{itemize}

We analyze the water excitation using the state of art, 3D non-LTE radiative
transfer code `$\beta$3D'. The main conclusions from the analysis are:

\begin{itemize}
\item Multiple ISM components with different physical conditions are required to 
explain the observed ${\rm H_2O}$ line shapes and line intensities. 
We identify two ISM components which are present in all galaxies - an extended
cold (T$_{\rm dust}\sim20-30$\,K) region (ER) and a warm (T$_{\rm dust}\sim50-70$\,K) 
component. The cold ER is only excited by collisions and has significant emission line 
intensities only in the ground-state and a few low-excitation lines. This gas 
component is also often responsible for the detected low-excitation absorption 
features seen in our targets. 
The warm component contributes almost all the emission of 
medium-excitation lines. 
For the two ultraluminous IR galaxies in our sample
(Mrk 231 and Arp 220), our models suggest the presence of an even warmer gas 
component (hot gas) to explain the emission and absorption in the high-excitation 
H$_2$O lines (E$_{\rm up}>350$\,K).

\item The multiple ISM model also allows to explain the observed dust SED
and CO SLED in our target galaxies.  The cold ER contributes mainly to 
millimeter and submm dust continuum. The warm component dominates the
total IR luminosity with its dust SED peaking at FIR regimes, suggesting
that its generated medium-excitation H$_2$O lines are excellent probes to study the 
kinematic of the FIR emitting regions.
The middle/high$-J$ ($7 \le J_{\rm up} \le 12$) CO emission lines
mainly arise from the warm component while the low$-J$ CO transitions
mainly come from the cold ER.  The CO SLEDs of these two components
peak at $4 \le J_{\rm up} \le 6$ and $8 \le J_{\rm up} \le 10$, 
respectively.  The hot component, if present,
contributes large amounts of IR emission in MIR regimes (which however
is usually attenuated by foreground dust) and has a significant effect
on the CO SLED at levels with $J_{\rm up} \ge 10$.

\item IR pumping dominates the excitation of high-excitation energy levels
of water  (with 250 - 350\,K $\le E/k_{\rm B}  \le$ 500 - 700\,K in warm component 
and $600 - 800$\,K $\le E/k_{\rm B}  \le$ 1300 - 1600\,K in hot component), and 
drive their level population
towards a Boltzmann distribution close to the dust temperature.
While collision dominates the excitation of low-excitation energy levels
(with $E/k_{\rm B}  \le$ 100 - 150 K, $E/k_{\rm B}  \le$ 250 - 350\,K and  
$E/k_{\rm B}  \le$ 600 - 800\,K in cold, warm and hot components, respectively), 
and drive some low-excitation level population ($E/k_{\rm B}  \le$ 150 - 200\,K and  
$E/k_{\rm B}  \le$ 400 - 600\,K in warm and hot components, respectively) 
towards a Boltzmann distribution at the kinetic gas temperature.
Our observed low-excitation lines (with 
$E_{\rm up} <  130$ K) and most of the observed medium-excitation lines (with 
$130 < E_{\rm up} <  250 - 350$ K) are excited dominantly by collision.
IR pumping becomes more and more important in exciting our observed 
medium-excitation lines with $E_{\rm up} \ge 200 - 300$ K.

\item The gas phase abundance of ${\rm H_2O}$ varies from 
$10^{-9}-10^{-8}$ in the cold ER, to $10^{-8} - 10^{-7}$ 
in the warm component and increases to $10^{-6}-10^{-5}$ in the 
hot component.
Therefore, our results suggest that the water abundance is typically larger 
in the higher density and warmer regions.

\end{itemize}

\bigskip
\section*{Acknowledgements}
We thank the anonymous referee for their carful reading of our manuscript and
for their constructive comments and suggestions, which improved the paper.
HIFI has been designed and built by a consortium of institutes and university 
departments from across Europe, Canada, and the United States under the 
leadership of SRON Netherlands Institute for Space Research, Groningen, 
The Netherlands, and with major contributions from Germany, France, 
and the US. Consortium members are: Canada: CSA, U.Waterloo; 
France: CESR, LAB, LERMA, IRAM; Germany: KOSMA, MPIfR, MPS; 
Ireland: NUI Maynooth; Italy: ASI, IFSI-INAF, Osservatorio 
Astrofisico di Arcetri – INAF; Netherlands: SRON, TUD; Poland: CAMK, 
CBK; Spain: Observatorio Astronmico Nacional (IGN), 
Centro de Astrobiologia (CSIC-INTA). 
Sweden: Chalmers University of Technology – MC2, RSS \& GARD; 
Onsala Space Observatory; Swedish National Space Board, 
Stockholm University – Stockholm Observatory; Switzerland: ETH Zurich, 
FHNW; USA: Caltech, JPL, NHSC. 
LJ and YG acknowledge support by NSFC grants 
\#11311130491, \#11420101002 and CAS Key Research Program of Frontier
Sciences B program \#XDB09000000. 

\begin{figure*}[h]
\includegraphics[width=\textwidth]{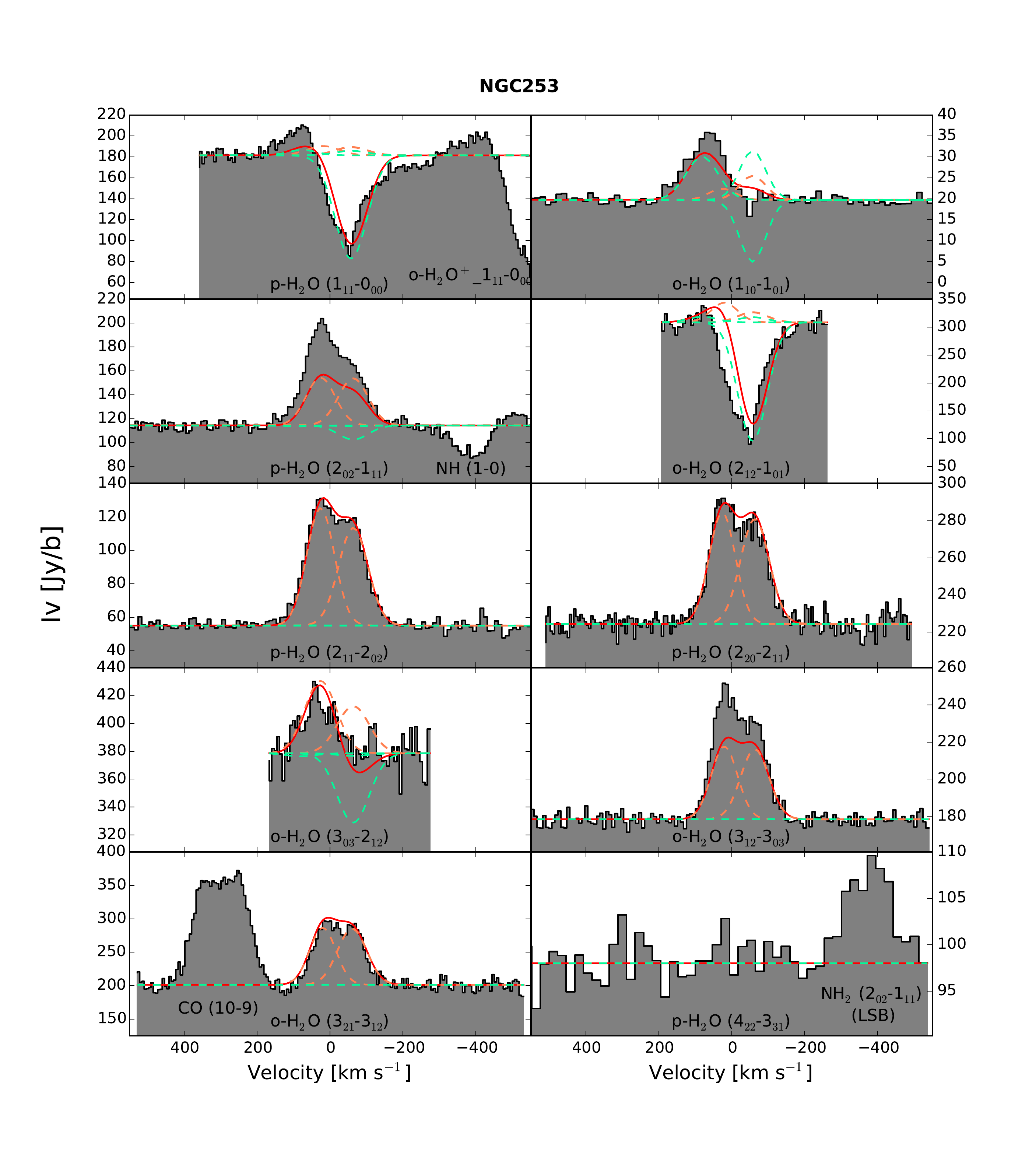}
\caption{The observed ${\rm H_2O}$ spectra (in grey colour) and the 
modelling results (in red solid curve) of NGC 253.
For explanation of the symbols used in the plots, 
see Fig.\ \ref{figure:m82_spectra}.
\label{figure:ngc253_spectra}}
\end{figure*}

\begin{figure*}[h]
\includegraphics[width=\textwidth]{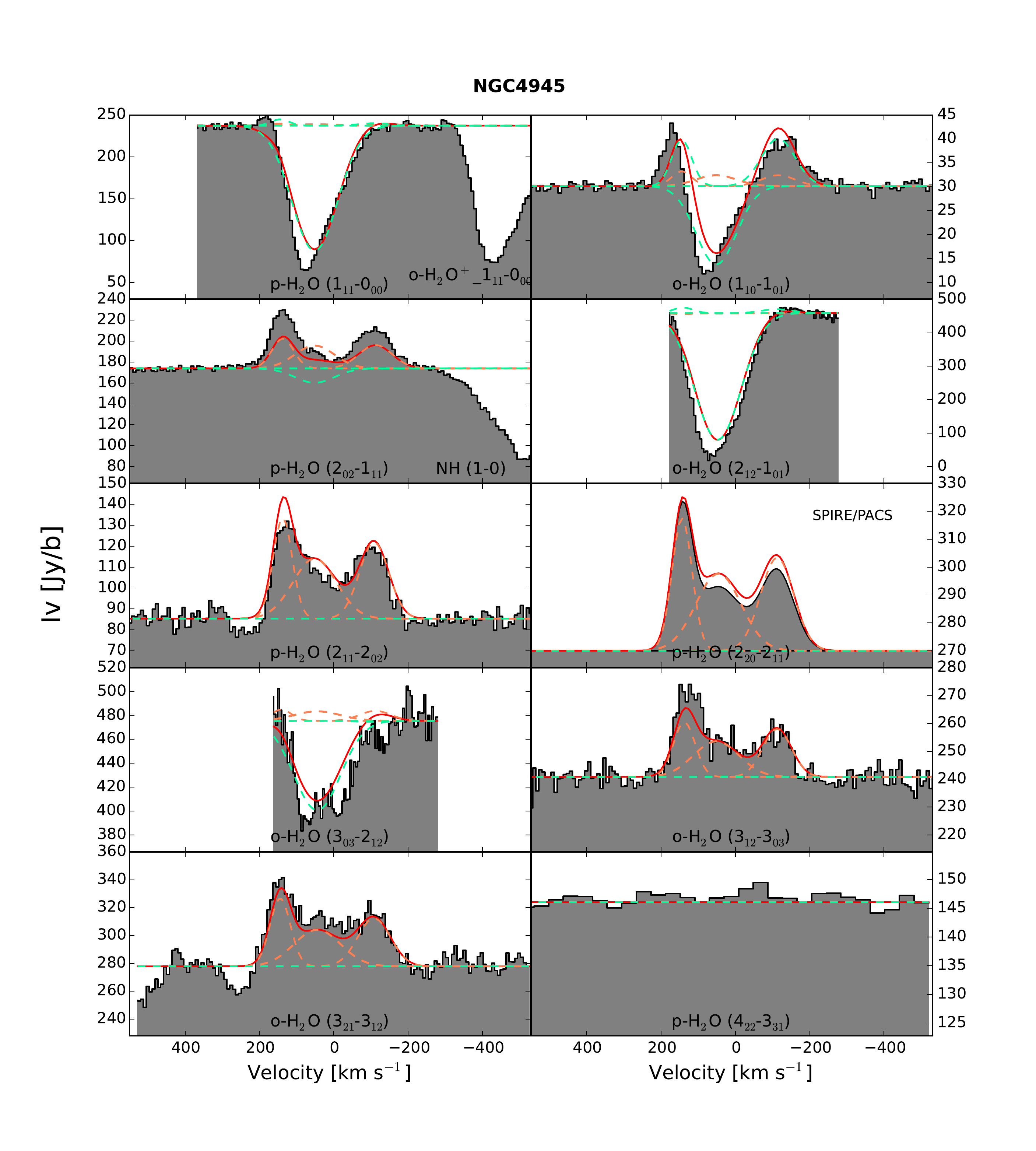}
\caption{The observed ${\rm H_2O}$ spectra (in grey colour) and the 
modelling results (in red solid curve) of NGC 4945.
The black-solid line in the subpanel of p-${\rm H_2O}$ ($2_{20}-2_{11}$)
with text `SPIRE/PACS'
presents the 
 \textit{SPIRE}/\textit{PACS} data whose line profile is
 estimated from  \textit{HIFI}-detected medium-excitation lines.
For explanation of the other symbols used in the plots, 
see Fig.\ \ref{figure:m82_spectra}.
\label{figure:ngc4945_spectra}}
\end{figure*}

\begin{figure*}[h]
\includegraphics[width=\textwidth]{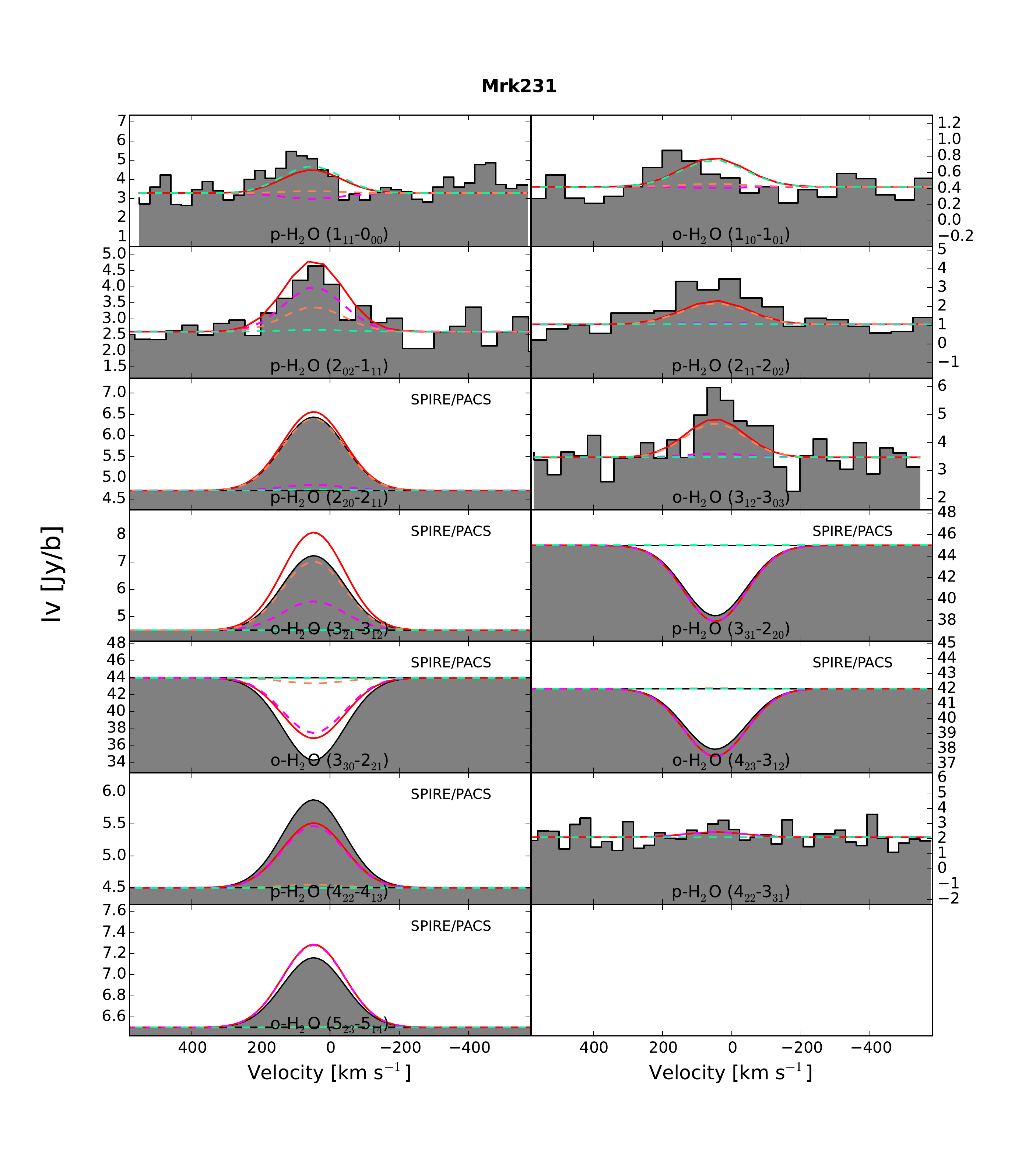}
\caption{The observed ${\rm H_2O}$ spectra (in grey colour) and the 
modelling results (in red solid curve) of Mrk 231.
The individual contributions from the cold ER, warm component and
hot component are displayed by green, orange and magenta colors, 
respectively.
The black-solid lines in the subpanels with text `SPIRE/PACS'
present the estimated line profiles 
for \textit{SPIRE}/\textit{PACS}-detected 
lines, whose line shapes are assumed to be same as
the \textit{HIFI}-detected medium-excitation
lines.
For explanation of the other symbols used in the plots, 
see Fig.\ \ref{figure:m82_spectra}.
\label{figure:mrk231_spectra}}
\end{figure*}

\begin{figure*}[h]
\includegraphics[width=\textwidth]{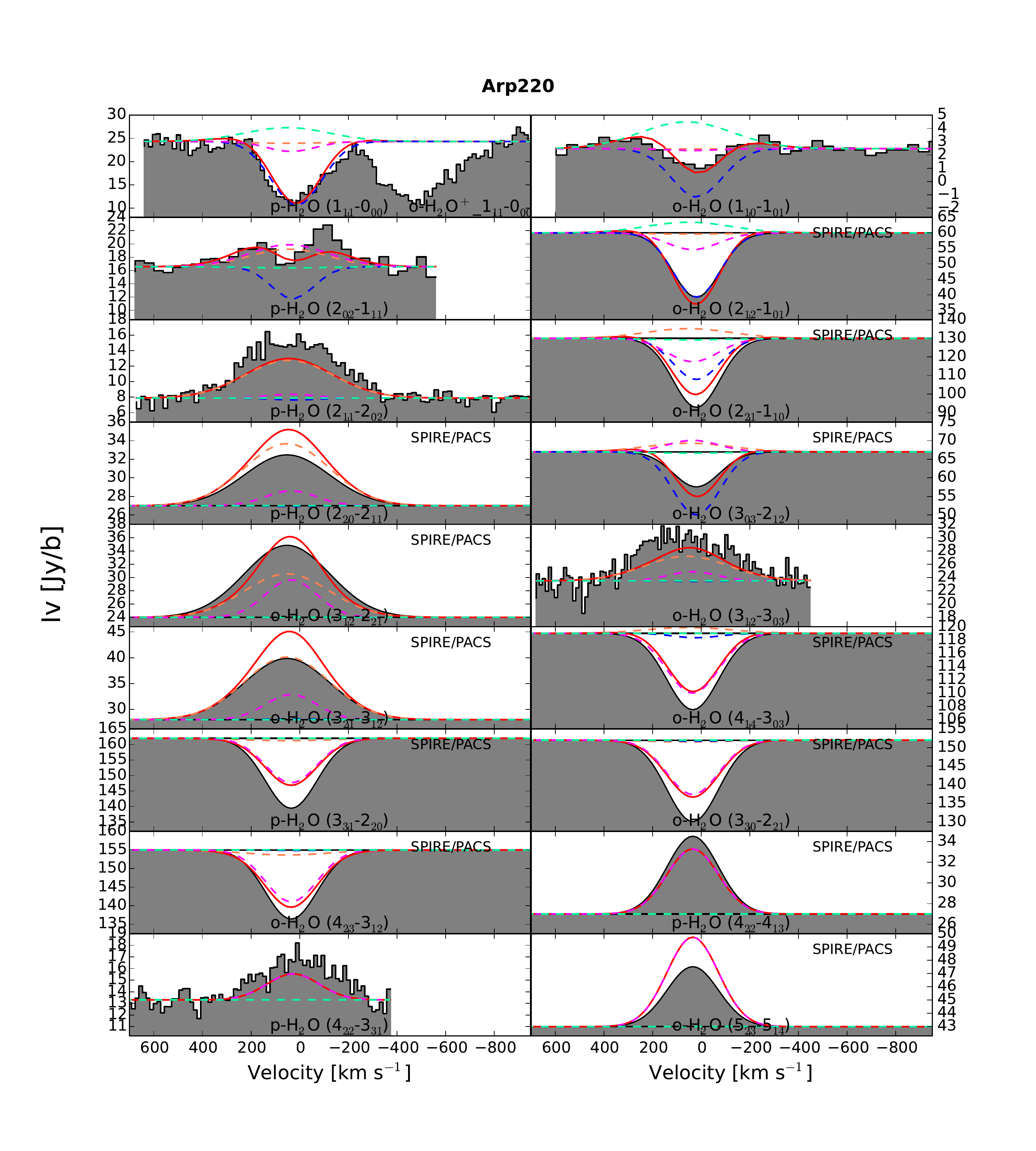}
\caption{The observed ${\rm H_2O}$ spectra (in grey colour) and the 
modelling results (in red solid curve) of Arp 220.
The blue dashed curves denote the absorption of outflow gas
towards the dust continuum of galaxy nuclei.
The black-solid lines in the subpanels with text `SPIRE/PACS'
present the estimated line profiles 
for \textit{SPIRE}/\textit{PACS}-detected 
lines, whose line shapes are assumed to be same as
the \textit{HIFI}-detected 
 high-excitation p-H$_2$O ($4_{22}-3_{31}$) line.
For explanation of the other symbols used in the plots, 
see Fig.\ \ref{figure:m82_spectra}.
\label{figure:arp220_spectra}}
\end{figure*}

\begin{figure*}[h]
\includegraphics[width=18.5cm]{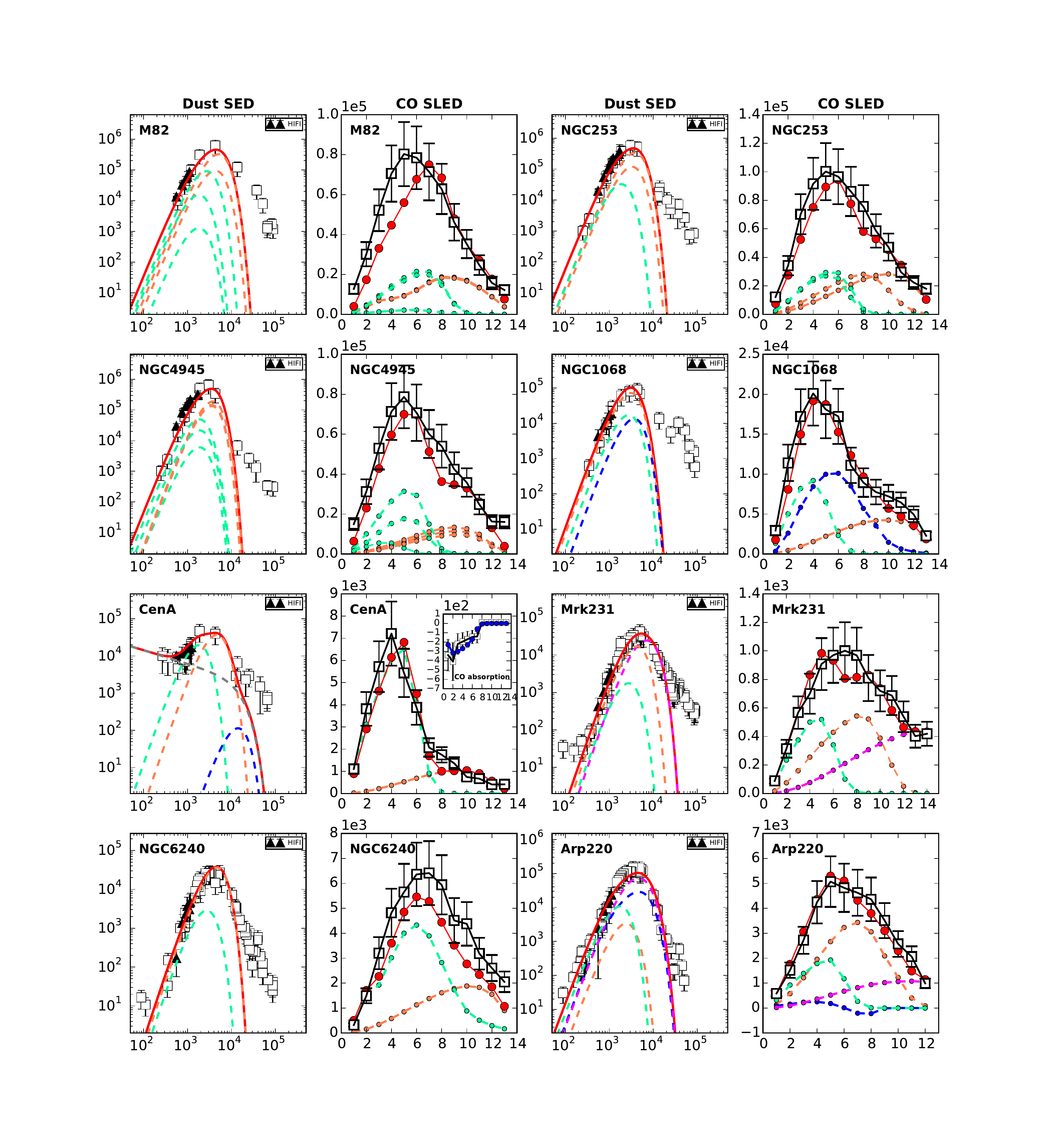}
\caption{The observed and modelled dust SED (in unit of 
mJy/beam) and 
CO SLED (in unit of ${\rm Jy~km~s^{-1}}$) for our 
sample galaxies. The black squares denote the observed dust continuum 
or CO emission fluxes, and the black solid triangles denote 
our \textit{HIFI} measured dust continuum fluxes.
The green, orange and magenta dashed lines indicate the best-fit model 
results for cold ERs, warm components and hot components,
respectively,  whose velocities are given in Table\ \ref{table: model parameters}.
The red solid lines denote the total modelling results of  all components.
The blue dashed lines in subplots of NGC 1068, Cen A and Arp 220
denote the modelling results of outflow gas, absorbing
gas (case I) and outflow absorbing gas, respectively.
The grey dashed line in dust SED plot of Cen A indicates the power-law
continuum emission of the radio core.
The small figure in CO SLED subplot of Cen A presents the 
observed (black open squares) and the modelled (blue solid circles)
low-$J$ CO absorptions towards the radio core.
 }
\label{figure:galaxy_co_IR_summary}
\end{figure*}

\clearpage
\LongTables 
\begin{deluxetable*}{llcccl} 
\tabletypesize{\small}
\renewcommand{\arraystretch}{0.4}
\setlength{\tabcolsep}{1pt}
\tablecolumns{6}
\tablewidth{0pc}
\tablecaption{Line parameters derived from Gaussian decomposition} 
\tablehead{ \colhead{Source} & \colhead{Line} & \colhead{$\delta v$} & \colhead{FWHM} 
 & \colhead{I} & \colhead{$I_{\nu}$ cont.} \\
 & & \colhead{${\rm [km~s^{-1}]}$} & \colhead{${\rm [km~s^{-1}]}$}  & \colhead{${\rm [Jy 
~beam^{-1}~km~s^{-1}]}$} & \colhead{${\rm [Jy~beam^{-1}]}$}  }
\startdata
M82 & o-H$_2$O (1$_{10}$-1$_{01}$) & 86 $\pm$ 2 & 122 $\pm$ 6  & 320 $\pm$ 0 & 13.6 $\pm$ 3.4 (538 $\mu$m) \\ 
  &  & -73 $\pm$ 2 & 103 $\pm$ 5  & 333 $\pm$ 25 &  \\ 
  &  & 26 $\pm$ 2 & 65 $\pm$ 1  & -137 $\pm$ 21 &  \\ 
  & p-H$_2$O (1$_{11}$-0$_{00}$) & 86 $\pm$ 2 & 122 $\pm$ 6  & $\le$ 153 & 85.0 $\pm$ 13.6 (269 $\mu$m) \\ 
  &  & -73 $\pm$ 2 & 103 $\pm$ 5  & $\le$ 140 &  \\ 
  &  & 26 $\pm$ 2 & 65 $\pm$ 1  & -869 $\pm$ 34 &  \\ 
  & p-H$_2$O (2$_{02}$-1$_{11}$) & 86 $\pm$ 2 & 122 $\pm$ 6  & 518 $\pm$ 76 & 59.7 $\pm$ 27.9 (303 $\mu$m) \\ 
  &  & -73 $\pm$ 2 & 103 $\pm$ 5  & 701 $\pm$ 70 &  \\ 
  &  & 26 $\pm$ 2 & 65 $\pm$ 1  & $\le$ 139 &  \\ 
  & p-H$_2$O (2$_{11}$-2$_{02}$) & 86 $\pm$ 2 & 122 $\pm$ 6  & 458 $\pm$ 54 & 33.2 $\pm$ 5.0 (398 $\mu$m) \\ 
  &  & -73 $\pm$ 2 & 103 $\pm$ 5  & 584 $\pm$ 41 &  \\ 
  &  & 26 $\pm$ 2 & 65 $\pm$ 1  & $\le$ 83 &  \\ 
  & o-H$_2$O (3$_{12}$-3$_{03}$) & 86 $\pm$ 2 & 122 $\pm$ 6  & 209 $\pm$ 69 & 85.3 $\pm$ 9.6 (273 $\mu$m) \\ 
  &  & -73 $\pm$ 2 & 103 $\pm$ 5  & 545 $\pm$ 63 &  \\ 
  &  & 26 $\pm$ 2 & 65 $\pm$ 1  & $\le$ 102 &  \\ 
  & p-H$_2$O (4$_{22}$-3$_{31}$) & - & - & $\le231$ & 53.0 $\pm$ 8.5 (327 $\mu$m) \\ 
\cline{1-6} \\
NGC 253 & o-H$_2$O (1$_{10}$-1$_{01}$) & 23 $\pm$ 3 & 91 $\pm$ 11  & 639 $\pm$ 92 & 19.7 $\pm$ 4.1 (538 $\mu$m) \\ 
  &  & -60 $\pm$ 3 & 96 $\pm$ 11  & -106 $\pm$ 40 &  \\ 
  &  & 80 $\pm$ 3 & 94 $\pm$ 9  & 1316 $\pm$ 75 &  \\ 
  & p-H$_2$O (1$_{11}$-0$_{00}$) & 23 $\pm$ 3 & 91 $\pm$ 11  & -2024 $\pm$ 234 & 181.5 $\pm$ 34.9 (269 $\mu$m) \\ 
  &  & -60 $\pm$ 3 & 96 $\pm$ 11  & -8548 $\pm$ 160 &  \\ 
  &  & 80 $\pm$ 3 & 94 $\pm$ 9  & 3629 $\pm$ 245 &  \\ 
  & o-H$_2$O (2$_{12}$-1$_{01}$) & 23 $\pm$ 3 & 91 $\pm$ 11  & -4179 $\pm$ 170 & 308.7 $\pm$ 70.8 (179 $\mu$m) \\ 
  &  & -60 $\pm$ 3 & 96 $\pm$ 11  & -18579 $\pm$ 214 &  \\ 
  &  & 80 $\pm$ 3 & 94 $\pm$ 9  & $\le$ 905 &  \\ 
  & p-H$_2$O (2$_{02}$-1$_{11}$) & 23 $\pm$ 3 & 91 $\pm$ 11  & 8001 $\pm$ 454 & 114.5 $\pm$ 26.4 (303 $\mu$m) \\ 
  &  & -60 $\pm$ 3 & 96 $\pm$ 11  & 4640 $\pm$ 460 &  \\ 
  &  & 80 $\pm$ 3 & 94 $\pm$ 9  & $\le$ 300 &  \\ 
  & p-H$_2$O (2$_{11}$-2$_{02}$) & 23 $\pm$ 3 & 91 $\pm$ 11  & 6524 $\pm$ 155 & 55.0 $\pm$ 14.0 (398 $\mu$m) \\ 
  &  & -60 $\pm$ 3 & 96 $\pm$ 11  & 5927 $\pm$ 161 &  \\ 
  &  & 80 $\pm$ 3 & 94 $\pm$ 9  & $\le$ 268 &  \\ 
  & p-H$_2$O (2$_{20}$-2$_{11}$) & 23 $\pm$ 3 & 91 $\pm$ 11  & 5496 $\pm$ 181 & 224.4 $\pm$ 30.3 (243 $\mu$m) \\ 
  &  & -60 $\pm$ 3 & 96 $\pm$ 11  & 4997 $\pm$ 250 &  \\ 
  &  & 80 $\pm$ 3 & 94 $\pm$ 9  & $\le$ 372 &  \\ 
  & o-H$_2$O (3$_{03}$-2$_{12}$) & 23 $\pm$ 3 & 91 $\pm$ 11  & 4171 $\pm$ 215 & 378.6 $\pm$ 237.7 (174 $\mu$m) \\ 
  &  & -60 $\pm$ 3 & 96 $\pm$ 11  & $\le$ 990 &  \\ 
  &  & 80 $\pm$ 3 & 94 $\pm$ 9  & $\le$ 944 &  \\ 
  & o-H$_2$O (3$_{12}$-3$_{03}$) & 23 $\pm$ 3 & 91 $\pm$ 11  & 5450 $\pm$ 123 & 178.4 $\pm$ 36.5 (273 $\mu$m) \\ 
  &  & -60 $\pm$ 3 & 96 $\pm$ 11  & 4692 $\pm$ 151 &  \\ 
  &  & 80 $\pm$ 3 & 94 $\pm$ 9  & $\le$ 302 &  \\ 
  & o-H$_2$O (3$_{21}$-3$_{12}$) & 23 $\pm$ 3 & 91 $\pm$ 11  & 6316 $\pm$ 382 & 201.3 $\pm$ 32.0 (257 $\mu$m) \\ 
  &  & -60 $\pm$ 3 & 96 $\pm$ 11  & 8092 $\pm$ 404 &  \\ 
  &  & 80 $\pm$ 3 & 94 $\pm$ 9  & $\le$ 592 &  \\ 
  & p-H$_2$O (4$_{22}$-3$_{31}$) & - & - & $\le1143$ & 98.0 $\pm$ 24.3 (327 $\mu$m) \\ 
\cline{1-6} \\
NGC 4945 & o-H$_2$O (1$_{10}$-1$_{01}$) & -112 $\pm$ 3 & 98 $\pm$ 7  & 1122 $\pm$ 38 & 30.1 $\pm$ 5.1 (538 $\mu$m) \\ 
  &  & 48 $\pm$ 3 & 136 $\pm$ 10  & -2348 $\pm$ 66 &  \\ 
  &  & 141 $\pm$ 3 & 63 $\pm$ 5  & 769 $\pm$ 46 &  \\ 
  & p-H$_2$O (1$_{11}$-0$_{00}$) & -112 $\pm$ 3 & 98 $\pm$ 7  & $\le$ 295 & 237.4 $\pm$ 36.6 (269 $\mu$m) \\ 
  &  & 48 $\pm$ 3 & 136 $\pm$ 10  & -25771 $\pm$ 286 &  \\ 
  &  & 141 $\pm$ 3 & 63 $\pm$ 5  & 1579 $\pm$ 138 &  \\ 
  & o-H$_2$O (2$_{12}$-1$_{01}$) & -112 $\pm$ 3 & 98 $\pm$ 7  & $\le$ 909 & 458.7 $\pm$ 69.4 (179 $\mu$m) \\ 
  &  & 48 $\pm$ 3 & 136 $\pm$ 10  & -68706 $\pm$ 291 &  \\ 
  &  & 141 $\pm$ 3 & 63 $\pm$ 5  & $\le$ 729 &  \\ 
  & p-H$_2$O (2$_{02}$-1$_{11}$) & -112 $\pm$ 3 & 98 $\pm$ 7  & 4301 $\pm$ 87 & 173.9 $\pm$ 24.8 (303 $\mu$m) \\ 
  &  & 48 $\pm$ 3 & 136 $\pm$ 10  & 2032 $\pm$ 84 &  \\ 
  &  & 141 $\pm$ 3 & 63 $\pm$ 5  & 3890 $\pm$ 60 &  \\ 
  & p-H$_2$O (2$_{11}$-2$_{02}$) & -112 $\pm$ 3 & 98 $\pm$ 7  & 3430 $\pm$ 207 & 85.3 $\pm$ 15.4 (398 $\mu$m) \\ 
  &  & 48 $\pm$ 3 & 136 $\pm$ 10  & 3939 $\pm$ 336 &  \\ 
  &  & 141 $\pm$ 3 & 63 $\pm$ 5  & 2285 $\pm$ 198 &  \\ 
  & o-H$_2$O (3$_{03}$-2$_{12}$) & -112 $\pm$ 3 & 98 $\pm$ 7  & $\le$ 891 & 475.5 $\pm$ 228.1 (174 $\mu$m) \\ 
  &  & 48 $\pm$ 3 & 136 $\pm$ 10  & -12823 $\pm$ 231 &  \\ 
  &  & 141 $\pm$ 3 & 63 $\pm$ 5  & $\le$ 714 &  \\ 
  & o-H$_2$O (3$_{12}$-3$_{03}$) & -112 $\pm$ 3 & 98 $\pm$ 7  & 1548 $\pm$ 120 & 240.8 $\pm$ 39.2 (273 $\mu$m) \\ 
  &  & 48 $\pm$ 3 & 136 $\pm$ 10  & 2362 $\pm$ 281 &  \\ 
  &  & 141 $\pm$ 3 & 63 $\pm$ 5  & 1842 $\pm$ 169 &  \\ 
  & o-H$_2$O (3$_{21}$-3$_{12}$) & -112 $\pm$ 3 & 98 $\pm$ 7  & 4135 $\pm$ 304 & 277.9 $\pm$ 31.5 (257 $\mu$m) \\ 
  &  & 48 $\pm$ 3 & 136 $\pm$ 10  & 5633 $\pm$ 342 &  \\ 
  &  & 141 $\pm$ 3 & 63 $\pm$ 5  & 3658 $\pm$ 255 &  \\ 
  & p-H$_2$O (4$_{22}$-3$_{31}$) & - & - & $\le593$ & 146.1 $\pm$ 17.4 (327 $\mu$m) \\ 
\cline{1-6} \\
NGC 1068 & o-H$_2$O (1$_{10}$-1$_{01}$) & -18 $\pm$ 2 & 211 $\pm$ 10  & 557 $\pm$ 18 & 4.3 $\pm$ 2.4 (538 $\mu$m) \\ 
  & p-H$_2$O (1$_{11}$-0$_{00}$) & -18 $\pm$ 2 & 211 $\pm$ 10  & 1552 $\pm$ 72 & 18.9 $\pm$ 10.9 (269 $\mu$m) \\ 
  & p-H$_2$O (2$_{02}$-1$_{11}$) & -18 $\pm$ 2 & 211 $\pm$ 10  & 1554 $\pm$ 94 & 17.1 $\pm$ 11.4 (303 $\mu$m) \\ 
  & p-H$_2$O (2$_{11}$-2$_{02}$) & -18 $\pm$ 2 & 211 $\pm$ 10  & 1264 $\pm$ 74 & 8.7 $\pm$ 3.4 (398 $\mu$m) \\ 
  & p-H$_2$O (4$_{22}$-3$_{31}$) & - & - & $\le137$ & 13.6 $\pm$ 6.0 (327 $\mu$m) \\ 
\cline{1-6} \\
Cen A & o-H$_2$O (1$_{10}$-1$_{01}$) & 0 $\pm$ 2 & 235 $\pm$ 11  & 682 $\pm$ 58 & 10.1 $\pm$ 2.6 (538 $\mu$m) \\ 
  &  & 16 $\pm$ 2 & 70 $\pm$ 3  & -384 $\pm$ 32 &  \\ 
  & p-H$_2$O (1$_{11}$-0$_{00}$) & 0 $\pm$ 2 & 235 $\pm$ 11  & 384 $\pm$ 138 & 16.9 $\pm$ 7.0 (269 $\mu$m) \\ 
  &  & 16 $\pm$ 2 & 70 $\pm$ 3  & -358 $\pm$ 75 &  \\ 
  & p-H$_2$O (2$_{02}$-1$_{11}$) & 0 $\pm$ 2 & 235 $\pm$ 11  & 631 $\pm$ 65 & 12.4 $\pm$ 8.3 (303 $\mu$m) \\ 
  &  & 16 $\pm$ 2 & 70 $\pm$ 3  & $\le$ 84 &  \\ 
  & p-H$_2$O (2$_{11}$-2$_{02}$) & - & - & $\le606$ & 11.6 $\pm$ 5.8 (398 $\mu$m) \\ 
  & p-H$_2$O (2$_{20}$-2$_{11}$) & - & - & $\le619$ & 17.1 $\pm$ 11.4 (243 $\mu$m) \\ 
  & o-H$_2$O (3$_{12}$-3$_{03}$) & - & - & $\le489$ & 15.0 $\pm$ 6.6 (273 $\mu$m) \\ 
  & o-H$_2$O (3$_{21}$-3$_{12}$) & - & - & $\le734$ & 18.4 $\pm$ 12.2 (257 $\mu$m) \\ 
  & p-H$_2$O (4$_{22}$-3$_{31}$) & - & - & $\le425$ & 11.6 $\pm$ 7.7 (327 $\mu$m) \\ 
\cline{1-6} \\
Mrk 231 & o-H$_2$O (1$_{10}$-1$_{01}$) & 50 $\pm$ 2 & 211 $\pm$ 10  & 66 $\pm$ 77 & 0.4 $\pm$ 0.3 (538 $\mu$m) \\ 
  & p-H$_2$O (1$_{11}$-0$_{00}$) & 50 $\pm$ 2 & 211 $\pm$ 10  & 362 $\pm$ 44 & 3.3 $\pm$ 2.2 (269 $\mu$m) \\ 
  & p-H$_2$O (2$_{02}$-1$_{11}$) & 50 $\pm$ 2 & 211 $\pm$ 10  & 376 $\pm$ 54 & 2.6 $\pm$ 1.7 (303 $\mu$m) \\ 
  & p-H$_2$O (2$_{11}$-2$_{02}$) & 50 $\pm$ 2 & 211 $\pm$ 10  & 588 $\pm$ 67 & 1.1 $\pm$ 0.7 (398 $\mu$m) \\ 
  & o-H$_2$O (3$_{12}$-3$_{03}$) & 50 $\pm$ 2 & 211 $\pm$ 10  & 390 $\pm$ 63 & 3.5 $\pm$ 2.3 (273 $\mu$m) \\ 
  & p-H$_2$O (4$_{22}$-3$_{31}$) & - & - & $\le191$ & 2.1 $\pm$ 1.4 (327 $\mu$m) \\ 
\cline{1-6} \\
Antennae & o-H$_2$O (1$_{10}$-1$_{01}$) & - & - &  $\le 68$ & 0.90 $\pm$ 0.5 (538 $\mu$m) \\ 
                 & p-H$_2$O (1$_{11}$-0$_{00}$) & - & - &  $\le 126$ & 4.65 $\pm$ 3.0 (269 $\mu$m) \\ 
                 & p-H$_2$O (2$_{02}$-1$_{11}$) & - & - &  $\le 233$ & 1.96 $\pm$ 1.2 (303 $\mu$m) \\ 
                 & p-H$_2$O (2$_{11}$-2$_{02}$) & - & - &  $\le 125$ & 1.71 $\pm$ 1.5  (398 $\mu$m) \\ 
                 & o-H$_2$O (3$_{12}$-3$_{03}$) & - & - &  $\le 171$ &  3.00 $\pm$  2.0 (273 $\mu$m) \\ 
                 & p-H$_2$O (4$_{22}$-3$_{31}$) & - & - &  $\le 148$ & 2.25  $\pm$ 1.9  (327 $\mu$m)  \\
\cline{1-6} \\
NGC 6240 & o-H$_2$O (1$_{10}$-1$_{01}$) & -10 $\pm$ 2 & 282 $\pm$ 14  & 175 $\pm$ 33 & 0.5 $\pm$ 0.3 (538 $\mu$m) \\ 
  & p-H$_2$O (1$_{11}$-0$_{00}$) & - & - & $\le334$ & 4.5 $\pm$ 3.0 (269 $\mu$m) \\ 
  & p-H$_2$O (2$_{02}$-1$_{11}$) & -10 $\pm$ 2 & 282 $\pm$ 14  & 660 $\pm$ 116 & 4.6 $\pm$ 3.0 (303 $\mu$m) \\ 
  & p-H$_2$O (2$_{11}$-2$_{02}$) & -10 $\pm$ 2 & 282 $\pm$ 14  & 592 $\pm$ 81 & 1.3 $\pm$ 0.9 (398 $\mu$m) \\ 
  & o-H$_2$O (3$_{12}$-3$_{03}$) & -10 $\pm$ 2 & 282 $\pm$ 14  & 253 $\pm$ 53 & 3.1 $\pm$ 2.1 (273 $\mu$m) \\ 
  & o-H$_2$O (3$_{21}$-3$_{12}$) & - & - & $\le546$ & 4.8 $\pm$ 3.2 (257 $\mu$m) \\ 
  & p-H$_2$O (4$_{22}$-3$_{31}$) & - & - & $\le375$ & 2.3 $\pm$ 1.6 (327 $\mu$m) \\ 
\cline{1-6} \\
Arp 220 & o-H$_2$O (1$_{10}$-1$_{01}$) & 52 $\pm$ 5 & 412 $\pm$ 32  & 812 $\pm$ 133 & 2.5 $\pm$ 1.6 (538 $\mu$m) \\ 
  &  & 20 $\pm$ 5 & 226 $\pm$ 18  & -850 $\pm$ 98 &  \\ 
  &  & 35 $\pm$ 5 & 235 $\pm$ 18  & $\le$ 96 &  \\ 
  & p-H$_2$O (1$_{11}$-0$_{00}$) & 52 $\pm$ 5 & 412 $\pm$ 32  & $\le$ 449 & 24.3 $\pm$ 14.3 (269 $\mu$m) \\ 
  &  & 20 $\pm$ 5 & 226 $\pm$ 18  & -3486 $\pm$ 141 &  \\ 
  &  & 35 $\pm$ 5 & 235 $\pm$ 18  & $\le$ 339 &  \\ 
  & p-H$_2$O (2$_{02}$-1$_{11}$) & 52 $\pm$ 5 & 412 $\pm$ 32  & 3162 $\pm$ 469 & 16.6 $\pm$ 11.1 (303 $\mu$m) \\ 
  &  & 20 $\pm$ 5 & 226 $\pm$ 18  & -1369 $\pm$ 477 &  \\ 
  &  & 35 $\pm$ 5 & 235 $\pm$ 18  & $\le$ 288 &  \\ 
  & p-H$_2$O (2$_{11}$-2$_{02}$) & 52 $\pm$ 5 & 412 $\pm$ 32  & 3481 $\pm$ 91 & 7.9 $\pm$ 4.2 (398 $\mu$m) \\ 
  &  & 20 $\pm$ 5 & 226 $\pm$ 18  & $\le$ 132 &  \\ 
  &  & 35 $\pm$ 5 & 235 $\pm$ 18  & $\le$ 134 &  \\ 
  & o-H$_2$O (3$_{12}$-3$_{03}$) & 52 $\pm$ 5 & 412 $\pm$ 32  & 3021 $\pm$ 156 & 23.5 $\pm$ 11.3 (273 $\mu$m) \\ 
  &  & 20 $\pm$ 5 & 226 $\pm$ 18  & $\le$ 232 &  \\ 
  &  & 35 $\pm$ 5 & 235 $\pm$ 18  & $\le$ 237 &  \\ 
  & p-H$_2$O (4$_{22}$-3$_{31}$) & 52 $\pm$ 5 & 412 $\pm$ 32  & $\le$ 188 & 13.3 $\pm$ 8.3 (327 $\mu$m) \\ 
  &  & 20 $\pm$ 5 & 226 $\pm$ 18  & $\le$ 139 &  \\ 
  &  & 35 $\pm$ 5 & 235 $\pm$ 18  & 811 $\pm$ 92 &  
\enddata
\tablecomments{\mbox{The errors of line centre ($\delta v$) and line width (${\rm FWHM}$) 
are not the real fitted errors, but the ranges by}
\mbox{which the parameters are allowed to vary. We allow the position of each Gaussian 
component to change by $\pm~2-3$}
\mbox{ ${\rm km~s^{-1}}$ ($\pm~5$ ${\rm km~s^{-1}}$ in Arp 220) and the line width 
by $\sim \pm~5\%$ ($\sim \pm~8\%$ in Arp 220). The upper limits to water }
\mbox{intensities in Antennae are derived by adopting a FWHM 
$=200~{\rm km~s^{-1}}$ from CO observation by \citet{gao2001}.}}
\label{table:gaussian_decomposition}
\end{deluxetable*}

\clearpage
\begin{deluxetable*}{llcccccccc}
\tabletypesize{\small}
\renewcommand{\arraystretch}{0.4}
\setlength{\tabcolsep}{1pt}
\tablecolumns{8}
\tablewidth{0pc}
\tablecaption{Physical parameters derived from best-fit models} 
\tablehead{&  & \colhead{$\delta v$} & \colhead{$Rs$} & \colhead{$n$(H)} & 
\colhead{$T_{\rm k}$} & \colhead{$T_{\rm dust}$} & \colhead{$N({\rm H})$} & 
\colhead{$X_{\rm H_{2}O}$} & \colhead{$M({\rm H})$} \\
\colhead{Source} & \colhead{Component} & \colhead{${\rm [km~s^{-1}]}$} & 
\colhead{[pc]} & \colhead{[${\rm cm^{-3}}$]} & \colhead{[K]} & \colhead{[K]} & 
\colhead{[${\rm 10^{23} ~ cm^{-2}}$]} & & \colhead{[${\rm 10^6~M_{\odot}\,}$]}}
\startdata
M82 & warm & -73 & 17 - 25 & $1 \times 10^4 - 10^5 $ & $80 - 160$ & $50 - 70$ & 
         $ 10 - 20 $ & $\sim 1 \times 10^{-8}$ & $7 - 30$ \\
        & warm & 86 & 17 - 25 & $\sim 1 \times 10^5 $ & $80 - 160$ & $\sim 50$ & 
         $ 10 - 20 $ & $\sim 1 \times 10^{-8}$ & $7- 30$ \\
         & ER  & -73, 86 & 60 - 90 & $\sim 1 \times 10^{5}$ & $20 - 30$ &  $20 - 30$ &
     $6 - 10$ & $\sim 1 \times 10^{-8}$ & $55-200$ \\
         & absorbing gas & 26 & $10 - 15$ & $\sim 1 \times 10^{4}$ & $40 - 160$ &  $20 - 30$ &
        $3 - 10$ & $1\times 10^{-9} - 10^{-8}$ & $0.7-6$ \\
\cline{1-10} \\
NGC 253 & warm & 23, -60 & 30 - 45 & $1 \times 10^5 - 10^6$ & $50 - 70$ & $40 - 50$ & 
         $ 10 - 20$ & $\sim 1 \times 10^{-7}$ & $22 - 100$ \\
      & ER & 80, -60 & 160 - 200 & $\sim 1 \times 10^{5}$ & $\sim 20$ &  $\sim 20$ &
     $3 - 6$ & $\sim 1 \times 10^{-7}$ & $190 - 600$\\
        & absorbing gas & -60 & 50 - 60 & $\sim 1 \times 10^4 $ & $\sim 20$ &  $\sim 20$ &
     $6 - 10$ & $\sim 1 \times 10^{-7}$ & $35 - 90$ \\
\cline{1-10} \\
NGC 4945 & warm & -112, 48, 141 & 30 - 40 & $\sim 1 \times 10^6$ & 50 - 70 & 50 - 60
& $20 - 40$ & $\sim 1 \times 10^{-7}$ & $45 - 160$\\
		& ER & -112, 141 & 120 - 180 & $1 \times 10^4 - 10^5$ & $\sim 20$ & $\sim 20$
& $6 - 10$ & $\sim 1 \times 10^{-8}$ & $200 - 800$\\
		& absorbing gas   & 48 & 50 - 70 & $1 \times 10^4 - 10^5$ & $\sim 20$ & $\sim 20$
& $6 - 10$ & $\sim 1 \times 10^{-8}$ & $40 - 130$ \\
\cline{1-10}  \\
NGC 1068 & warm & -18 & 70 - 100 & $\sim 1 \times 10^6$ & $50 - 60$ & $40 - 50$ & 
         $ 40 - 60$ & $\sim 1 \times 10^{-7}$ & $500 - 1500$\\
		 & ER  & -18 & 400 - 450 & $1 \times 10^3 - 10^4$ & $20 - 30$ &
$20 - 30$ & $ \sim 1$ & $\sim 1 \times 10^{-7}$ & $400 - 500$ \\
	       & outflow  & -18 & 150 - 200 & $\sim 1 \times 10^4$ & $120 - 180$ &
$30 - 40$ & $ \sim 1$ & $\sim 1 \times 10^{-7}$ & $55 - 100$ \\
\cline{1-10}  \\
Cen A   & warm & 0 & 7 - 15 & $1 \times 10^5 - 10^6$ & $50 - 80$ 
& $40 - 60$ & $ 10 - 60$ & $1 \times 10^{-7}-10^{-6}$ & $1-35$ \\
        & ER        & 0 & 70 - 100 & $\sim 1 \times 10^4$ & $\sim 20$ 
& $\sim20$ & $ 6 - 10$ & $\sim 1 \times 10^{-8}$ & $70 - 250$ \\
& absorbing gas & 16 & $\leq 1$ pc & $\sim 1 \times 10^3 $ & $120 - 180$ 
& $20 - 180$ & $ 1 - 3$ & $\sim 1 \times 10^{-7}$ &  $\le 0.007$ \\
& (case I) & & & & & & & & \\
& absorbing gas & 16 & $\leq 1$ pc & $1 \times 10^4 - 10^5 $ & $20 - 30$ 
& $20 - 30$ & $ 1 - 3$ & $\sim 1 \times 10^{-7}$ & $\le 0.007$ \\
& (case II) & & & & & & & & \\
\cline{1-10}  \\
Mrk 231    & warm & 50 & 300 - 550 & $\sim 1 \times 10^5 $ & $50 - 70$ & $
50 - 60$ & $ 10 - 20 $ & $\sim 1 \times 10^{-7}$ & $2 - 15 \times 10^3$\\
		& ER  & 50 & 1000 - 1500 & $\sim 1 \times 10^4 $ & $30 - 50$ & $
\sim 30$ & $ 1 - 3$ & $\sim 1 \times 10^{-7}$ & $3 - 17 \times 10^3$ \\
                 & hot & 50 & 60 - 80 & $\sim 1 \times 10^6 $ & $180 - 200$ & $160 - 180$ & 
         $40 - 60$ & $\sim 1 \times 10^{-6}$ & $4 - 10 \times 10^2$ \\
\cline{1-10} \\
NGC 6240 & warm & -10 & 250 - 350 & $\sim 1 \times 10^6$ & $60 - 70$ & $60 - 70$ & 
         $ 10 - 20$ & $\sim 1 \times 10^{-7}$ & $15 - 60 \times 10^2$\\
		& ER  & -10 & 800 - 1000 & $1 \times 10^3 - 10^4$ & $120
- 400$ & $20 - 30$ & $ 1 - 3$ & $\sim 1 \times 10^{-7}$ & $16 - 75 \times 10^2$\\
\cline{1-10} \\
Arp 220     & warm & 52 & 120 - 150 & $1 \times 10^4 - 10^5$ & $50 - 60$ 
      & $30 - 40$ & $ 40 - 60$ & $1 \times 10^{-7} - 10^{-6}$ & $14 - 34 \times 10^2$ \\
		& ER  & 52 & 1000 - 1200 & $\sim 1 \times 10^4$ & $\sim 20$ 
        & $\sim 20$ & $ 10 - 20$ & $\sim 1 \times 10^{-7}$ & $25 - 75 \times 10^3$\\
                  & absorbing gas & 20 & 70 - 100 & $1 \times 10^4 - 10^5$ & $100 - 200$ 
& $60 - 80$ & $300 - 600$ & $1 \times 10^{-8} - 10^{-7}$ & $35 - 150 \times 10^2$ \\
                & hot & 35 & 70 - 100 & $1 \times 10^5 - 10^6 $ & $100 - 180$ & $100 - 180$ & 
$\sim 100$ & $1 \times 10^{-6} - 10^{-5}$ & $12 - 25 \times 10^2$ 
\enddata
\tablecomments{\mbox{The physical parameters are given for a ISM component at a single 
velocity, and thereby the total size and gas mass}
\mbox{of a ISM component are sum of the values
 derived for different velocities ($\delta v$). The continuum covering factor of the absorbing} 
\mbox{gas  can be estimated as the total size of the 
absorbing gas compared to the total size of the warm background gas.}
}
\label{table: model parameters}
\end{deluxetable*}

\bigskip
\newpage
\appendix
\section{A. The Methods of Line Modelling}
\subsection{A.1 Dust SED Model} \label{appendix: dust SED model}
The dust and gas are assumed to be coexistent in our model 
and their mass ratio is set to be the Galactic value of  $1 : 100$ \citep[e.g.][]{draine2007}.
Therefore, the frequency-dependent dust emission arising from a  
single ISM component with uniform physical properties 
can be calculated with a modified blackbody spectrum:
\begin{equation}
	S_{\nu}=B_{\nu}(T_{d})(1-e^{-\tau(\nu)})\Omega_{s},
\end{equation}	
where $B_{\nu}$ is the Planck function, $\tau(\nu)$ is the 
total dust optical depth and $\Omega_{s}$ is the solid angle of the ISM component determined by 
$\Omega_{s}=\pi R_{s}^{2}/d^2$ ($d$ is the distance of galaxy).
The dust optical depth $\tau(\nu)$ was computed by:
\begin{equation}
	\tau(\nu)=\kappa_{d}(\nu)N_{\rm H}\mu_{H}/100,
\end{equation}	
where $N_{H}$ is the total proton column density in ${\rm cm}^{-2}$,
$\mu_{H}$ is the mass of a hydrogen nucleus in g and $\kappa_{d}(\nu)$ is the 
frequency dependence of the dust absorption coefficient 
in a form of $\kappa_{d}(\nu)=0.4(\nu/250~{\rm GHz})^\beta$ in unit of ${\rm
cm^2~g^{-1}}$ \citep{priddey2001}. 
The dust emissivity index $\beta$ is set to 1.5.

The combination of dust SEDs from multiple ISM
components depends on their physical distribution relative to each other.
If they do not spatially overlap, the total dust SED is simply
the sum of the individual dust SEDs from all ISM components.
Otherwise, the absorption of dust continuum emission of the background component
by the foreground dust needs to be taken into account:
\begin{equation}
	S_{\nu}=B_{\nu}(T_{d})(1-e^{-\tau(\nu)})\Omega'_{s} e^{-\tau'(\nu)},
\end{equation}	
where $\Omega'_{s}$ is the solid angle of overlapping region and 
$\tau'(\nu)$ is the dust opacity of the foreground ISM component.

\subsection{A.2 The $\beta$3D: the Extended Escape Probability Method} 
\label{appendix: the extended escape probability method}
The level populations of molecular gas are determined by collisions and radiation
through the equations of statistical equilibrium:
\begin{equation}
	n_{u}\sum_{l}(A_{ul}+B_{ul}\langle J_{\nu}\rangle_{ul}+C_{ul})=
	\sum_{l} n_{l} (A_{lu}+B_{lu}\langle J_{\nu}\rangle_{lu}+C_{lu}) ,
\label{equation:statistical equilibrium equations}
\end{equation}
where $l=45$ which is the total number of levels included, $A_{ul}$ and 
$B_{ul}$ are Einstein coefficients, and $C_{ul}$ are collisional
excitation ($u<l$) and de-excitation ($u>l$) rates.
The collision rates of para- and ortho- ${\rm H_2O}$ with
 para- and ortho- ${\rm H_{2}}$ (whose ortho-to-para ratio
in thermal equilibrium is estimated by Equation (1) in 
\citet{mumma1987}) 
have been taken from LAMDA database \citep{schoeier2005},
which are originally calculated by \citet{daniel2011}.

The profile-averaged mean radiation intensity, $\langle J_{\nu}\rangle_{ul}$, is given by:
\begin{equation}
	\langle J_{\nu}\rangle_{ul}=(1-\epsilon_{ul})S_{L}
	+(\epsilon_{ul}-
		\eta_{ul})B(T_{d})+\eta_{ul}B(T=2.7~K),
\label{equation:Jv}
\end{equation}	
where $S_{L}$ is the source function, $B(T_{d})$ is the Planck function
at the dust temperature $T_{d}$ and $B(T=2.7~K)$ is the 2.7\,K cosmic microwave
background. The method has introduced two profile-averaged photon
escape probabilities $\epsilon_{ul}$ and $\eta_{ul}$. 
The $\epsilon_{ul}$ is the probability that a photon
escapes line absorption, and $\eta_{ul}$ is the probability that a photon
escapes dust absorption as well as line absorption and therefore contributes
to the observed line emission by a distant observer.
The first term of the right-hand side of the Equation\ \ref{equation:Jv} can be explained as the
contribution of line photons, while the second term as the contribution of 
dust photons, and the third term as the amount of external radiation 
reaching the test point.

Under the Large Velocity Gradient (LVG) assumption and our assumption 
that the excitation of the molecular
gas at a given point is only connected with the gas and dust of the same clump, 
the two escape probabilities 
$\epsilon_{ul}$ and $\eta_{ul}$ can be simplified into the following forms:
\begin{equation}
	\epsilon_{ul}=\frac{\kappa_{d}}{\kappa_{L}+\kappa_{d}}+
	\frac{\kappa_{L}}{\kappa_{L}+\kappa_{d}}\int\frac{d\Omega}{4\pi}[
	\frac{1-e^{-(\tau_{L}+\tau_{d})}}{\tau_{L}+\tau_{d}}],
\end{equation}
\begin{equation}
	\eta_{ul}=\int\frac{d\Omega}{4\pi}
    [\frac{1-e^{\tau_{L}}}{\tau_{L}}]e^{-\tau_{d}},
\label{equation:eta}	
\end{equation}
where $\kappa_{L}$ is the line absorption coefficient, 
$\kappa_{d}$ is the dust absorption coefficient 
with a form of $\kappa_{d}(\nu)=0.4(\nu/250~{\rm GHz})^\beta$ in unit of ${\rm
cm^2~g^{-1}}$ \citep{priddey2001}, 
$\tau_{L}$ and $\tau_{d}$ are the line and dust optical depths, respectively,
from the test point to the edge of the clump along a line of sight.
The $\kappa_{L}$ is calculated by a form of 
$\kappa_{L}=[(n_{l}B_{lu}-n_{u}B_{ul})\frac{h\nu_{ul}}{4\pi}]\varphi(\nu)$,
where $\varphi(\nu)$ is the normalized absorption line profile 
contributed by turbulence and thermal speed of gas:
\begin{equation}
\varphi(\nu)=\frac{1}{(\Delta\upsilon_{D}\frac{\nu_{0}}{c})\sqrt{\pi}}
e^{-(\nu-\nu_{0})^2/(\Delta\upsilon_{D}\frac{\nu_{0}}{c})^2},
\end{equation}
where $\nu_{0}$ is the line rest frequency,  
$\Delta\upsilon_{D}$ is the Doppler velocity width in a form of
$\Delta\upsilon_{D}=(\upsilon_{turb}^2+\frac{2k_{\rm B} T_{\rm k}}
{\mu})^{1/2}$ ($\upsilon_{turb}$ is the turbulence velocity, $k_{\rm B}$ is  the 
Boltzmann constant, $\mu$ is the molecular weight and
$T_{\rm k}$ is the gas kinetic temperature).
The turbulence velocity $\upsilon_{turb}$ used to calculate the Doppler 
velocity width is estimated from molecular
cloud linewidth-size scaling relations where a $3\times$
larger normalization value is adopted (i.e., $\sigma/R^{0.38}=3.3$), 
since starburst galaxies are often reported to deviate the local classical molecular
cloud scaling relations (Larson's relations) 
\citep[e.g.,][]{swinbank2011, kruijssen2013}.
The grid size of clumps ($ngx \times ngy \times ngz$) are fixed to
 $20 \times 20 \times 20$, but the corresponding physical sizes of grid cell and 
clump ($R$) are determined by the column density 
$N_{\rm clump}$(H) of a clump and gas density n(H).
In the absence of dust emission and absorption, 
we obtain the usual expression for the escape probability
\citep[e.g.][]{castor1970, hollenbach1979}.
Finally, the critical density is calculated by:
\begin{equation}
n_{\rm cr} = \frac{\sum_{u>l}\epsilon_{\rm ul}A_{\rm ul}}
{\sum_{u>l}C_{\rm ul}}.
\label{equation:ncr}
\end{equation}
Due to the large Einstein $A$ coefficients, the critical densities of water
lines are usually in the order of $10^8 - 10^9~{\rm cm^{-3}}$ under
the optically thin case ($\epsilon_{\rm ul} \simeq 1$) \citep{poelman2007a}.

Note that the angle dependence in the above equations is replaced by
the summation over a fixed number of directions in our model, i.e.,
$\int\frac{d\Omega}{4\pi}=\sum_{\kappa=1}^{N}$.
The number of directions is arbitrary, but a 6-ray approximation is
implemented, i.e., $N=6$, 
to represent the six different orthogonal directions in
a three-dimensional Cartesian grid.
The line and dust optical depths ($\tau_{L}$ and $\tau_{d}$) are calculated from the 
test point up to the edge of clump cube in each of six directions.
Hence, the probability for a photon to escape is connected through
all the grid points along six directions.
Equation\ \ref{equation:statistical equilibrium equations} - \ref{equation:eta}
constitute the core of our method to calculate level populations 
including dust radiation.

\subsection{A.3 The Ray Tracing Approach} 
\label{appendix: the ray tracing approach}

The ISM component, which is an ensemble of uniform clumps in a rectangular box,
forms the basic unit in our ray tracing.
We perform the ray tracing in a three-dimensional Cartesian grid,
with $z-$axis along the line of sight.
For each pixel ($x$, $y$) in the projected surface (i.e., $xy-$plane at $z=0$) ,
we calculate the line and dust continuum fluxes emerging from this pixel
at different frequency channels by:

\begin{equation}
	I(x,y)=\sum_{z'=1}^{Z}\Lambda(x,y,z')exp[-\sum_{i=1}^{z'-1}
	(\tau_{L}(x,y,i)+\tau_{d}(x,y,i))]
\label{equation:Iv}
\end{equation}	

\begin{equation}
	\Lambda(x,y,z')=[\frac{\kappa_{L}} {\kappa_{L}+\kappa_{d}}S_{L}(x,y,z')
	+\frac{\kappa_{d}}{\kappa_{L}+\kappa_{d}} B(T_{d}(x,y,z'))]
	(1-e^{-(\tau_{L}(x,y,z')+\tau_{d}(x,y,z'))})
\label{equation:Lambda}
\end{equation}	
where $z'$ labels all the grid cells in backward order along the ray,
and $i$ indicates grid cells locating in front of the $z'$-th grid cell
in backward order.
The $\tau_{L}(x,y,z')$ and $\tau_{d}(x,y,z')$ indicate integrated line and dust optical depth
of a single grid cell at ($x,y,z'$), respectively.
And $Z$ is the total number of grid cells along the
$z-$direction, which is determined by the clump 
size ($ngz = 20 $) and
number of clumps along the line of sight
(i.e., $Z$ = $N(H)/N_{\rm clump}(H) \times ngz$). 
A simple physical interpretation of Equation \ref{equation:Iv} -
\ref{equation:Lambda} is that $\Lambda(x,y,z')$ equals the 
amount of photons generated by molecular gas and 
dust at the grid cell ($x,y,z'$), and $exp[-\sum_{i=1}^{z'-1} 
(\tau_{L}(x,y,i)+\tau_{d}(x,y,i))]$ is the probability
that a photon at grid cell ($x, y, z'$) escapes both dust and line absorption 
along the $z-$direction.

We attribute the clumps inside an ISM component a random normal distributed 
global velocity field following the parameters derived by Gaussian decomposition of 
the H$_2$O spectra.
The final output of our ray tracing approach is a 3D line
data cube, with the first two dimensions along the spatial $xy$-plane
and the third dimension along frequency (or velocity).
We then integrate the line cube along the $xy$-plane to derive a
single global spectrum.
In our ray tracing approach, the surface size (i.e., the area along the $xy-$plane)
of an ISM component is not important,  since the modelled line intensity per unit area and 
the global line shape will not change with surface size as long as 
the given global velocity field can be fully sampled.
We determine the physical size ($R_{\rm s}$) of an ISM component
by scaling the modelled line intensity per unit area to the observed value. 

The spectra from multiple ISM components
is simply the sum from individual ISM components
if they do not spatially overlap.  Otherwise, we need to integrate the
emission through all the gas and dust of overlapped ISM components
(i.e., $Z = \sum_{j=1}^{n}Z_{j}$ in Equation.\ \ref{equation:Iv}, $n$ is 
the total number of overlapped ISM components),
in order to consider the line and dust absorption of the front ISM component
against the background ISM component.

\subsection{A.4 Select Best Models} 
\label{appendix: select best models}

In summary, our galaxy (nucleus) model is a 
combination of multiple ISM components.
An ISM component is constrained by six parameters: $N_{\rm clump}({\rm H})$, 
$n({\rm H})$, $T_{\rm k}$, $T_{\rm d}$, $X_{\rm H_{2}O}$ 
and $N$(H).
The first five parameters constrain the excitation condition of the
molecule (${\rm H_{2}O}$ and CO).
The last parameter $N$(H), together with the parameters (line centres and line widths) 
derived from Gaussian decomposition of our ${\rm H_{2}O}$ spectra, 
further determines the integrated line intensity 
and global line shape.

We model each velocity component of the ${\rm H_{2}O}$ spectra
separately.
For each velocity component, we derive the first and second ISM components
by fitting the medium-excitation and ground-state/low-excitation
emission lines, respectively.
Additional ISM components are then added 
to fit the ground-state/low-excitation absorption lines and/or
high-excitation lines where necessary.
We first remove the underlying continuum and fit
only the ${\rm H_{2}O}$ line intensities 
(negative integrated areas are fitted in the cases of absorption lines).
The best-fit parameters for an ISM component
are searched by minimizing the sum of chi-squared residuals.
We then do a second pass to  discard the models
that significantly overestimate ($\ge 120\%$)  the observed total
continuum fluxes and CO ($J_{\rm up} = 1 -13$) fluxes.

We utilize the ISM component dominating the FIR luminosity (normally the 
warm component derived by fitting medium-excitation
lines) as the background continuum source of the ground-state/low-excitation 
absorptions. For each derived possible model of background warm component 
(or background hot component in the case of Arp 220),
we search for the best-fit models of the absorption component
by allowing its physical size to vary from zero to the maximum coverage.
It's worth mentioning that some models of the warm component (or hot 
component in the case of Arp 220)
fail to produce satisfactory absorptions that match the observations, 
and we have discarded these models. 

Since our models of individual ISM components
are derived by fitting only certain ${\rm H_{2}O}$ transitions,
they may significantly overestimate or underestimate 
line intensities of the other transitions.
Therefore, we add up the modelled ${\rm H_{2}O}$ spectra from all the sets of ISM 
components at different velocities, and do a third pass by discriminate 
the combined models whose predicted global ${\rm H_{2}O}$ spectra
show large discrepancies with the observations.
The combined models with predicted global dust continuum or CO line 
intensities exceeding $120\%$ of the observed values were also excluded. 
The derived final parameters of our models are
given in Table.\ \ref{table: model parameters}.

\section{B. Detailed Models for Individual Galaxies} 
\label{appendix: detailed models for individual galaxies}

\subsection{M82}
M82 is a nearby, almost edge-on galaxy ($i \simeq 77$ deg), notable for its 
spectacular bipolar outflow \citep[e.g.,][]{kamenetzky2012}.
High-resolution CO maps of the central $\sim 1$ kpc disk indicate that the 
molecular gas is largely concentrated in three areas: a north-east lobe, south-west lobe, 
and to a lesser extent,  a central region \citep{weiss2001, weiss2010, walter2002}.
Overall, M82 shows weaker ${\rm H_{2}O}$ line emission relative to CO.
For example, the CO($J=3-2$)/${\rm H_{2}O} (2_{11}-2_{02})$ ratio in M82 is much larger 
($\simeq 40$) compared to that of NGC 4945 ($\simeq 5.5$) and NGC 253 ($\simeq 5.6$).
Our models suggest that the faintness of H$_2$O in M82 is mainly due to 
the lower ${\rm H_{2}O}$ abundance 
(see Table.\ \ref{table: model parameters}). 
The blue-shifted and red-shifted ${\rm H_{2}O}$ emissions are associated
with the south-western ($\delta v=-73~{\rm km~s^{-1}}$) and
north-eastern ($\delta v=86~{\rm km~s^{-1}}$) molecular lobes,
respectively.  Our model for M82 requires a warm and
a more extended cold component to match the observed H$_2$O line
intensities, where the warm gas dominates
the medium-excitation H$_2$O lines, dust FIR continuum and 
high-$J$ ($J \geq 8$) CO lines while the cold gas mainly contributes
to the low-excitation ${\rm H_{2}O}$, the long wavelength submm dust
continuum and low/middle-$J$ ($J \leq 8$) CO emission lines.  The
blue-shifted emission from the south-western lobe is stronger than
the red-shifted emission from the north-eastern lobe. 
In our model this differences arise from
the slightly elevated dust temperature in the south-western lobe compared to the
north-eastern lobe (${\rm T_{dust} \simeq 50 - 70}$\,K vs $\simeq 40 - 50$ K).  
The absorption seen in ground-state lines occurs close to
the systemic velocity ($\delta v=26~{\rm km~s^{-1}}$) and does not
have any correspondence to the emission features seen in medium-excitation 
H$_2$O lines.  This implies that the absorptions arise from a different region
than the warm component, possibly from the front cold gas. 
 As the absorption in M82 is
observed to be much narrower and shallower than those in NGC 4945 and
NGC 253, our model suggests that the cold gas in M82 covers only a
small fraction of the continuum ($F_{cc} \le 0.25$).  In a more detailed
study of the H$_2$O absorption in M82, \citet{weiss2010} compared the
absorption profile with high spatial resolution ($3.5\arcsec$) CO
observations. They find that the CO line profile is in very good agreement with
the water absorption profile within only a small region at the galaxy center 
which associates with a strip orthogonal to the molecular disk of M82.

\subsection{NGC253}
NGC 253 is a nearby barred spiral galaxy with a compact nuclear
starburst and a weak AGN in its center \citep{muellersanchez2010,
aalto2011}. The nuclear starburst drives a $\sim$\,100 pc-scale
molecular gas outflow/wind detected in CO \citep{mitsuishi2012,
bolatto2013}. As in the case of M82 our model reveals that the
medium-excitation ${\rm H_{2}O}$ emission lines and most of
dust continuum arise from a warm ($T_{\rm dust} \simeq 50$ K), dense
($n({\rm H}) \sim 10^5 - 10^6~{\rm cm^{-3}}$) and compact ($Rs \simeq 50 -
60$ pc) region which is very likely associated with the inner nuclear
starburst disk. The line profiles of this warm gas component displays a double
Gaussian and follows qualitatively the shape of the  CO (3-2) line
profile (see Fig. \ref{figure:Comparison_CO}). It is, however,
significantly narrower than the  CO (3-2) emission profile and lacks
emission at the terminal velocity of the CO spectrum. In contrast both
ground transitions of o-H$_2$O and p-H$_2$O show emission significantly
red wards ($\delta v \approx 60~{\rm km~s^{-1}}$) of the medium-excitation
lines in good agreement with the shape of the red wing of the CO line
profile. This demonstrates convincingly that the ground-state lines
originate from a different physical volume than the
medium-excitation lines and it is tempting to speculate that
the H$_2$O ground transitions are at least partly associated with the
outflowing molecular material in NGC 253. This interpretation 
is also supported by the blue shifted ($\delta v=-60~{\rm km~s^{-1}}$) 
absorption seen in the p-H$_2$O ground transition which forms together 
with the red shifted emission a P Cygni profile. 
We have modelled the blue and red line wings of the outflow using similar physical 
parameters resembling those found for the cold gas in NGC 4945.
The two outflows generate almost equal emissions in ground-state o-${\rm H_{2}O}$ 
($1_{10}-1_{01}$) line (green dashed curves in Fig.\ \ref{figure:ngc253_spectra}).
The reason for asymmetrical line shapes found in low-excitation lines of 
NGC 253 is simply due to their different spatial geometry
with respect to the nuclear disk.
That is the red-shifted outflow locates behind the nuclear disk,
while the blue-shifted outflow resides in front of, and almost completely 
covers ($F_{\rm cc} \simeq 1$), the nuclear disk (see Fig.\ \ref{figure:galaxy_models}), 
in agreement with the high spatial
resolution CO maps from \citet[][]{bolatto2013}.
Therefore, the part of blue-shifted outflow locating right in front of the nuclear disk 
produces ground-state and low-excitation absorption lines, while the rest produces ground-state line emission.
The absorption in o-${\rm H_{2}O}$ ($1_{10}-1_{01}$) and o-${\rm H_{2}O}$ ($3_{03}-2_{12}$) 
lines happen to compensate for the blue-shifted line emission, making these two lines partly invisible.

\subsection{NGC4945}
NGC4945 is a nearby, almost edge-on, disk galaxy.  The central region
of this galaxy contains an inclined nuclear starburst disk
\citep{moorwood1996, marconi2000, chou2007} and a heavily enshrouded
AGN \citep{iwasawa1993, guainazzi2000, perez-beaupuits2011}. The
central region is obscured by a strongly absorbing rotating
circumnuclear star-forming ring seen nearly edge-on
\citep{marconi2000, spoon2003, chou2007}.  Our modelling of NGC 4945
suggests that the strong emission seen in medium-excitation
lines come from a  dense ($n({\rm H}) \sim 10^6~{\rm cm^{-3}}$),
 warm ($T_{\rm dust} \sim 50 - 60$ K) region with a size of 
 $Rs \sim 50 - 70$ pc (this total size is sum of the sizes of three 
warm components at different velocities given by Table\ \ref{table: model parameters}). 
The warm component is most likely associated with the nuclear starburst disk,
whose size has been determined to be $\sim 100$ pc from
high-resolution HST-NICMOS observations of the Pa$\alpha$ line
\citep{marconi2000} and CO interferometric maps \citep{chou2007}.
Since the warm nuclear disk material does not produce enough line
emission in the low-excitation transitions, we added another ISM component to
match these low-excitation emissions. We find the
corresponding emission component is relatively cold ($T_{\rm dust} \sim 20$
K), less dense ($n({\rm H}) \sim 10^5~{\rm cm^{-3}}$) and much more extended 
($Rs \sim 200 - 300$ pc).
By fitting the absorption features seen in ground-state/low-excitation lines,
we obtain an absorption component with similar physical properties of the
cold emission component except for its much smaller size ($Rs \sim 50 - 70$ pc).
Thus, we speculate that the cold emission and absorption components arise from the
same physical region which is very likely to be the surrounding molecular ring.  
In this picture, the ground-state emission seen at 
$\delta v=141~{\rm km~s^{-1}}$ and $\delta v=-112~{\rm km~s^{-1}}$ arise
from the edges of molecular ring, whose sight-lines do not
intercept with the warm nuclear gas and are therefore free of absorption.  
The absorption feature at $\delta v=48~{\rm km~s^{-1}}$ arises from the
cold gas of the molecular ring which is located right in front of the
warm nuclear disk.  The warm nuclear disk is almost entirely covered
by the cold gas of the molecular ring (i.e., $F_{cc} \simeq 1$), given by the deep-to-bottom 
absorption seen at the velocity center.  The terminal velocities in the low-excitation and
medium-excitation spectra are almost identical (see Fig.\
\ref{figure:ngc4945_spectra}), implying that the nucleus disk spans approximately 
the same range of velocities as the surround molecular ring.  This interpretation is 
consistent with high spatial resolution CO maps which show that the rotation curve 
flattens beyond the radius of nuclear disk ($R \sim 100$ pc) \citep{chou2007}.  
Note that the source size of the cold molecular ring (ER) given in Table\
\ref{table: model parameters} denotes the projected area at the two sides
of the molecular ring rather than the real radius of the ring structure.  The
gas kinetic temperatures are found to be close to the dust
temperatures within both the nucleus disk and molecular ring.

\subsection{NGC1068}
NGC 1068 is a nearby and bright Seyfert galaxy seen at an inclination
angle of $i \leq 40 $ deg.  Molecular gas (CO and HCN) observations
towards its center have shown a central circumnuclear disk (CND) with
a radius around $100 - 150$ pc \citep[e.g.,][]{krips2011,
spinoglio2012}, and a star-forming ring/spiral at a radius of $1.0 -
1.5$ kpc \citep{tacconi1994, schinnerer2000}. As the brightest
prototype Seyfert-2 AGN, NGC 1068 exhibits pronounced radio jets
extending out to several kpc from the center \citep{gallimore2004,
krips2006}.  Following the above modelling approach, we utilized a
warm component plus a cold component to match the
medium-excitation and low-excitation ${\rm H_{2}O}$ emission
lines, respectively.  According to the derived physical parameters,
the warm component is very likely associated with the CND, while the
cold component should be at least partly related to the star-forming
ring/spiral.  In contrast to higher inclination galaxies in our sample,
the low-excitation H$_2$O lines in NGC 1068 do not show absorption,
indicating that no cold gas is located in front of the warm CND.  The
model of above two components are able to fit most of ${\rm H_{2}O}$
lines very well, but fail to explain a large part of 
the o-${\rm H_{2}O}$ ($2_{12}-1_{01}$) and middle-$J$ CO ($ 4 \le
J_{\rm up} \le 8$) line intensities which are very luminous in NGC
1068.  Therefore, we added another component to match the remaining
intensities of these lines.  We find that this additional component
contains a substantial amount of relatively low density 
($\sim 10^4~{\rm cm^{-3}}$), low opacity ($N_{\rm H} \sim 1 \times
10^{23}~{\rm cm^{-2}}$) but very hot gas ($T_{\rm k} \sim 120 - 180$ K)
within a radius of $\sim 150 - 200$ pc.  Therefore, we speculate that
this gas is probably associated with the part of molecular gas in the CND which
is shock heated to high kinetic temperatures 
as a consequence of an interaction between the radio jet and
the CND \citep{krips2011}.  The strong interaction between the radio
jet and the CND has been indicated by many recent observations: the CO
and HCN line observations display a complex kinematic behaviour of the gas
\citep[e.g.,][]{krips2011}; MIR observations reveal hot and
ionized gas following the orientation of the radio jet
\citep[e.g.,][]{poncelet2008,mullersanchez2009}; and X-ray
observations show that clouds in many locations are strongly affected
by shock heating \citep[e.g.,][]{wang2012}.  The strong collisional
excitation of the shocked gas leads to enhanced emission of 
o-${\rm H_{2}O}$ ($2_{12}-1_{01}$) line, and line ratios of CO middle$-J$ 
to $1-0$ \citep[e.g., CO ($3-2$)/($1-0$), ][]{wang2014} in NGC 1068.
With this supplemented component of hot shocked gas, we are able to
fit all the observed ${\rm H_{2}O}$ lines, dust SED and CO SLED very
well.

\subsection{Centaurus A, NGC5128}
Cen A is the closest giant elliptical and a powerful radio galaxy
\citep[for a review, see][]{israel1998}. Cen A exhibits a powerful and
variable radio nucleus \citep{meisenheimer2007, israel2008}.
Observations of molecular gas towards its central region ($\leq 3$ kpc) 
has revealed several different prominent components: a nuclear disk ($\sim 30$
pc) containing ionized and molecular gas, a compact circumnuclear molecular
disk (CND, within the inner $400$ pc), as well as an even larger extended thin
molecular disk \citep{espada2009, espada2013, israel2014, salome2016}.
According to our model, the ground-state emissions
arise from a cold ($T \sim 20$ K), extended ($Rs \sim 70 - 100$ pc) 
region, which is likely related to dense gas in the CND or the outer molecular disk.
Although we have only detected one line above ground-states, 
we are still able to roughly constrain physical
parameters of warm component in Cen A  by utilizing the upper limits of those undetected 
${\rm H_{2}O}$ lines, together additional information from 
dust SED and CO SLED.
The warm component in Cen A is found to be very compact (R$_s\sim7-15$\,pc),
implying that only a small fraction of the gas is excited to high levels.

In order to model the absorption features seen in ground-state lines, we first
adopted the dust continuum of warm component as the background. With this assumption we
derived a cold ($T_{\rm dust} \simeq 20$ K) absorbing gas component
with extremely high optical depth ( $N_{\rm H} \geq 6 \times 10^{25}~{\rm
cm^{-2}}$), which we consider as unrealistic as it would result in a dust SED
inconsistent with the observation.  We noticed that 
o-${\rm H_{2}O}$ ($1_{10} - 1_{01}$) line in Cen A has a almost comparable (or even a slightly higher) 
absorption depth compared with p-${\rm H_{2}O}$ ($1_{11} - 0_{00}$) 
line, which is very unusual as the ratios of their absorption intensities are much
smaller ($10-25 \%$) in all other cases.
Considering these matters, the background source should be more likely 
the radio core rather than warm dust.  
Recent works found that at the lowest frequencies
($\leq 500$ GHz) observed with \textit{SPIRE} and \textit{HIFI}, the
continuum flux of Cen A is still completely dominated by the radio
core \citep[][and Fig.\ \ref{figure:galaxy_co_IR_summary}]{israel2014}.
We adopted an index of -0.36 \citep[$F_{\nu} \propto \nu^{-0.36}$,][]{meisenheimer2007} 
and a normalisation value of $F_{\rm 461 GHz} = 8.2$ Jy \citep{israel2014} 
to model the radio power-law spectrum.  
As the radio core is unresolved even at very high spatial
resolution \citep[$\sim 0.1 \arcsec$ or $\sim 1.8$ pc,][]{israel1998, espada2009}, 
we adopt different values from 0.01 pc to 10 pc for its physical size ($R_{\rm core}$).  
Our models suggest that the absorptions arise from either less 
dense ($\sim 10^3~{\rm cm^{-3}}$)
but high kinetic temperature ($T_{\rm k} \sim 120 - 180$ K) gas, or relatively
denser ($\sim 10^4~{\rm cm^{-3}}$) but cold ($T \sim 20 - 30$ K) gas.
The former may associate with shocked gas around the base
of the radio jet \citep{ott2013}, while the latter could be related to 
cold dense gas in the circumnuclear molecular disk or the outer
molecular gas disk. 
Our model can also explain the absorptions detected in low$-J$ CO lines
(see the top subplot in panel of Cen A 
of Fig.\ \ref{figure:galaxy_co_IR_summary}, where the observed CO data
points are from \citet{israel2014}).  
We obtain same model results for $R_{\rm core} \leq 1$ pc, however,
no satisfactory solutions can be found any more if the radio core 
becomes larger than a few pc.

\subsection{Mrk 231}
Mrk 231 is the most luminous infrared galaxy in the local universe,
with $L_{\rm TIR} \simeq 3.2 \times 10^{12}~\Lsolar$ \citep{sanders2003}.
Multi-wavelength observations reveal a QSO-like 
nucleus \citep[e.g.,][]{soifer2000, gallagher2002} and a 
compact starburst disk \citep[e.g.,][]{carilli1998, tacconi2002},
embedded in a more extended ($\sim 1$ kpc scale) diffuse 
disk \citep{taylor1999}.
Our favored model for Mrk 231 consists of a cold ($T_{\rm dust} \sim 30$ K),
extended ($R_s \simeq 1000 - 1500$ pc) diffuse halo, 
a warm ($T_{\rm dust} \sim 50 - 60$ K) starburst disk of radius 
$\sim 300-550$ pc and a highly obscured, compact ($Rs \sim 60 - 70$ pc) hot 
($T_{\rm dust} \sim 160 - 180$ K) dense ($n({\rm H}) \simeq 10^6$ ${\rm cm^{-3}}$) component.
Our derived sizes are in good agreements with iterferometric imaging 
of CO($1-0$) and ($2-1$) lines,  which shows that a starburst disk with $R_{\rm s} \simeq 500$ pc 
\citep[][]{downes1998} embedding in a $\sim$kpc extended diffuse disk
\citep[][]{taylor1999}.
Unlike normal LIRGs of our sample, Mrk 231 shows intense 
high-excitation ${\rm H_{2}O}$ features  (with $E_{\rm up} \sim 600 - 700$ K)
in both emission and absorption, which can not be 
explained by the warm component.
Our model reveals these high-excitation lines emerge from a very compact component 
with extremely high dust temperature and gas density,
which is most likely related to the AGN dominated region.
The contributions of hot component to ${\rm H_{2}O}$ line intensities
are denoted by magenta dashed curves in Fig.\ \ref{figure:mrk231_spectra}.

The observed and modelled dust SED and CO SLED of Mrk 231 are presented 
in Fig.\ \ref{figure:galaxy_co_IR_summary}.
As one can see from the figure, the bulk of far-IR continuum 
arises from the warm disk
(with $\tau_{100{\rm \mu m}} \sim 0.5 - 1.0$), while large amounts of 
mid-IR continuum fluxes arise from the hot component 
(with $\tau_{100{\rm \mu m}} \sim 1.0 - 1.5$
and $\tau_{25{\rm \mu m}} \sim 8 - 10$).
The hot component of Mrk 231 accounts for about $\sim 25\%$ of the total infrared
luminosity, but contains only $0.04\%$ of the total dust mass.
Note that the hot component should be deeply buried into the 
warm starburst disk and halo.
Otherwise, it will severely overshine dust SED in mid-IR wavelength regime.
In our model, the mid-IR infrared emission from the hot component 
is attenuated by foreground dust with $\tau_{25{\rm \mu m}} \simeq 1.2 - 2.5$.
The CO SLED of Mrk 231 displays an approximately flat distribution over 
rotational ladders above $J_{\rm up}=10$, due to the strong high-$J$ CO emissions
from the hot component. Finally, our model for Mrk 231 is broadly consistent with
the model derived by \citet[][]{GA2010}.

\subsection{The Antennae Galaxies - NGC4038/NGC4039}
The Antennae is an early-stage merger between two
gas-rich spiral galaxies NGC 4038 and NGC 4039 \citep{schirm2014}.
Its majority of molecular gas and star formation is occurring
within two nuclei and the overlap region
\citep[e.g.,][]{gao2001, schirm2014, schirm2016}.
Our \textit{HIFI} single-pointing observations were
positioned at the overlap region, which is the strongest continuum
source of the Antennae.  We do not have any detections within
this region, which we suspect is due to its low mass of dense
gas.  If assuming the overlap region has a warm component that is similar to
those found in other galaxies, we estimate 
its physical size to be around $R_{\rm s} = 40-50$ pc and dense gas 
mass around a few $10^8~\Msolar$.
The dust SED fitting predicts that its dust temperature is around $\sim 40 - 50$\,K and
dust mass around $\sim 2 \times 10^6~\Msolar$.
Our derived low mass of warm dense gas is consistent with recent studies based
on \textit{SPIRE}-FTS CO observations using a non-LTE analysis
\citep{schirm2014} and high angular-resolution observations of 
dense molecular gas (HCN and HNC) with {\it ALMA} \citep{schirm2016},
which showed that the fraction of molecular gas in dense and warm
phase in the Antennae is only around $0.2-0.3\%$.

\subsection{NGC6240}
NGC 6240 is an early-stage
merger hosting two nuclei that are separated by $\sim 1.5 \arcsec$
\citep{beswick2001}.  CO observations reveal that most of
molecular gas in NGC 6240 is located in the overlap region of 
the two nuclei \citep[e.g.,][]{iono2007, engel2010, feruglio2013b,
feruglio2013a}.  We utilized a warm component to model 
medium-excitation ${\rm H_{2}O}$ emissions (which is likely
associated with the gas concentration between the two nuclei),
and an extended low-excitation component (i.e., ER) to account for 
ground-state ${\rm H_{2}O}$ emissions.  
The latter is likely associated with the
large-scale ($\sim$ kpc) gas distribution detected in CO
\citep{downes1998}.  While the warm gas component in NGC 6240 shows
similar physical properties with those derived for other galaxies, the
low-excitation gas is inconsistent with the parameters found in other
systems but exhibits a much higher kinetic temperature (${\rm T_k}
\simeq 120 - 400$ K, see Table.\ \ref{table: model parameters}).
This high kinetic temperature is required to match the
exceptionally high line-to-continuum ratios ($L'_{\rm line}/L_{\rm IR}$) found
in ground-state ${\rm H_{2}O}$ and low-$J$ CO ($J \leq 7$) lines.
The line-to-continuum ratio ($L'_{\rm H_2O}/L_
{\rm IR}$) of ground-state ${\rm H_{2}O}$ transitions for NGC 6240 is 
around $3 \times 10^{-4}$, 
which is significantly higher than the value found e.g. for Mrk 231
($L'_{\rm H_{2}O}/L_{\rm IR} \simeq 7 \times 10^{-5}$).
The high line-to-continuum ratios of CO lines has been investigated in
several recent publications and have been attributed to galaxy-wide shocks due to 
the advance merger state of NGC 6240 \citep[e.g.,][]{meijerink2013, papadopoulos2014}.
This picture also explains the strong ${\rm H_2}$($1-0$) S(1) emission
at 2.12 $\mu$m \citep{goldader1995}, the spectacular butterfly-shaped
emission-line nebula seen in HST H$\alpha$ images \citep{gerssen2004}
and Chandra X-ray image \citep{feruglio2013a}, and the massive
molecular outflow detected by the high-resolution interferometry CO
maps \citep{feruglio2013b} and HCN/CS maps \citep{scoville2015}.  Our
model with high kinetic temperatures but normal dust
temperatures gives further support to this picture.

Note that we use the integrated fluxes from
\textit{Herschel}/\textit{SPIRE} for p-${\rm H_{2}O}$
($1_{11}-0_{00}$) and o-${\rm H_{2}O}$ ($3_{21}-3_{12}$) lines since
they are not detected by our \textit{HIFI} observations.  
\textit{Herschel}/\textit{PACS} observations show that 
high-excitation ${\rm H_2O}$ lines of NGC 6240 are much weaker than those of
Mrk 231 or Arp 220. This suggests that a significant hot component,
similar to the one in Mrk 231 or Arp 220,  may not exist in NGC 6240. 
This is in agreement with our dust SED fitting which suggests that the size 
of such a hot component, if it exists, can not be larger than $\sim 20$ pc.

\subsection{Arp 220}
Arp 220 is the nearest ultra-luminous infrared galaxy.
It is a late-stage merger with two counter-rotating disks
separated by roughly $\simeq 1\arcsec$ (or $\simeq 400$ pc),
orbiting within an extended kpc-scale 
molecular gas disk \citep[e.g.][]{scoville1997, downes1998}.
A highly dust-obscured AGN in the western nucleus of Arp 220 is 
suggested by mounting evidence \citep[e.g.][]{downes2007}.
A large reservoir of molecular gas has been observed to be concentrated
in both nuclear disks ($FWHM \simeq 200 - 250$ pc
from CO(6--5) map) \citep{scoville1997, tunnard2015, rangwala2015}.
High-resolution {\it ALMA} imaging of dense gas (CO(6--5), HCN, CS) have shown, that the
two nuclei have almost the same line centroids and velocity dispersions
(with velocity offset only being
$\Delta V \sim 20 - 100~{\rm km~s^{-1}}$) \citep{scoville2015, rangwala2015}.
Thus, we can not kinematically separate the two nuclei in our ${\rm H_{2}O}$ spectra
and thereby we model the two nuclei together as a single region. 
The predicted size and line intensities of each nucleus, therefore,
are half of our modelled values, if they are assumed to be equal. 
We note that recent high-spatial resolution
observations of H$_2$O with {\it ALMA}, however, showed that 2/3 of the high-excitation water emission is emitted by the western nucleus \citep[][]{koenig2016}. 
It is worth noting that strong self-absorption has been detected at the centers
of both nuclei, where CO $J=6-5$, HCN, and CS lines show double 
peaked velocity profiles in both nuclei \citep{scoville2015, rangwala2015, wang2016}.
So it is possible that the double peaked line shapes seen in 
our ${\rm H_{2}O}$ spectra are also due to self-absorption.
We thereby utilized a broad Gaussian emission plus a narrow absorption 
to fit the double peaked ${\rm H_{2}O}$ lines.

We first model the medium-excitation and high-excitation ${\rm
H_{2}O}$ lines of Arp 220.  We found the medium-excitation
lines come from a warm ($T_{\rm dust} \sim 40$ K) region ($N_{\rm H}
\sim 5 \times 10^{24}~{\rm cm^{-2}}$, $Rs \sim 120 - 150$ pc), which
is very likely to associate with the outer starburst regions of two nuclear disks.  
The high-excitation lines arise from a very compact ($Rs \sim 70 -
100$ pc), hot ($T_{\rm dust} \sim 100 - 180$ K) region whose column
density is extremely high ($N_{\rm H} \sim 1 \times 10^{25}~{\rm
cm^{-2}}$).  The hot component is very likely associated with the
compact AGN- or starburst-dominated region at the nuclear center. 
To match the absorptions seen in ground-state and low-excitation lines,
we add a third component in front of the hot component.
The gas absorber is located in front of the hot component rather than the
warm component because it is the hot component which dominates dust continuum
emission and displays a velocity dispersion ($235 \pm 18~{\rm
km~s^{-1}}$) comparable to the line width of the absorptions 
($226 \pm 18~{\rm km~s^{-1}}$).
Our model reveals that the ground-state/low-excitation absorptions arise from 
an ISM component with warm dust temperature ($T_{\rm dust} \sim 60 - 80$ K), high gas
temperature ($T_{\rm k} \sim 100 - 200$ K) and large column density ($N_{\rm
H} \sim 5 \times 10^{25}~{\rm cm^{-2}}$), whose line intensities 
are denoted by blue dashed curves in Fig.\
\ref{figure:arp220_spectra}.  The absorption component is not likely to arise
from the part of warm disk, given by its very distinct physical properties.
Considering its high gas temperature, the absorption component is very likely to associate
with massive outflows driving from the nuclei.  The molecular outflows
have been suggested by many molecular gas (like ${\rm OH^{+}}$, ${\rm
H_2O^{+}}$ and ${\rm HCO^{+}}$) observations, which show P Cygni
absorption line profiles in both nuclei \citep{sakamoto2009, tunnard2015}.
The last component required to account for the broad ground-state ${\rm H_{2}O}$ 
emissions of Arp 220, is a kpc-size ($Rs \sim 1000 - 1200$ pc) 
molecular disk with large amounts of cold ($T_{\rm dust} \sim 20$ K), less dense 
($\sim 10^4~{\rm cm^{-3}}$) gas. 
 
\begin{figure}[t]
\begin{center}
\includegraphics[width=0.6\textwidth]{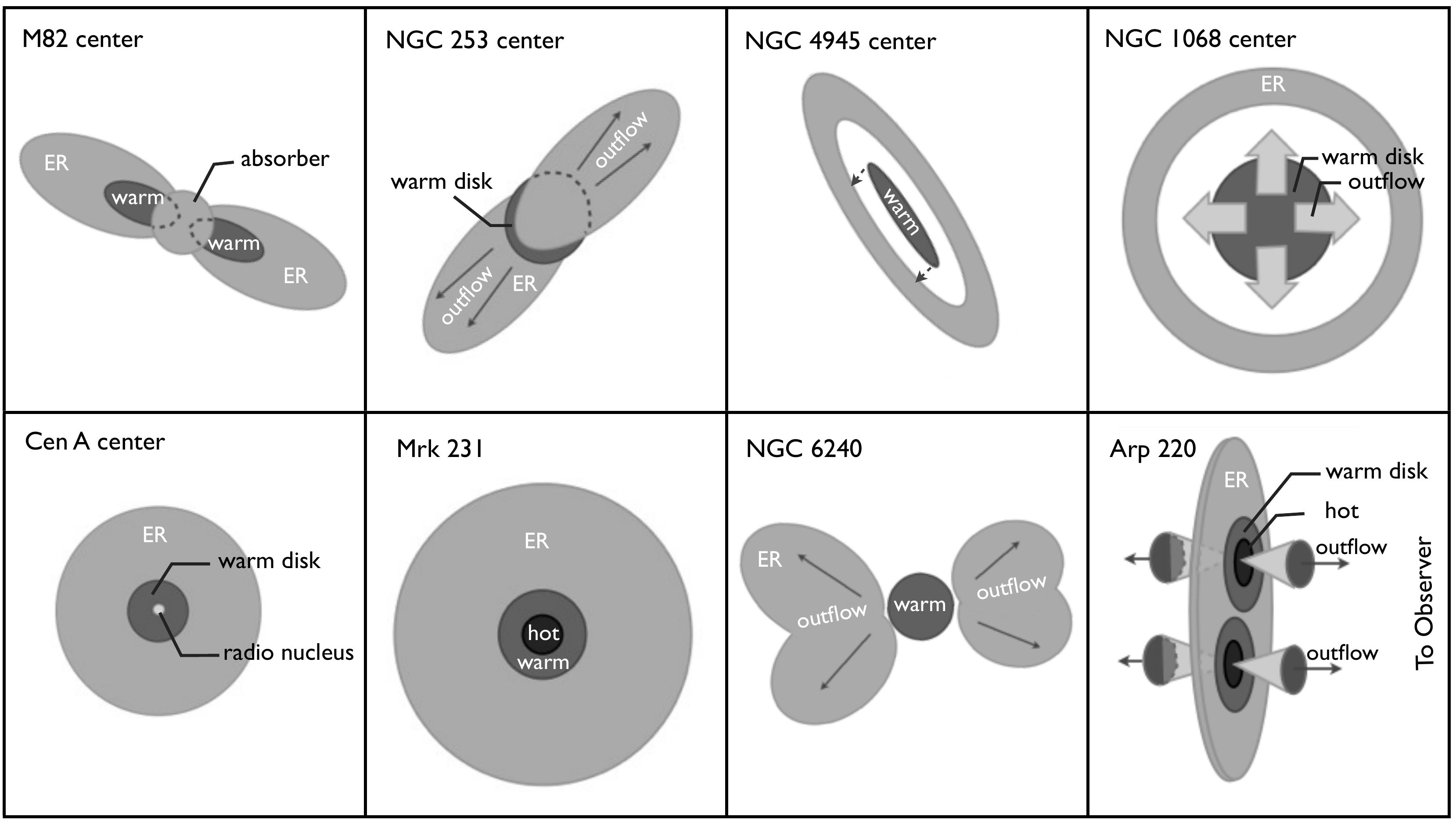}
\caption{Schematic representation of our modelled sources, showing 
possible geometries for our sample galaxies. The detailed descriptions
of each modelled source are given in the text.}
\label{figure:galaxy_models}
\end{center}
\end{figure}

Our model for Arp 220 is also capable of explaining the observed dust
SED and CO SLED (see Fig.\ \ref{figure:galaxy_co_IR_summary}).  
The massive kpc-size molecular disk, where the two nuclei are embedded in, 
contributes most of submm
continuum flux and low$-J$ ($J \leq 3$) CO emissions (green lines).
The warm component (i.e., the outer starburst regions of two nuclei) barely 
contributes to total dust continuum,
but it produces the major part of dense gas emissions such as
medium-excitation ${\rm H_{2}O}$ and CO $5 \le J_{\rm
up} \le 9$ lines (orange lines). 
The hot component and its massive molecular outflow 
dominate dust SED in FIR/MIR regime,
whose size ($D \sim 140 - 200$ pc) matches the observed 
dust continuum size ($FWHM \simeq 140 - 220$ pc) well.  
We therefore predict that dust continuum comes from a more compact region 
than dense gas emission, which has been confirmed by several 
observations \citep[e.g.,][]{rangwala2015}.  
Our model also suggests strong self-absorptions in CO $5 \le J_{\rm up} \le 8$ lines
by the molecular outflow (blue points in Fig.\ \ref{figure:galaxy_co_IR_summary}).
As the absorption intensities (blue points) are comparable to (or even larger
than) CO emissions from the hot component (magenta points),
CO $5 \le J_{\rm up} \le 8$ lines should be close to (or even go below) 
zero continuum level,  at the center region 
where hot component and its molecular outflow reside in.  
This picture corresponds well with high resolution {\it ALMA}
CO (6-5) observation which displays a double-peaked line profile
at the centers of both nuclei but shows blue/red line peaks further outside
\citep{rangwala2015}.
The high$-J$ ($J_{\rm up} \geq 11$) CO lines, however, arise mainly from the hot component
and they shine through molecular outflows without absorptions.
A schematic representation of our model for Arp 220 is given in 
Fig.\ \ref{figure:galaxy_models}. 
Finally, we compare our model of Arp 220 with that in \citet{GA2004, GA2012}.
Overall, our model predicts a roughly similar hot component,
but results in a denser, more compact and opaque warm component.  
Our model also requires a molecular outflow rather than an 
extended halo \citep{GA2004, GA2012} to match the absorptions 
seen in low-excitation lines. 

\section{C. Water Excitation Under Different Physical Conditions} 
\label{appendix: water excitation under different physical conditions}

\begin{figure}[t]
\begin{center}
\includegraphics[width=0.7\textwidth]{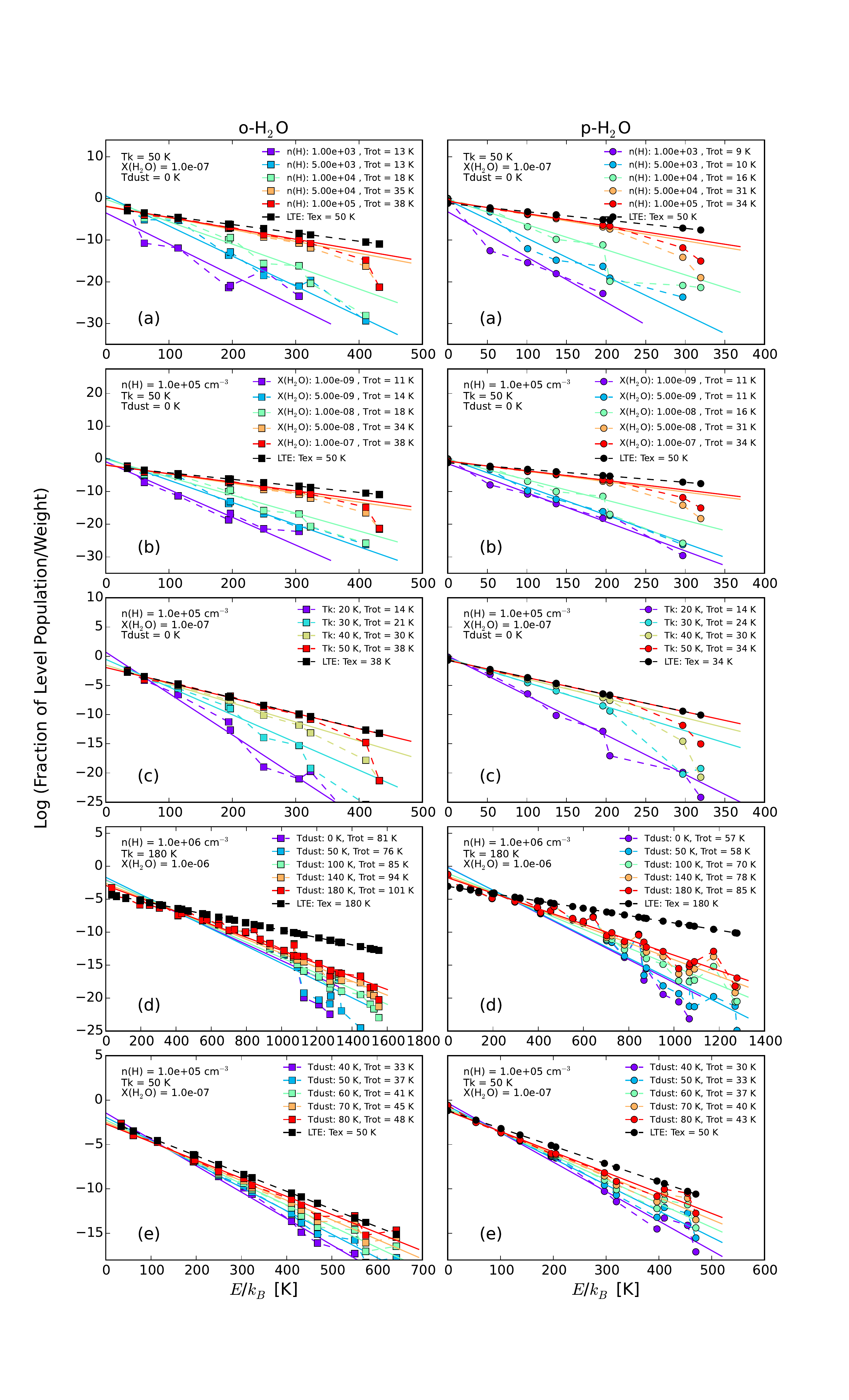}
\caption{The level populations of water under different physical conditions.
The values for gas density (n(H)), gas temperature ($T_{\rm K}$), water
abundance (X(${\rm H_2O}$)) and dust temperature ($T_{\rm dust}$) are
given in the plots.
For explanation of the symbols used in the figure, see 
Fig.\ \ref{figure:warm_Tdust_level_populations}.
}
\label{figure:level_populations_summary}
\end{center}
\end{figure}

We present the level populations of water under different physical
conditions in Fig.\ \ref{figure:level_populations_summary}.  The
subplot (a - b) in Fig.\ \ref{figure:level_populations_summary} show
that water gas is sub-thermally excited when gas density is not
exceeding $1 \times 10^4~{\rm cm^{-3}}$ and water abundance is smaller
than $1 \times 10^{-8}$.  The subplot (c) shows that the excitation
temperature of water increase with increasing kinetic temperatures
since the low/moderate-excited levels are mainly collisionally excited
under the physical conditions of $n(H) \simeq 1 \times 10^5~{\rm
cm^{-3}}$ and $X({\rm H_2O}) \simeq 1 \times 10^{-7}$.  Once the gas
density, kinetic temperature and water abundance increase to extremely
large values ($n({\rm H}) \ge 10^6~{\rm cm^{-3}}$, $T_{\rm k}
\ge 100$\,K and $X({\rm H_2O}) \ge 10^{-6}$) as the case of hot component
found in ULIRGs, the water level population can be thermally excited up to 
$E/k_{\rm B} \simeq 500 - 800$\,K as shown in subplot (d).
When the dust temperature is larger than the kinetic temperature, the thermalized
levels start to be overpopulated by IR pumping, i.e., IR pumping
starts to play a role in exciting the thermalized lines.
For example, the level populations of o-${\rm H_2O}$ 
with $E_{\rm up} \sim 200 - 300$\,K 
(and $E_{\rm up} \sim 100 - 200$\,K for p-${\rm H_2O}$) in the warm component
will not vary with the dust temperature if $T_{\rm dust} \le T_{\rm k}$ (see 
Fig.\ \ref{figure:warm_Tdust_water_excitation}).
However, once $T_{\rm dust} > T_{\rm k}$,  more and more
water will be populated within these levels with increasing dust temperature 
(see subplot (e) in Fig.\ \ref{figure:level_populations_summary}).

\clearpage

\bibliographystyle{apj}
\bibliography{references}
\clearpage
\end{document}